\documentclass[a4paper,fleqn,usenatbib]{mnras}
\pdfoutput=1

\usepackage[T1]{fontenc}
\usepackage{ae,aecompl}
\usepackage{listings}

\usepackage{graphicx}    
\usepackage{amsmath}    
\usepackage{amssymb}    
\usepackage{color}      
\usepackage{float}
\usepackage{caption}
\usepackage{subcaption}
\usepackage{lscape}

\usepackage{newtxtext,newtxmath}
\graphicspath{{./Graphs/}}

\def\Msol{\mathrel{\rm M_{\odot}}} 
\def\Msolyr{\mathrel{\rm M_{\odot}yr^{-1}}} 

\definecolor{purple}{RGB}{175,0,175}
\definecolor{red}{RGB}{255,0,0}
\definecolor{darkblue}{RGB}{0,0,175}
\definecolor{lime}{RGB}{0,255,0}

\title[Re-examining evidence for star-formation suppression in QSOs]{The impact of ionised outflows from z$\sim$2.5 quasars is not through instantaneous in-situ quenching: the evidence from ALMA and VLT/SINFONI}

\author[J. Scholtz, et al.]{\parbox[h]{\textwidth}{ 
J.\ Scholtz$,^{\! 1,2}$\thanks{E-mail: honzascholtz@gmail.com}
C.M. Harrison,$^{\! 3}$\thanks{E-mail: christopher.harrison@newcastle.ac.uk}
D.J. Rosario,$^{\! 2}$
D.M. Alexander,$^{\! 2}$
K.K. Knudsen,$^{1}$
F. Stanley,$^{1}$
Chian-Chou Chen,$^{4}$
D. Kakkad,$^{5}$
V. Mainieri,$^{6}$
J. Mullaney$^{7}$
}
\\
\\
$^{1}$ Chalmers University of Technology, Department of Space, Earth and Environment, Onsala Space Observatory, 439 92 Onsala, Sweden\\
$^{2}$ Centre for Extragalactic Astronomy, Durham University, Department of Physics, South Road, Durham, DH1 3LE, UK\\
$^{3}$ School of Mathematics, Statistics and Physics, Newcastle University, Newcastle upon Tyne, NE1 7RU, UK \\
$^{4}$ Academia Sinica Institute of Astronomy and Astrophysics (ASIAA), No. 1, Sec. 4, Roosevelt Rd., Taipei 10617, Taiwan \\
$^{5}$ European Southern Observatory, Alonso de Cordova 3107, Vitacura, Casilla 19001, Santiago de Chile, Chile \\
$^{6}$ European Southern Observatory, Karl-Schwarzschild-Strasse 2, Garching bei München, Germany \\
$^{7}$ Department of Physics and Astronomy, University of Sheffield, Sheffield, S3 7RH, UK\\
}

\date{XYZ}

\pubyear{2021}

\begin{document}
\label{firstpage}
\pagerange{\pageref{firstpage}--\pageref{lastpage}}
\maketitle

\begin{abstract}
We present high-resolution ($\sim$2.4\,kpc) ALMA band 7 observations (rest-frame $\lambda \sim 250\mu$m) of three powerful z$\sim$2.5 quasars ($L_{\rm bol}=10^{47.3}$-$10^{47.5}$ ergs s$^{-1}$). These targets have previously been reported as showing evidence for suppressed star formation based on cavities in the narrow H$\alpha$ emission at the location of outflows traced with [O~{\sc iii}] emission. Here we combine the ALMA observations with a re-analysis of the VLT/SINFONI data to map the rest-frame far-infrared emission, H$\alpha$ emission, and [O~{\sc iii}] emission. In all targets we observe high velocity [O~{\sc iii}] gas (i.e., W80$\sim$1000--2000\,km\,s$^{-1}$) across the whole galaxy. We do not identify any H$\alpha$ emission that is free from contamination from AGN-related processes; however, based on SED analyses, we show that the ALMA data contains a significant dust-obscured star formation component in two out of the three systems. This dust emission is found to be extended over $\approx$1.5--5.5\,kpc in the nuclear regions, overlaps with the previously reported H$\alpha$ cavities and is co-spatial with the peak in surface brightness of the [O~{\sc iii}] outflows. In summary, within the resolution and sensitivity limits of the data, we do not see any evidence for a instantaneous shut down of in-situ star formation caused directly by the outflows. However, similar to the conclusions of previous studies and based on our measured star formation rates, we do not rule out that the global host galaxy star formation could be suppressed on longer timescales by the cumulative effect of quasar episodes during the growth of these massive black holes.

\end{abstract}




\section{Introduction}\label{QSO_Intro}

It is now accepted that inside every massive galaxy resides a supermassive black hole (\citealt{Kormendy13}). During the growth of these supermassive black holes, via accretion events, these objects become visible as active galactic nuclei \citep[AGN; ][]{Soltan82,Merloni04}. Current theoretical models of galaxy evolution require AGN to inject significant energy into their host galaxies in order to reproduce the basic properties of local galaxies and the intergalactic medium (IGM), such as the steep mass function, the black hole--spheroid relationships, increased width of specific star formation rate distributions as a function of stellar mass, galaxy sizes, galaxy colour bi-modality, AGN number densities and enrichment of the IGM by metals \citep[e.g.,][]{Silk98,DiMatteo05,Alexander12,Dubois13,Dubois13b,Vogelsberger14,Hirschmann14, Crain15,Segers16,Beckmann17,Harrison17,Choi18,Scholtz18}. In these models, the AGN are required to regulate the cooling of the interstellar medium (ISM) and/or intracluster medium (ICM) or to eject gas out of the galaxy through galactic wide outflows. This process is usually referred to as ``AGN feedback'' and it is believed to regulate the rate at which stars can form. However, from an observational perspective, it is still not clear what role AGN play in regulating star formation in the overall galaxy population, especially at high redshift and for the most radiatively powerful AGN called quasars \citep[e.g.,][]{Harrison17,Cresci18}. 

Over the past decade, many studies have focused on identifying and characterising multiphase outflows in galaxies \citep[see e.g.,][]{Fiore17,Harrison18,Cicone18,Veilleux20}. Indeed, there is now significant evidence that energetic ionised, atomic and molecular outflows are commonly found in AGN host galaxies across a wide range of cosmic epochs \citep[e.g.][]{Veilleux05,Morganti05, Ganguly08,Sturm11,Cicone12,Harrison12,Cicone14,Zakamska14,Balmaverde15,Carniani15,Brusa15,Harrison16,Woo16,Leung17,Brusa18,ForsterSch18b,Lansbury18,Fluetsch19,Perna19,RamosAlmeida19,Husemann19}. Furthermore, these AGN-driven outflows have been identified on a large range of spatial scales \citep[between tens of parsecs to tens of kiloparsecs; e.g.,][]{StorchiBergmann10,Veilleux13,Carniani15,Cresci15,Feruglio15,Kakkad16,McElroy16,Rupke17,Jarvis19, Kakkad20}. Given that these outflows can be located on these scales, they may have the potential to impact upon the ISM and, consequently, the star formation inside the host galaxies.

Despite observations showing that AGN outflows are common, and can extend over large scales, the impact that they have on star formation is still open to debate. For example, statistical, non spatially resolved, studies in the literature provide contradictory conclusions on the relationship between ionised outflows and the star formation rates of the host galaxies \citep[][]{Balmaverde16,Wylezalek16,Woo17,Woo20}. In many of the spatially-resolved studies, the most powerful outflows are capable of removing or destroying at least some of the star-forming material at a rate faster than it can be formed into stars \citep[e.g., see][]{Fiore17,Harrison18}. However, in the majority of observations, there are considerable uncertainties in these calculations due to uncertain spatial extents and the many assumptions that have to be considered to convert emission-line luminosities and velocities into mass outflow rates \citep[e.g.][]{Karouzos16,VillarMartin16,Husemann16,Rose18,Harrison18,Davies20}. On the other hand, measurements can be more accurate for more nearby sources \citep[e.g.][]{Baron18,Revalski18,Venturi18,Fluetsch19,Perna20} and star formation has been detected {\em inside} outflows in some local AGN host galaxies, which may be a form of `positive' feedback \citep{Maiolino17,Gallagher19,Perna20}. However, samples of nearby sources typically lack the most powerful AGN, which are more common at high redshift (i.e., $z\gtrsim$1) and are thought to be most important for influencing galaxy evolution \citep[e.g.,][]{McCarthy11}.  

At high redshift, observational evidence for a direct impact by AGN on star formation is particularly scarce. Indirect evidence may have been found in some AGN host galaxies with depleted molecular gas reservoirs; however, there are contradictory claims on this topic in the literature \citep[e.g.][]{Kakkad17,Perna18,Kirkpatrick19, Fogasy20}.
A potentially powerful approach to determine the impact of AGN-driven outflows, which has been applied to high-redshift AGN studies, has been to use integral field units (IFUs) to map narrow H$\alpha$ emission to trace star formation and to map [O{\sc iii}] emission to trace AGN outflows. Most relevant for this work are the studies of four high-redshift AGN \citep[presented in][]{Canodiaz12,Cresci15,Carniani16} where it was suggested that there was no star formation at the location of the [O~{\sc iii}] outflows and potentially increased star formation along the edges the outflows \citep[see][for similar conclusions based on a $z$=1.4 radio-loud quasar]{Vayner17}. 

In \citet{Scholtz20} we demonstrated some of the challenges in interpreting the aforementioned results of suppressed H$\alpha$ at the location of [O~{\sc iii}] outflows. For example, although narrow H$\alpha$ emission can be used as star formation tracer, \citep[e.g.][]{Hao11, Murphy11}, it is sensitive to dust obscuration and can be contaminated by AGN photo-ionisation. Indeed, most of the star formation in high-redshift galaxies is obscured \citep{Madau96,Casey14,Whitaker14} and, although obscuration corrections can help \citep[][]{AlaghbandZadeh16}, sometimes the H$\alpha$ emission can be {\em completely} hidden (e.g., \citealt{Hodge16,Chen17,chen20}). This dust absorbed emission is re-radiated at far-infrared (8--1000\,$\mu$m; FIR) wavelengths and, consequently, the FIR emission is sensitive to on-going {\em obscured} star formation \citep[for reviews see][]{Kennicutt12, Calzetti13}. Importantly for this work, high-redshift AGN and quasar host galaxies have been shown to host significant levels of star-formation obscured by dust \citep[e.g.,][]{Whitaker12, Burgarella13, Stanley15, Stanley18}.

Using additional ALMA continuum observations, \citet{Scholtz20} re-analysed the source from \citet{Cresci15} with the addition of 7 other moderate luminosity X-ray AGN at z=1.5-2.5 (L$_{\rm X}$=$10^{43}$--$10^{45.5}$ ergs s$^{-1}$). We found no direct evidence for [O~{\sc iii}]-traced outflows suppressing or enhancing star formation. Nonetheless, the targets in that work were moderate luminosity AGN (i.e., L$_{\rm bol}$=$10^{44}$--$10^{46.5}$ ergs s$^{-1}$, assuming an X-ray to bolometric conversion factor of 10) and do not represent the most powerful sources, or those with the most extreme outflows. Therefore, in this current work we perform similar analyses, i.e., combining ALMA measurements of the rest-frame far infrared emission with IFU data, on the very powerful z$\sim$2.4 quasars (i.e., L$_{\rm bol}\sim10^{47.5}$ ergs s$^{-1}$) presented in \citet{Canodiaz12} and \citet{Carniani16}. These objects were pre-selected to be bolometrically luminous and host high velocity [O~{\sc iii}] outflows. These studies reported cavities in narrow H$\alpha$ emission, that they associated with star formation, at the same spatial location of the extreme [O~{\sc iii}] outflows, interpreting these results as evidence of negative AGN feedback in the location of the outflow and positive feedback on the edges of the outflow.

In \S2 we describe our targets, and the observations used in our study, \S3 describes the data analyses of the ALMA and IFU observations, including spectral fitting and constructing FIR, narrow H$\alpha$ emission and outflow maps.
In \S4 we present and discuss our results and in \S5 we draw our conclusions. 
In all of our analyses we adopt the cosmological parameters of
$\rm H_0 = 67.3 \ km \, s^{-1}$, $\rm \Omega_M = 0.3$, $\rm
\Omega_\Lambda = 0.7$ \citep{Planck13} and assume a \citet{Chabrier03} initial mass
function (IMF). We publish all the scripts for the data analyses \href{https://github.com/honzascholtz/Three_QSOs}{here}.

\section{Target description and Data}
The primary objective of this work is a study of three quasars in which earlier studies have claimed that AGN-driven outflows suppress in-situ host galaxy star formation. We use new ALMA band 7 observations to obtain FIR continuum images which are compared to the distribution of narrow H$\alpha$ and [O~{\sc iii}] emission derived from archival VLT/SINFONI observations. In \S\,\ref{sec:Sample} we describe the selection of our sample, in \S\,\ref{sec:ALMA_data} and \S\,\ref{sec:IFU_data} we describe the ALMA and IFU data we used in the analyses.

\subsection{Target description}\label{sec:Sample}

We selected three z$\sim$2.5 type 1 quasars for this work: 2QZJ002830.4-2817 from \citet{Canodiaz12} and; LBQS0109+0213 and HB89 0329-385 from \citet{Carniani16}. We note for the latter two objects the original [O~{\sc iii}] analyses were presented in \citet{Carniani15}. We will refer to these objects in this work as 2QZJ00, LBQS01 and HB8903, respectively. Originally, 2QZJ00, LBQS01 and HB8903 sources were selected as quasars with large [O~{\sc iii}] equivalent widths ($>$ 10\AA \space in the rest frame) and bright in $H$-band ($<$ 16.5 mag) from a sample of \citet{Netzer04} and \citet{Shemmer04} (for more information see \citealt{Carniani15}). These selection criteria were designed to select the most bolometrically luminous quasars during the peak epoch of galaxy and black hole growth \citep[e.g.,][]{Madau14} that could easily have their [O~{\sc iii}] kinematics mapped with an IFU.  Furthermore, the high equivalent width [O~{\sc iii}] lines can be easily de-blended from the emission associated with the broad line region (H$\beta$ and FeII emission; see \S~\ref{sec:Gal1D}). 

We compiled the black-hole masses and bolometric luminosities from \citet{Carniani15}. The black-hole masses are in the range of $10^{9.9}-10^{10.1}$ M$_{\odot}$, with bolometric luminosities of $10^{47.3}$--$10^{47.5}$ ergs s$^{-1}$, placing them at the extreme end of the quasar population (these values are discussed further in Section~\ref{sec:qso_implications}). The objects' names, sky coordinates, redshifts, black-hole masses and bolometric luminosities are summarised in Table \ref{Table:Sample_qso}.

\begin{figure}
    \centering
    \includegraphics[width=1.05\columnwidth]{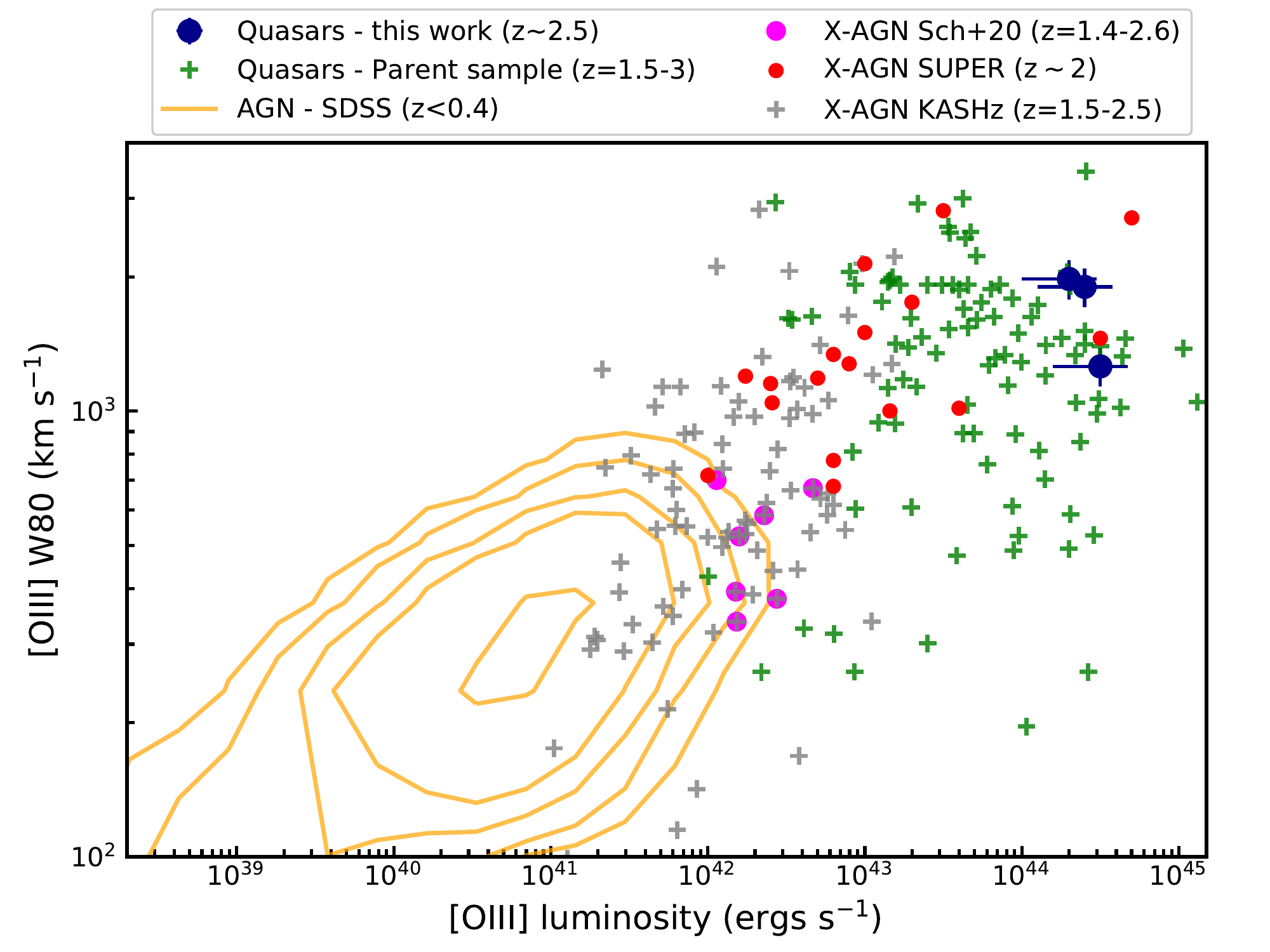}
   \caption{ Plot of [O~{\sc iii}] emission-line width (W80) versus [O~{\sc iii}] luminosity. We compare the quasars in this study (blue circles) with quasars and AGN populations from the following works: the parent sample of these quasars (green crosses; \citealt{Shemmer04} and \citealt{Netzer04}), spectroscopically identified $z<0.4$ AGN from \citet{Mullaney13} (orange contours), X-ray AGN from the KASHz survey \citep[][grey crosses, z=1.2-2.5]{Harrison16}, Type-1 X-ray AGN from the SUPER survey (red circles, \citealt{Kakkad20}) and the X-ray AGN from \citet{Scholtz20} (magenta circles, z=1.4-2.6). The three quasars studied here represent the most extreme objects in terms of bolometric power and [O~{\sc iii}] velocities.}
   \label{fig:OIII_Lbol}
\end{figure}

Figure~\ref{fig:OIII_Lbol} places our targets within the context of the overall AGN and quasar population in terms of their [O~{\sc iii}] luminosities and [O~{\sc iii}] line widths (W80; velocity width containing 80\% of the total flux; measured in \S~\ref{sec:EL_analyses}). We compare our sources with: the parent sample of 104 $z$=1.5-3 quasars from \citet{Shemmer04} and \citet{Netzer04}; the $\approx$24,000 $z<0.4$ spectroscopically selected AGN from \citet{Mullaney13}, the 86 moderate luminosity X-ray AGN with IFU observations from the KMOS Survey at High-$z$ \citep[KASHz;][ and Harrison et al in prep.]{Harrison16}, and the 18 type-1 X-ray AGN with IFU observations from the SINFONI Survey for Unveiling the Physics and Effect of Radiative feedback  \citep[SUPER;][]{Circosta18,Kakkad20} \footnote{For QSOs from \citet{Shemmer04}, \citet{Netzer04} and \citet{Mullaney13}, we used reported emission line profile parameters (FWHM of both Gaussian components and their velocity offset) to recalculate the W80 values.}. We also highlight the moderate luminosity X-ray AGN from \citet{Scholtz20}, where we performed a similar experiment to this one, combining ALMA data and IFU data. We note that the sources from that study were less extreme in both luminosity and [O~{\sc iii}] emission-line width than the three targets presented here. Indeed, the three quasars in our sample have a [O~{\sc iii}] luminosity $\sim 2\times 10^{44.3}$ ergs\,s$^{-1}$, a factor of $\sim$ 100 larger than a typical AGN population at z = 1--2.5 \citep[]{Harrison16, Scholtz20} and have extreme [O~{\sc iii}] W80 values of 1260--1980 kms$^{-1}$.

\begin{table*}
\caption{A table of the basic properties of our quasar sample.
(1) Object name used in this work;
(2) Object ID from \citet{Canodiaz12,Carniani15};
(3,4) Optical coordinates of the objects; 
(5) Redshift measured from the peak of the narrow emission lines (see \S\,\ref{sec:Gal1D}); 
(6) Bolometric luminosity from \citet{Shemmer04}; 
(7) Star formation rates derived from total FIR emission estimated in \S\,\ref{sec:SED};
(8) Black hole masses compiled from \citet{Carniani15};
(9) Original paper analysing the VLT/SINFONI data; 
(10) On-source exposure time in the IFU $K$-band (H$\alpha$) observations.
(10) On-source exposure time in the IFU $H$-band ([O~{\sc iii}]) observations.}
\centering
\resizebox{0.85\paperwidth}{!}{\begin{tabular}{@{}lcllccccccccc@{}} 
\hline 
\hline 
(1) & (2)       & (3)      & (4)       & (5)  &  (6)                        & (7)                  &(8) & (9) & (10) & (11) \\
ID  & Full name & RA       &  DEC      &  z   & log$_{10}$                  & SFR(FIR)             & log$_{10}$                &  Original    & IFU K-band exposure & IFU H-band exposure \\
    &           & (optical)& (optical) &      &  L$_{\rm Bol}$/ergs s$^{-1}$ & M$_{\odot}$yr$^{-1}$& (M$_{\rm BH}$/M$_{\odot}$)&  work          & time (ks) & time (ks) \\
\hline 
HB8903     &  HB89 0329-385       & 52.776542 & -38.401389 & 2.445 & 47.5  & $<59$ & 9.9$\pm$0.3 &    (b)       & 14.4 &12.6 \\
LBQS01     &  LBQS0109+0213       & 18.070417 &   2.496389 & 2.352 & 47.5 & 69$_{-31}^{+170}$ & 10.0$\pm$0.3 &  (b)       & 14.4 &14.4 \\
2QZJ00     &  2QZJ002830.4-281706 & 7.126833  & -28.284667 & 2.401 & 47.3  & 56$_{-25}^{+196}$ & 10.1$\pm$0.3 &  (a)       & 2.4 &12.6\\
\hline 
\end{tabular}}
\par (a) \citet{Canodiaz12}; (b) \citet{Carniani16}
\label{Table:Sample_qso}
\end{table*}

\subsection{ALMA observations and imaging}\label{sec:ALMA_data}

To map the rest-frame FIR emission (i.e., $\approx$ 250 $\mu$m) for our quasar host galaxies, we use ALMA band 7 data (870 $\mu$m, PI: Harrison, programme ID 2017.1.00112.S) with a native resolution of $\sim$0.3 arcseconds and a maximum recoverable scales of $\sim$4.3 arcseconds. The observations were performed using 45--49 antennae with a baselines range of 15--800 m. We discuss the origin of this emission in \S\,\ref{sec:SED}. 

We calibrated all the data and created the measuring sets using standard ALMA scripts using a version of Common Astronomy Software Application (CASA) applicable to the Cycle of the observations (Cycle 5; CASA v5.1.2). We performed additional checks (i.e. checking the flags, phase diagrams) to see if all calibrations (such as phase corrections) and the pipeline flagging of bad antennae pairs worked correctly and checked that none of the spectral windows used to create the continuum image contains any visible strong emission lines. 

Since the ALMA band 7 observations were performed at a higher resolution than the IFU observations (0.3 arcsec compared to $\sim$0.5--0.6 arcsec; see \S \ref{sec:IFU_data}), we chose natural weighting for the imaging of the ALMA data. This increases sensitivity while decreasing the resolution. The final images had a resolution of $\sim 0.5$ arcseconds, enabling a resolution-matched comparison to the IFU data. The dirty images were consequently cleaned down to 3$\sigma$ by estimating the noise (root-mean-square; RMS) in the dirty maps. We put cleaning boxes around our primary science targets and any visible source in the image. The final RMS of the maps of 2QZJ00, LBQS01 and HB8903 is 0.05, 0.06 and 0.10 mJy/beam, respectively. This was calculated by masking any sources in the field and calculating the RMS in the primary beam. The sources are all significantly detected with peak signal-to-noise ratios (SNRs) of 28--36. We show the ALMA band 7 continuum maps in Figure~\ref{fig:QSO_ALMA_uv} and tabulate the properties of the images in Table~\ref{Table:spec_sf}.

The targets in our study were also part of an ALMA follow up programme using band 3 observations ($\lambda \sim$3 mm) to search for CO(3--2) emission lines \citep{Carniani17}. The corresponding 3 mm continuum measurements in these data provide a valuable additional data point for assessing the contamination to the ALMA band 7 data point from the radio synchrotron emission during our SED analyses in \S\,\ref{sec:SED}. These observations were taken as part of programme 2013.0.00965.S and 2015.1.00407.S with a maximum baseline of 1.5 km. The data was calibrated using CASA software (version 4.5). Similarly to our band 7 observations, we used natural weighting to maximise the sensitivity of our final images; however, we only select spectral windows without detected emission-lines. The final resolution of our images were $\sim$ 0.6 arcseconds with a sensitivity of 12-18 $\mu$Jy/beam. The targets were detected with peak SNRs of 11--350. We use the 3 mm continuum photometry in \S\,\ref{sec:SED}. For both band 3 and band 7 data, we assumed conservative flux calibration errors of 10\%.

\subsection{IFU Data}\label{sec:IFU_data}

Our targets were observed with the VLT/SINFONI integral field spectrograph, in the $K$-band and $H$-band to map the H$\alpha$ and [O~{\sc iii}] emission lines, respectively. The H$\alpha$ IFU data was first published in \citet{Canodiaz12} (2QZJ00)\footnote{We note that \citet{Williams17} presented adaptive optics assisted SINFONI IFU data for 2QZJ00. However, their study focuses on the properties of the broad line region and they find that the star-forming regions and diffuse spatially resolved [O~{\sc iii}] outflows are not detected in these data. We, therefore, focus on the data presented in \citet{Canodiaz12} for our re-assessment of the relationship between outflows and star formation in this quasar.} and \citet{Carniani16} (LBQS01 and HB8903) and were observed through ESO programmes ID 077.B-0218(A) and 091.A-0261(A). The H-band data we use were first presented in \citet{Carniani15} and were observed through ESO programme ID 086.B-0579(A). We note that initial results from shorter exposure $H$-band VLT/SINFONI data were presented earlier in \citealt{Canodiaz12}. 

Most of the observations were performed using the  8$\times$8\,arcsec field of view which is divided into 32 slices of width 0.25\, arcsec with a pixel scale of 0.125 arcsec. The exception is the $K$-band observations of 2QZJ00 that were performed using the smaller field of view of 3$\times$3\,arcsec which is divided into 32 slices of width 0.10\, arcsec with a pixel scale of 0.05 arcsec. All observations were performed using the seeing limited mode. SINFONI has a spectral resolution of R=4000 and 3000 in $K$-band and $H$-band, respectively. We measured the width of the skylines (in the vicinity of the science emission lines), and this width was subtracted in quadrature from the observed emission line widths to account for spectral broadening. The on-source exposure time for the $K$-band data varied between 14.4 ks for LBQS01 and HB8903 and 2.4 ks for 2QZJ00. The short exposure for 2QZJ00 results in noticeably poorer quality $K$-band data (see more in \S\,\ref{sec:SF_morph}). The on-source exposure time for the $H$-band data was 14.4 ks for LBQS01, 12.6 ks for HB8903 and 12.6 ks for 2QZJ00.

For as much consistency as possible between our work and the previous work we used the published $K$-band cubes from \citet{Carniani16}\footnote{Available to downloaded at http://vizier.u-strasbg.fr/viz-bin/VizieR?-source=J/A+A/591/A28}. Nonetheless, the $K$-band and $H$-band IFU data for 2QZJ00 and the $H$-band data for HB8903 and LBQS01 used in this work were reduced following the same basic methods. Here, we briefly outline the procedure. The IFU data reduction was carried out using the standard techniques within  {\sc esorex} \citep[ESO Recipe Execution Tool; ][]{Freudling13}. The individual exposures were stacked on the centroids determined from white-light images from the datacubes. The flux calibration solutions were derived using the  {\sc iraf} routines {\sc standard}, {\sc sensfunc} and {\sc calibrate} on the standard stars, which were observed on the same night as the science observations\footnote{All our reduced data cube are available \href{https://github.com/honzascholtz/Three_QSOs}{here}}. 

Following \citet{Scholtz20} we use the spatial profile of the unresolved H$\alpha$ and H$\beta$ broad line regions (BLR; see Appendix\,\ref{sec:app:Hal_sizes}) as a measure of the point spread-function (PSF) in each cube, since it is a measure directly from the observations (i.e., this method is preferred to using a non simultaneously observed star observation). The final FWHM of the PSF measured from the spatial profile of the unresolved BLR is 0.4 arcseconds for 2QZJ00 and 0.6 arcseconds for the LBQS01 and HB8903, for the $K$-band observations. For the $H$-band data, the FWHM of the PSF is $\sim$0.5 arcseconds, for all targets, consistent with reports by the original studies using the same data. 

\subsection{Astrometric alignment}
The goal of this study is to compare the location of the rest-frame FIR emission traced by ALMA band 7 with the rest-frame optical emission lines traced by the VLT/SINFONI observations, and hence, we need to accurately align the two astrometric frames. The nature of interferometric observations requires accurate astrometry of the targets to reliably calculate the phase differences from each of the antennae. The absolute astrometric accuracy of ALMA depends on the baseline and frequency of the observations. For our setup of baseline 800 m and 350 GHz, the accuracy of the astrometric frame is $\approx$ 20--30\,mas (ALMA Cycle 7 Technical Handbook\footnote{https://almascience.eso.org/documents-and-tools/cycle7/alma-technical-handbook}).

For the astrometric calibration of the IFU data we follow the basic procedure described in \citet{Scholtz20} \citep[also similar to][who also align ALMA and SINFONI IFU data]{Carniani17}. For our reference astrometry in this study, we use GAIA observations. Although there is a lack of sufficiently high spatial resolution {\em near infrared} (NIR) imaging (see \S \ref{sec:SED}) of our targets, the continuum emission is dominated by the point source quasar emission across all optical-NIR wavelengths. Therefore we can benefit from using the high-quality imaging available from GAIA to align our NIR cubes. Since our sources are sufficiently bright ($H$-band magnitude $<$16.5) they are easily detected in GAIA Data Release 2 \citep{Gaiacol18b} dataset. 

To determine the central position of the quasar in the data cubes, we collapse the cubes along the spectral channels to create a white-light (continuum) image (i.e., by excluding wavelengths contaminated by emission lines). We then proceeded with fitting a simple 2D Gaussian model to the resulting continuum image. The final RA and Dec of the central continuum positions are then determined from the corresponding GAIA positions. The NIR continuum is very bright in our data cubes, and the uncertainties on the final position are $\sim$0.1 arcsecond corresponding to $\sim$1 pixel \citep[see][for more discussion]{Scholtz20}. This astrometric alignment is sufficiently accurate for the largely qualitative conclusions in this work that compares the location of the [O~{\sc iii}], H$\alpha$ and rest-frame far infrared emission (\S~\ref{sec:FIR_Halpha}).

\section{Analyses}
In this study, we investigate three powerful quasars with extreme [O~{\sc iii}] emission-line widths (see Figure \ref{fig:OIII_Lbol}), that have previously been presented in the literature to have cavities in the narrow H$\alpha$ emission at the location of [O {\sc iii}] outflows \citep{Canodiaz12, Carniani16}. Specifically, we study the origin and spatial distribution of rest-frame FIR emission and re-assess the  origin and distribution of the narrow H$\alpha$ and [O {\sc iii}] outflows. In this section we present the data analyses of the ALMA data in \S\,\ref{sec:ALMA_analyses} and assess the origin of the rest-frame FIR emission in \S\,\ref{sec:SED}. In \S\,\ref{sec:EL_analyses} we present our analyses of the $H$-band and $K$-band VLT/SINFONI data.

\subsection{ALMA Data Analyses}\label{sec:ALMA_analyses}

In this section, we describe how we measure the flux densities and sizes of the rest-frame FIR emission traced by the ALMA band 7 observations. We note that we use the rest-frame FIR flux densities to assist in establishing the origin of this emission in \S\,\ref{sec:SED}. Following the approaches described in \citet{Scholtz20} we measure these quantities in both the image- and the \textit{uv}-plane. 

\begin{figure}
    \centering
    \includegraphics[width=1.0\columnwidth]{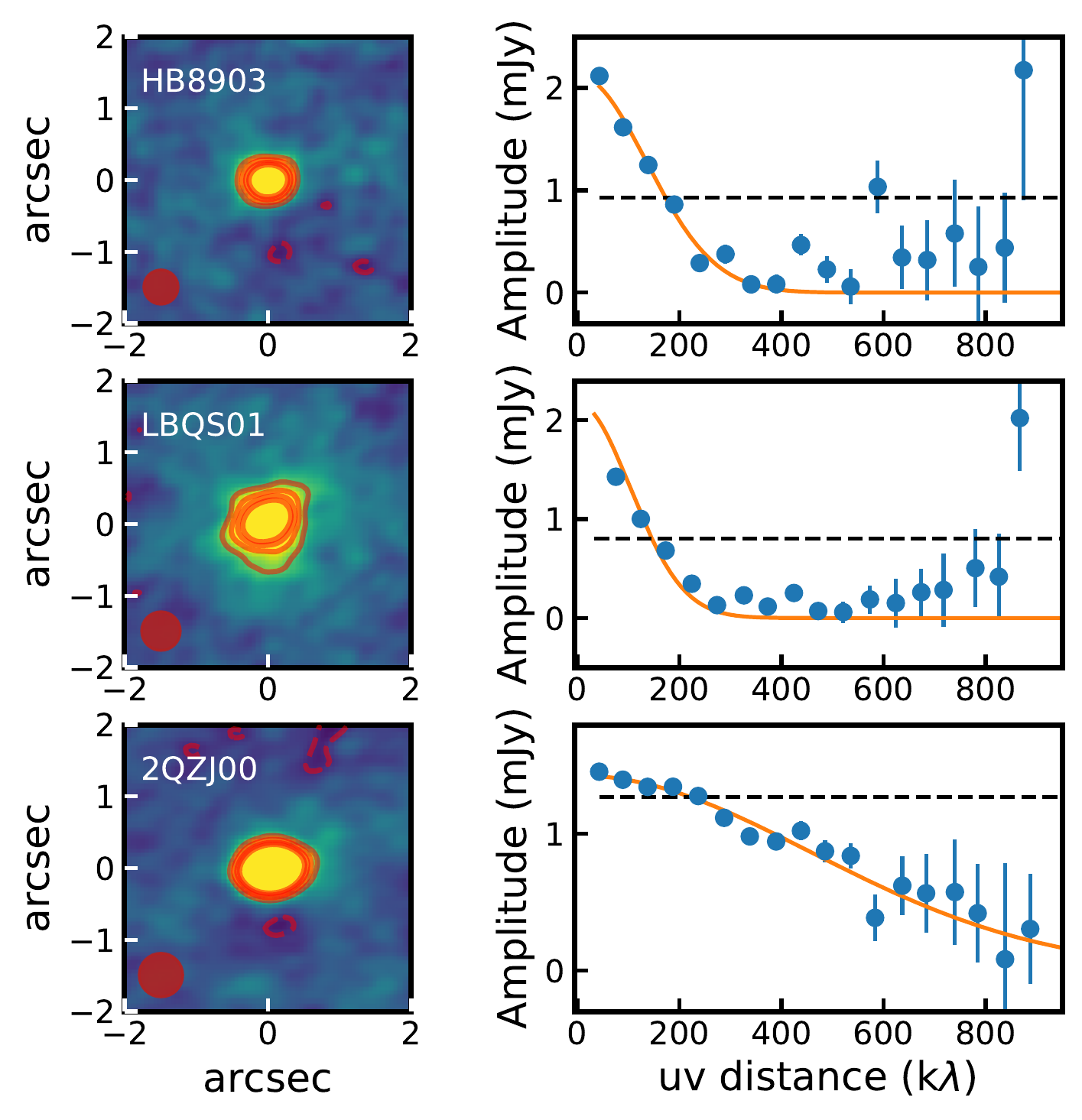}
   \caption{Summary of our analyses on the ALMA data. Left column: ALMA band 7 continuum images mapping rest-frame FIR emission. The red solid contours indicate 2, 3, 4, 5 $\sigma$ levels of the data, while the red dashed contours show -1, -2 $\sigma$ levels of the data. The red shaded ellipses represent the synthesised beams of the observations. Right column: The \textit{uv} amplitude vs \textit{uv} distance binned per 50k$\lambda$, after subtracting any companion sources in the field. The orange solid curves and black dashed curves show resolved and unresolved model fits, respectively. In all cases, the rest-frame FIR emission is resolved on 0.18--0.7 arcsec (i.e., $\approx$1.5--5.5 kpc) scales. 
  }
   \label{fig:QSO_ALMA_uv}
\end{figure}

To measure the sizes and fluxes of the ALMA band 7 emission in the image plane, we used the CASA \texttt{IMFIT} routine to fit a single elliptical Gaussian model, convolved with the synthesised beam, to the images shown in the left column of Figure~\ref{fig:QSO_ALMA_uv}. We used these models to measure the flux densities and the sizes of FIR continuum. To do the {\it uv}-plane fitting, we use \texttt{UVMultiFit} \citep[][]{UVMultiFit}, which simultaneously fits multiple objects in the field thus allowing us to take into account the background sources that contaminate the visibilities in the measurement sets. Using \texttt{IMFIT} on the images we found  that the primary sources and the background sources (one per field) are all resolved. Therefore we fit all sources with a Gaussian profile and we used the \texttt{IMFIT} results as the initial guesses on the parameters.

To visualise the visibilities, we investigated how the amplitude varies as a function of \textit{uv}-distance. To do this we first subtracted the background sources from the visibility data by using the results from the \texttt{UVMultiFit} fitting process described above.  Using  \texttt{fixvis} we then phase centred the data, now with the background sources removed, to the targets' positions. For each target, we extracted the visibility amplitudes from the measurement sets, binning the visibilities in bins of 50k$\lambda$. These are presented right column of Figure~\ref{fig:QSO_ALMA_uv}. We fitted these amplitudes as a function of $uv$ distance with two different functions, a constant, to represent an unresolved source \citep[e.g.,][]{Rohlfs96}, and a half Gaussian model centred on 0, to represent a resolved source with a Gaussian morphology (see curves in Figure~\ref{fig:QSO_ALMA_uv}). For each fit, we calculated the Bayesian Information Criterion (BIC).\footnote{The Bayesian Information Criterion (\citealt{Schwarz78}) uses $\Delta \chi^{2}$ but also takes into the account the number of free parameters, by penalising the fit for more free parameters. BIC is defined as BIC=$\Delta \chi^{2} + k \log(N)$, where N is the number of data points and k is the number of free parameters.} The $\Delta$BIC values between the point source and resolve source models are 230--1700, strongly confirming that the rest-frame FIR emission is resolved in all three objects.

We obtain very good consistency in the size and flux density measurements from all our three methods across the targets (i.e., using \texttt{UVMultiFit}, \texttt{IMFIT} and fitting the amplitudes as a function of {\it uv}-distance; see Table~\ref{Table:spec_sf}). We note that the flux densities are $\approx$1--2\,mJy and although they are not formally consistent using the three different methods (because of the small fitting uncertainties) they all agree within 3--22\%, which is sufficient for our purposes of using them to infer star formation rates and the dominant physical source of the FIR emission (where systematics dominate the uncertainties; see \S~\ref{sec:SED}). Furthermore the sizes across all methods yield results of FWHM$\approx$2--5\,kpc, and that agree within 1--2$\times$ of the uncertainties across the three methods, which is sufficient for our primary purpose of comparing to other galaxy populations at the same redshift (see \S~\ref{sec:SED} \& \ref{sec:FIR_Halpha}).

We use the results from the \texttt{UVMultiFit} method as the final, preferred, measurement of flux densities and sizes. We summarise the flux densities, and sizes (from our collapsed \textit{uv}-data fit, \texttt{IMFIT} and \texttt{UVMUltiFit}) of the rest-frame FIR emission in Table \ref{Table:spec_sf}.

We repeated these same sets of analyses on the ALMA band 3 data (see \S~\ref{sec:ALMA_data}) and we also use the results from \texttt{UVMultiFit}. Our measured band 3 continuum (tracing rest-frame millimetre emission) flux densities are summarised in Appendix Table \ref{Table:Phot} and they agree with those measured by \citet{Carniani17} within the errors.

\begin{table*}
\caption{A table of the basic rest-frame FIR emission (i.e., the ALMA band 7 data) properties of our quasar sample. (1) Object ID in this paper;
(2) The FWHM radii of the rest-frame FIR emission derived using \texttt{UVMultiFit} (see \S\,\ref{sec:ALMA_analyses}); 
(3) Flux density of the rest-frame FIR continuum measured using \texttt{UVMultiFit}
(4) The FWHM radii of the rest-frame FIR emission derived in the \textit{uv}-plane (see \S\,\ref{sec:ALMA_analyses}); 
(5) Flux density of the rest-frame FIR continuum measured in the \textit{uv} plane;
(6) The FWHM radii of the rest-frame FIR emission derived in the image-plane (see \S\,\ref{sec:ALMA_analyses}); 
(7) Flux density of the rest-frame FIR continuum measured in the image-plane; 
(8) SNR of the peak rest-frame FIR continuum measured from the maps;
(9) RMS of the rest-frame FIR maps;
(10) Beam size of the final rest-frame FIR maps.}
\centering
\resizebox{0.8\paperwidth}{!}{\begin{tabular}{@{}lccccccccc@{}} 
\hline 
\hline 
(1) & (2)                 & (3)           & (4)            & (5)         &  (6)    &    (7) & (8) & (9) & (10) \\
ID  &FIR FWHM (\texttt{UVMultiFit})& Flux FIR (\texttt{UVMultiFit})& FIR FWHM (uv)& Flux FIR (uv)  & FIR FWHM& Flux FIR     & FIR & RMS  & FIR Beam     \\
    &  (kpc) & (mJy) & (kpc)              & (mJy)         &  \texttt{IMFit} (kpc)   &  \texttt{IMFit} (mJy)& SNR & (mJy) & (arcsec) \\
\hline    
HB8903     & $3.7\pm0.5$  & $1.486\pm 0.079$  & $4.5\pm 0.7$  & 1.99$\pm$0.24 & $3.7\pm0.5$  & $1.486\pm 0.079$ & 28 & 0.10& $0.48\times 0.39$ \\
LBQS01     & $5.5\pm0.3$ & $2.00\pm0.16$.    & $5.4\pm 0.7$  & 1.9$\pm$0.13 & $4.5\pm0.7$  & $2.00\pm0.16$    &29 & 0.06 & $0.54\times 0.40$  \\
2QZJ00     & $1.7\pm 0.3$& $1.1 \pm 0.04$& $3.0\pm 0.5$  & 1.21$\pm$0.15 & $2.1\pm 0.3$ & $1.092 \pm 0.027$&36 & 0.05 & $0.61\times0.45$  \\
\hline 
\end{tabular}}

\label{Table:spec_sf}
\end{table*}

\subsection{SED fitting}\label{sec:SED}

Here we describe the de-composition the SEDs to establish the origin of the 870$\mu$m emission (rest-frame FIR) from the ALMA band 7 data and to estimate star formation rates. There are potentially three main sources of the FIR emission: (1) star-formation heated dust; (2) AGN-heated dust or (3) radio synchrotron emission \citep{Dicken08, Mullaney11, Falkendal19}.

Our quasars were selected to be brighter than 16 mag in $H$-band \citep{Carniani15} and, as a result, these quasars are bright enough to be detected in several all-sky surveys. We used the optical--mid-infrared photometry from 2MASS \citep{2MASS}, PanSTARRS \citep{Panstarrs}, VST-ATLAS \citep{Shanks15} and WISE \citep{WISE}. We also queried the 1.4\,GHz radio sky surveys of FIRST \citep{First}  and NVSS \citep{NVSS}. Additional radio photometry at 4--8\,GHz was obtained from \citet{Hooper95} and \citet{Healey07}. We include the 870 $\mu$m and 3 mm data from the ALMA band 7 and band 3 data described above. We present the compiled photometry used for our SED analyses in the Appendix, in Table \ref{Table:Phot}. 

\begin{figure}
    \centering
    \includegraphics[width=0.848\columnwidth]{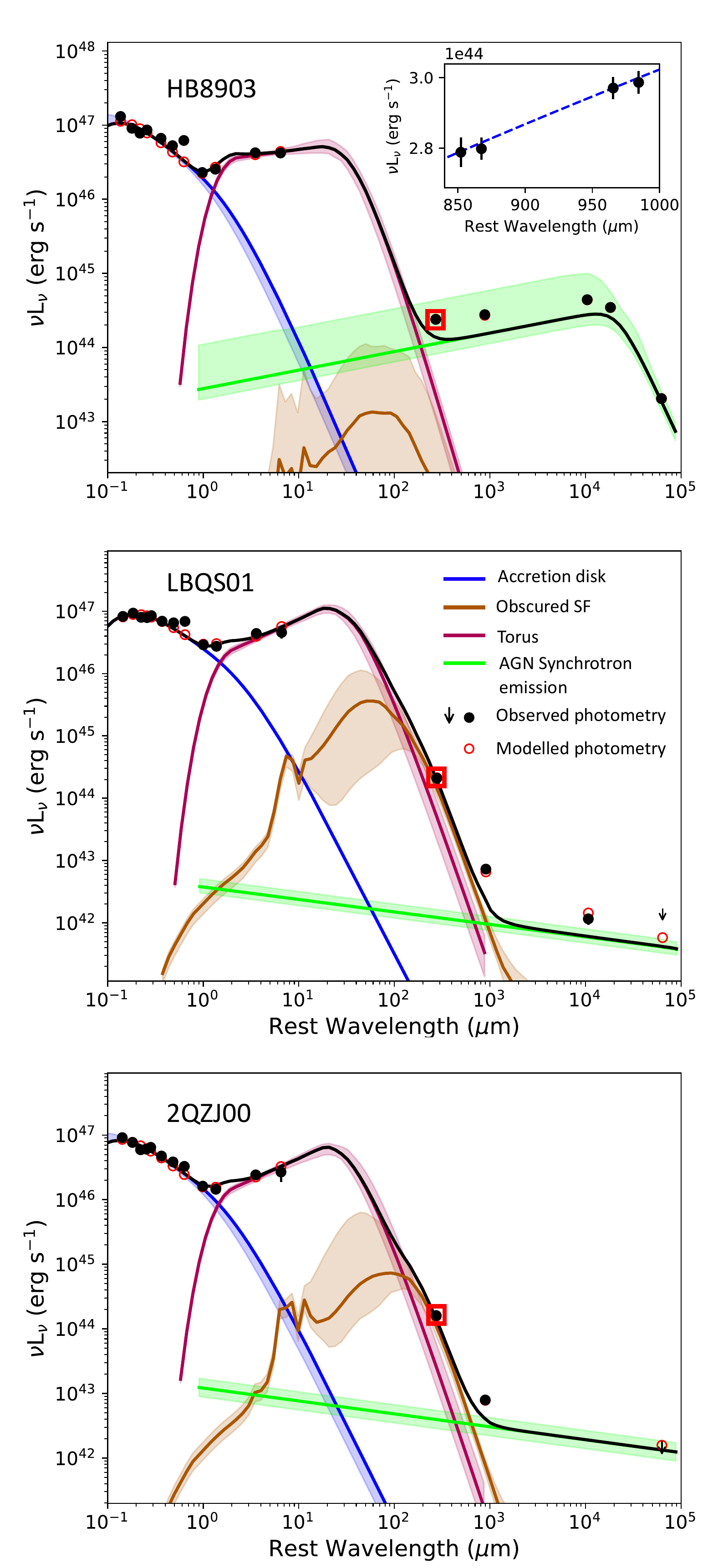}
   \caption{Results of our fitting to the optical to radio spectral energy distributions. The black, blue, red, brown and green curves show the total fit, accretion disk, dust heated by AGN (torus), dust heated by star formation (obscured star formation) and the radio synchrotron emission, respectively. Due to nature of the SED fitting, the template visualisation is sampled from a coarse grid.
   The black filled points show the measured photometry, while the red empty points indicate the modelled photometry by the SED code. The ALMA band 7 photometry is highlighted by red square. The inset Figure~in the top panel shows individual spectral windows of the ALMA band 3 observations. In this inset the blue dashed line shows the fit to these data that was used to constrain the spectral slope in the overall SED fit. The rest-frame FIR emission (250 $\mu$m) traces dust heated by star formation in LBQS01 and 2QZJ00, however, in HB8903 this emission is severally contaminated by radio synchrotron emission.}
   \label{fig:SED_FIR_radio}
\end{figure}

The multi-wavelength SEDs of the targets were modelled using the Bayesian SED code \texttt{FortesFit} \citep{Fortes}. Four SED components were used in the modelling:
\begin{itemize}

\item An AGN accretion disc with a range of spectral slopes (-1.1 to 0.75) as prescribed by the models of \citet{Slone12} with a variable extinction following a Milky Way law ($B-V$ reddening up to 1 mag).

\item An AGN dust emission component with a range of shapes as prescribed by the empirical templates from \citet{Mullaney11} (short-wavelength slope: -3 to 0.8, long-wavelength slope: -1 to 0.5, turnover wavelength: 20 to 60 microns). 

\item Dust emission heated by star-formation following the one-parameter template sequence (0--4) from \citet{dale14}.

\item Radio synchrotron emission described by a broken power law with two spectral indices $\alpha_{\rm low}$ at low frequencies and $\alpha_{\rm high}$ at high frequencies, and a variable break frequency. In the case of HB8903, we placed a tight prior on $\alpha_{\rm high}$ in the range of 1.0-2.0 and allowed the break frequency to vary in the GHz regime along with a variable $\alpha_{\rm low}$ (see details below). For the other two sources, there are no constraints on the power-law slopes or break frequency, so in order to derive an upper limit, we assumed a fixed spectral slope across the radio bands ($\alpha_{\rm low}$ = $\alpha_{\rm high}$ = 0.7). 

\end{itemize}

 The 1.4 GHz and 3 mm photometry of LBQS01 and 2QZJ00 indicate a very weak potential radio synchrotron emission. For these objects, we used a tight prior on radio spectral index of $\sim-0.7$, which is typical for extra-galactic synchrotron-powered radio sources at low radio frequencies. However, these photometric points indicate potentially strong radio synchrotron emission in the case of HB8903. Indeed HB8903 is `radio loud' with a radio luminosity of $L_{\rm 8.4GHz}=10^{27.7}$\,W\,Hz$^{-1}$ \citep[][]{Healey07}. It was, therefore, necessary to further constrain the SED in the sub-mm and mm regime for this source. This exercise showed that is was necessary to use two power-law components for this target. The 3 mm continuum observed in ALMA band 3 was detected at $\sim350 \sigma$, allowing us to split the data into the different spectral windows and obtain a flux density measurement in each one. Since the CO(3-2) emission was not detected in any spectral window (also see \citealt{Carniani17}), we were able to use all four spectral windows. Consequently, we were able to measure the spectral index ($\alpha$) to be 1.5 in this wavelength region by fitting to these four data points (see the inset plot in the top panel of Figure~\ref{fig:SED_FIR_radio}).
 
Probabilistic priors were used to constrain the luminosity of the accretion disc and AGN dust emission components based on the bolometric luminosity \citep[see][]{Rosario19}. \texttt{FortesFit} generates full marginalised posterior distributions of FIR luminosity from star formation ($L_{\rm IR,SF}$; over 8--1000$\mu$m) and AGN, as well as other parameters that are not used in this work. We present the individual SEDs and the resulting fits in Figure~\ref{fig:SED_FIR_radio}. Using these FIR luminosities (i.e., with the AGN contribution subtracted), we estimate star formation rates (SFR(FIR)) using the calibration from \citet{Kennicutt12}. The SFR(FIR) values are provided in Table \ref{Table:Sample_qso}, along with their uncertainties. We note that the degeneracies of fitting the different model components are fold in our uncertainties and are showed as coloured shaded regions in Figure \ref{fig:SED_FIR_radio}. They were calculated as 68\% confidence posterior distribution functions from the fitting shown in Figure \ref{fig:PDFs}, convolved with the systematic uncertainties (0.3 dex, \citealt{Kennicutt12}) on the conversion between $L_{\rm IR,SF}$ and SFR(FIR).

Our main concern here is to determine the dominant source of the ALMA band 7 photometric point: a) cold dust heated by star formation; hot dust heated by the quasar or c) radio synchrotron emission from the radio jets. The SED decomposition showed that the ALMA band 7 continuum observations have a significant component tracing the cold dust heated by star formation in objects LBQS01 and 2QZJ00. Based on the SED fitting, we estimated that the AGN contamination to the ALMA band 7 continuum for LBQS01 an 2QZJ00 is 12$^{+26}_{-11}$ and 14$^{+24}_{-13}$ \%, respectively, where the range accounts for 1 $\sigma$ confidence interval. In Appendix \ref{sec:app:SED_ver} we further tested that our SED de-composition procedure performs as expected. Specifically, we show that the inferred infrared luminosity from the AGN component is correlated with the UV luminosity of the AGN but the inferred far infrared luminosity from star formation is not. There is a synchrotron contribution to the band 7 ALMA photometric point of $\sim 3$ and $\sim 7$ \%, for LBQS01 and 2QZJ00, respectively. In object HB8903, the ALMA band 7 continuum emission is severally contaminated by the radio synchrotron emission from the AGN (over 99 \%). We therefore only have an upper limit (3$\sigma$) on the infrared luminosity due to star formation, and hence on the derived star formation rate.  

\begin{figure}
    \centering
    \includegraphics[width=0.99\columnwidth]{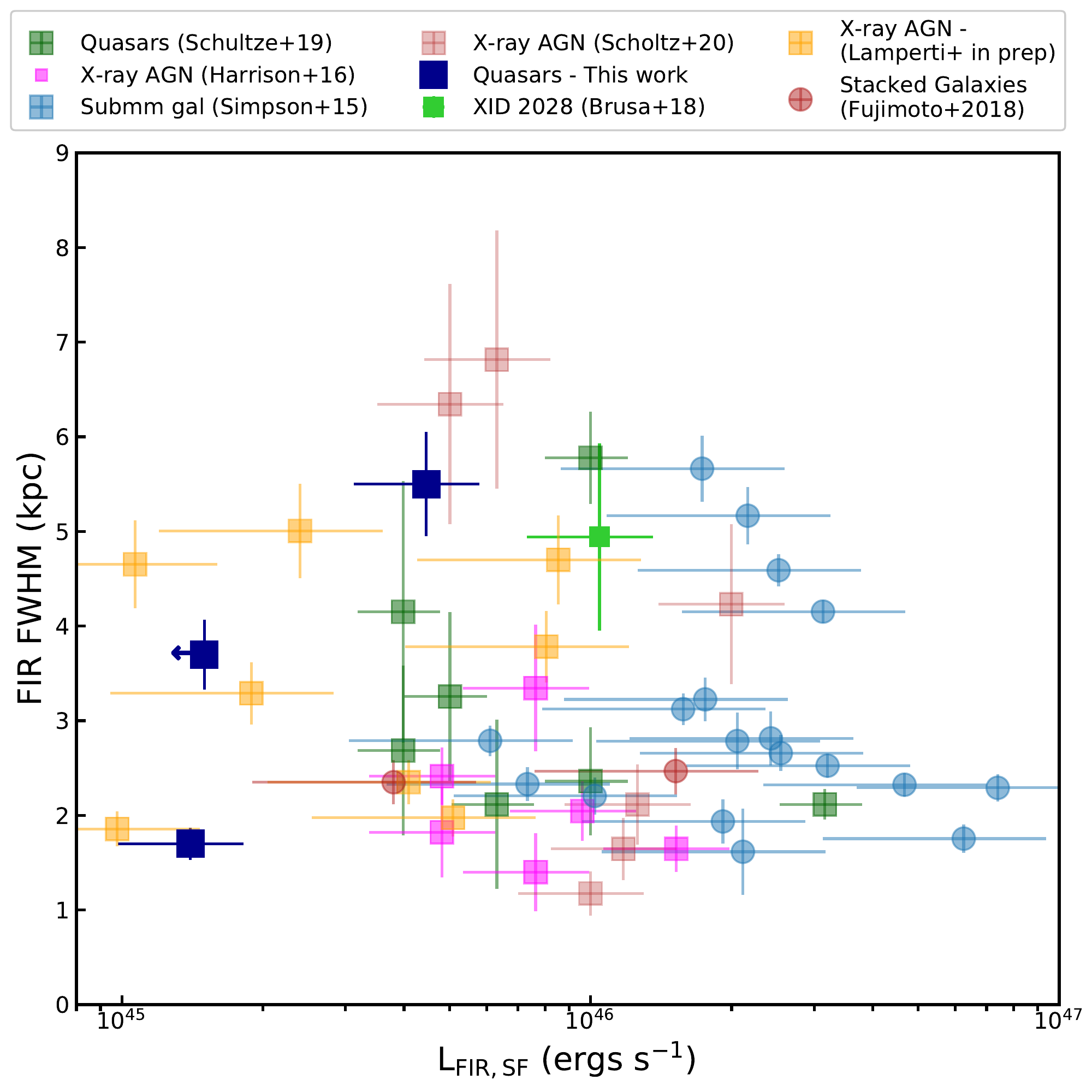}
   \caption{Plot of the total infra-red luminosity due to star formation (L$_{\rm IR,SF}$) as a function size of the FIR  emission for our quasar sample (dark blue squares), quasars from \citet{Schulze19} (dark green squares), X-ray AGN from \citet{Harrison16Alm} (magenta squares), \citet{Scholtz20} (brown squares) and Lamperti et al (in prep; orange squares), XID 2028 (green square, \citealt{Brusa18, Scholtz20}), submm galaxies from \citet{Simpson15} (light blue points) and stacked z$\sim$2 galaxies from  \citet{Fujimoto18}. Our objects FIR size to luminosity ratios consistent with other quasars and star-forming galaxies.}
   \label{fig:LIR_Size}
\end{figure}

In Figure \ref{fig:LIR_Size}, we compare our sample in the FIR size versus FIR star formation luminosity due to star formation with other high-redshift z$\sim 2$ galaxy samples with equivalent measurements. This includes quasars, moderate luminosity AGN and non-active galaxies. The quasars studied here fall within the scatter of these other high-redshift galaxy populations. In particular they have FIR sizes consistent with those seen in high-redshift star-forming galaxies.

Overall, we conclude that for LBQS01 and 2QZJ00 the ALMA band 7 continuum has a significant contribution from the cold dust, whilst for HB8903 is it completely dominated by synchrotron emission.

\subsection{IFU Data Analyses}\label{sec:EL_analyses}

For each of our targets, there are IFU $K$-band observations covering the H$\alpha$ and [N~{\sc ii}]6548,6583 emission lines and $H$-band observations covering the [O~{\sc iii}]4960,5008\AA \space emission-line double and the H$\beta$ emission lines (see \S~\ref{sec:IFU_data}). In this section we provide information about the: (a) modelling of the emission-line profiles (\S\,\ref{sec:Gal1D}); (b) extraction of spectra from a grid of spatial regions across the emission-line regions (\S\,\ref{sec:reg_spec});
(c) mapping of the narrow H$\alpha$ emission (see \S\,\ref{sec:mutli-fit}); and (d) creation of [O~{\sc iii}] velocity maps (\S\,\ref{sec:mutli-fit-oiii}).

\subsubsection{Nuclear spectrum and emission-line modelling}\label{sec:Gal1D}

We extracted an initial spectrum from the $K$-band and $H$-band cubes with the primary goal of determining the H$\alpha$+[N~{\sc ii}] and H$\beta$+[O~{\sc iii}] spectrum from the nuclear regions. We then used these to characterise the H$\alpha$ and H$\beta$ emission associated with the central quasar. To do this, we first determined the peak of the continuum emission (i.e., the central quasar location) in the IFU data cubes by collapsing the datacube in the wavelength direction, excluding any spectral channels contaminated by the emission lines or sky-lines. We fitted a two-dimensional Gaussian to the continuum map to find the centre of the continuum emission. Given that the continuum emission is dominated by the emission coming from the quasar a two-dimensional Gaussian is a sufficient model for this procedure. 

\begin{figure}
    \centering
    \includegraphics[width=0.99\columnwidth]{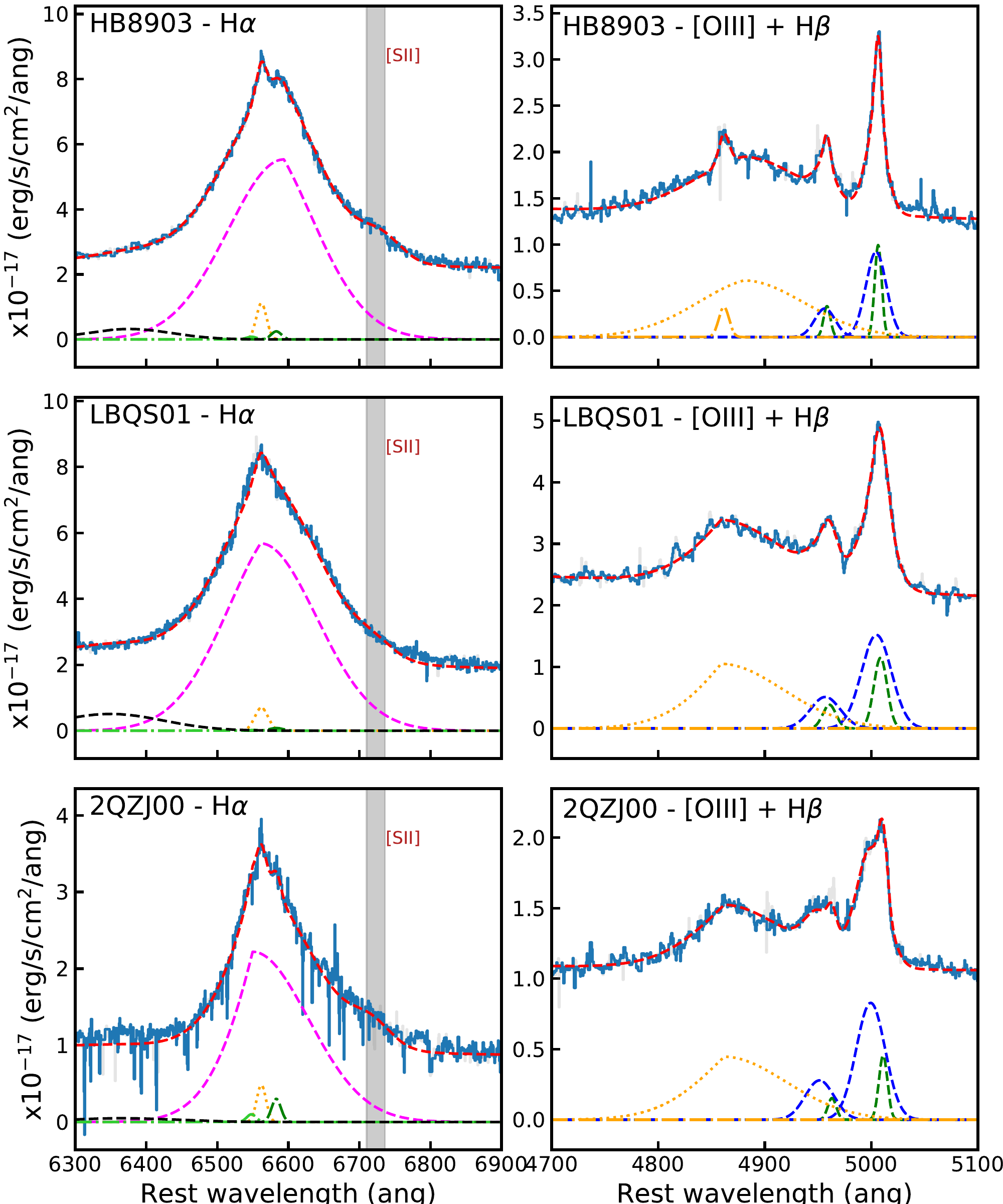}
   \caption{ H$\alpha$ \& [N~{\sc ii}] (left) and  [O~{\sc iii}] \& H$\beta$ (right) emission-line profiles extracted from the nuclear spectra for the three quasars in our sample. Light blue curves show the data and the grey curves show the masked sky-line residuals.
   Left column: H$\alpha$ spectral region. Overlaid on the H$\alpha$ profiles the yellow, magenta, green, black and red curves show the narrow H$\alpha$ (dominated by NLR emission), broad line H$\alpha$, [N~{\sc ii}], additional ''X'' component and the total fit, respectively. The grey shaded region shows the wavelength region of the [S~{\sc ii}] emission line. 
   Right column: [O~{\sc iii}] emission-line profiles. The green, dark blue and red dashed curves show the narrow and broad [O~{\sc iii}] line components and the total fit, respectively. The orange dashed and dotted lines show narrow and broad line H$\beta$, respectively.
   }
   \label{fig:QSO_Halpha_spec}
\end{figure}

For each of the targets, we extracted a spectrum with circular aperture centred on the continuum emission with a diameter of 0.5\,arcsec (i.e., the approximate size of the PSF) to determine the quasar spectrum. We present these spectra in Figure~\ref{fig:QSO_Halpha_spec}. This choice of aperture was chosen to be dominated by the central quasar emission whilst also maximising the signal to noise. However, we note that our conclusions do not change if we use a different aperture. 

To model the emission-line profiles observed by the IFU, each emission line was fitted with one or two Gaussian components (or modified Gaussian components; see below) with the centroid, FWHM and normalisation (flux) as a free parameter. In each case, the continuum is characterised by a linear function. 

To characterise the H$\alpha$ emission-line profile we simultaneously fitted the following components: (a) broad line H$\alpha$ (to describe the emission from the BLR); (b) narrow H$\alpha$ (which would include contributions from the AGN narrow line region [NLR], any outflows, and any contribution from star formation); and (c) [N~{\sc ii}]$6548$\AA, [N~{\sc ii}]$6583$\AA \space emission line doublet. At this point we employ a simple model and do not attempt to decompose any star formation, outflows and NLR components to the overall narrow H$\alpha$ emission. We will discuss this further in \S~\ref{sec:labelling} and Appendix \ref{sec:app:models}. To avoid degeneracies in the fit, the central wavelength and FWHM of the narrow H$\alpha$ and the [N~{\sc ii}]$6548$\AA, [N~{\sc ii}]$6583$\AA \space doublet were tied together, with rest-frame wavelengths of 6549.86\AA, 6564.61\AA \space and 6585.27\AA, respectively. This method assumes that the H$\alpha$ and [N~{\sc ii}] originate from the same gas, a commonly used assumption in high redshift observations \citep{FSchreiber09,Genzel14, Harrison16,ForsterSch18a}. During the fitting, the H$\alpha$ and [N~{\sc ii}]6583\AA \space fluxes were free to vary but the [N~{\sc ii}]6548\AA/[N~{\sc ii}]6583\AA \space flux ratio was fixed to be 3.06 (based on the atomic transition probability; \citealt{Osterbrock06}). 

Following the approach by other studies, we modelled the broad line H$\alpha$ emission, originating from the BLR, as a Gaussian component multiplied by a broken power-law \citep[][]{Netzer04, Nagao06, Cresci15, Carniani16, Kakkad20, Vietri20}. This model is a good description of the asymmetrical nature of these emission line profiles in high luminosity sources such as our quasars. We also tested the same BLR characterisation as done by the other authors, i.e., using one or two Gaussians (\citealt{Canodiaz12}, \citealt{Scholtz20}), without any changes to our conclusions on the spatial distribution of the narrow line emission.\footnote{We note that \citet{Williams17} characterise the broad line H$\beta$ with both a positive and negative Gaussian component, the latter of which they attribute to an ultra-dense fast outflow. This is an intriguing result; however, this different approach does not affect our final conclusions} In \S\ref{sec:SF_morph} \& Appendix \ref{sec:app:models} we discuss how more complex approaches to characterise the narrow H$\alpha$ (to attempt to account for the relative contributions from an AGN narrow line region, including possible outflow components, and from star formation) do not provide robust results. Our primary approach is to model the narrow H$\alpha$ emission as a single Gaussian component.

We note a small residual amount of emission above our fit over wavelengths 6300--6400 \AA \space in Figure~\ref{fig:QSO_Halpha_spec} and is likely attributed to a blend of multiple weak emission lines over this region such as: [O~{I}], [S~{\sc iii}] and [Si~{\sc ii}]. This was treated as an additional broad and weak Gaussian component ''X'' by \citet{Carniani16}. We include this component in high SNR spectra such as the nuclear aperture and regional spectra, but we did not include in the individual spaxel spectra where the SNR is not high enough to detect it. Importantly, including this additional component, or not, does not change the properties of the narrow H$\alpha$ emission which is the focus of our analyses. The [S~{\sc ii}] emission doublet is not detected with sufficiently high SNR to be reliably modelled as both emission lines at once, and therefore we modelled this emission line doublet as a single broad Gaussian, similarly to \citet{Carniani16}. Again, this does not affect our mapping of the narrow H$\alpha$ emission. 

For the $H$-band IFU observations, we simultaneously fit the [O~{\sc iii}]$4959$\AA, [O~{\sc iii}]$5007$\AA \space and H$\beta$ emission lines, using the respective rest-frame wavelengths of 4960.3\AA \space and 5008.24\AA \space and 4862.3\AA. We tied the line widths and central velocities of the [O~{\sc iii}] lines and fixed the [O~{\sc iii}]$\lambda$5008/[O~{\sc iii}]$\lambda$4960 \space flux ratio to be 2.99 \citep{Dimitrijevic07}. Given the complexity of the [O~{\sc iii}] line profile, and high SNRs, we found that a two Gaussian component model was a good description for each of the [O~{\sc iii}] emission lines. 

\begin{figure*}
    \centering
    \includegraphics[width=0.8\paperwidth]{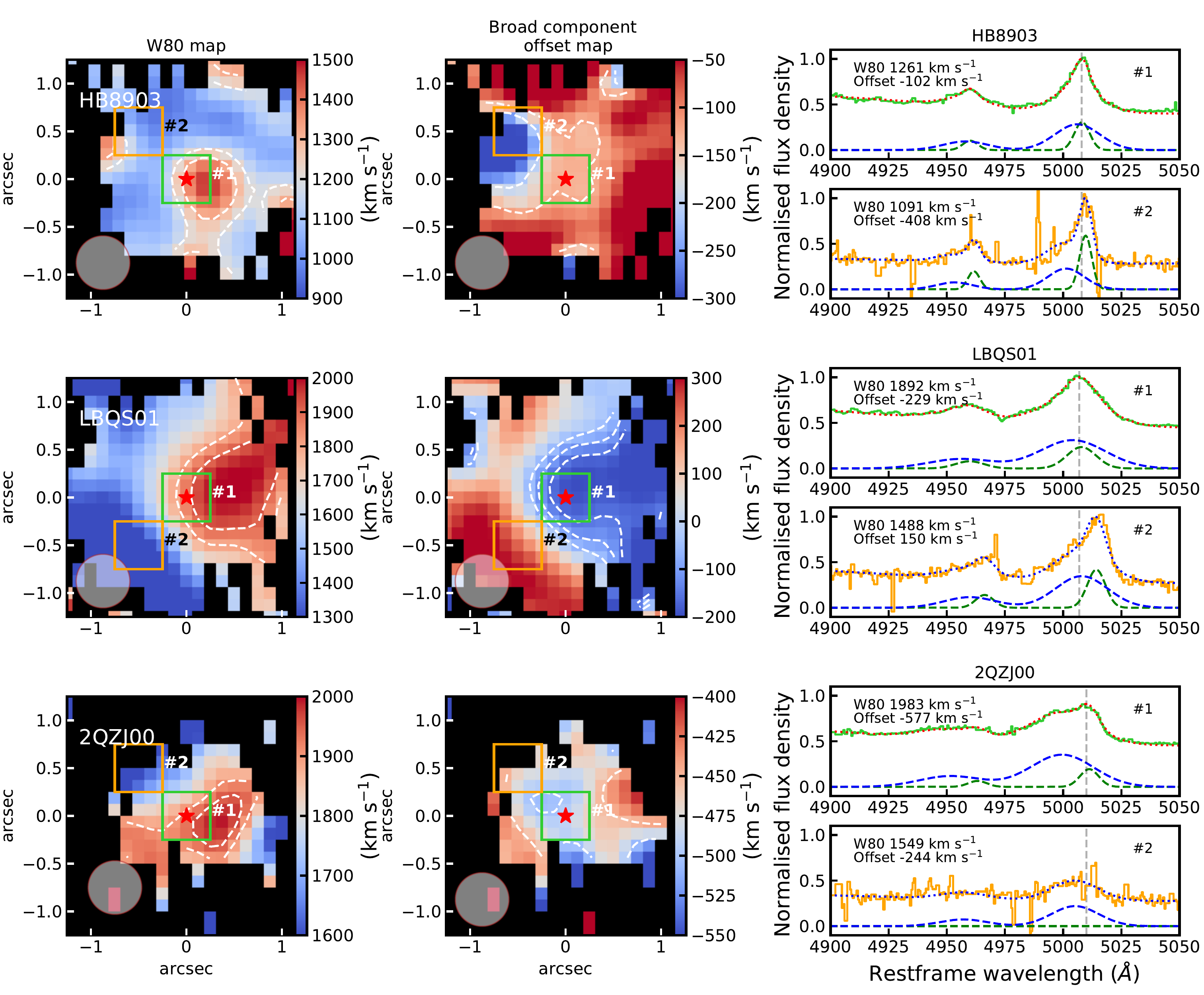}
   \caption{Summary of [O~{\sc iii}] spectral line analyses to map AGN-driven outflows in the quasar host galaxies. 
   Left Column:
   Maps of the W80 of the [O~{\sc iii}] emission line. The white contours show the velocities of 1200 and 1300 kms$^{-1}$ for HB8903; 1800 and 1900 kms$^{-1}$ for LBQS01  and; 1900 and 2000 kms$^{-1}$ for 2QZJ00. The red star shows the continuum centre (i.e., quasar position). 
   Middle Column:
   Maps showing the velocity of the broad component, with respect to the systemic. The contours show the velocity offsets of: -200 and -125 kms$^{-1}$ for HB8903;  -150,-100 and -50 kms$^{-1}$ for LBQS01 and; -500 and 450 kms$^{-1}$ for 2QZJ00. Again, the red star indicates continuum centre. The green and orange rectangles show the regions from which we extracted the emission-line profiles displayed in the right column.
   Right Column:
   Regional spectra extracted from the green (top sub-panel) and orange (bottom sub-panel) regions shown in the maps showing the [O~{\sc iii}]4960,5008 emission-line doublet. The dark green and blue dashed curves show the narrow and broad components of the [O~{\sc iii}] emission line. The total fit is showed in red (top panels) and blue (bottom panels) as dotted curves. The grey vertical dashed line represents the velocity of the quasar. We label the W80 width and velocity of the broad component for each emission-line profile for an easier comparison with the maps. The full set of regional spectra are shown in Figures \ref{fig:HB89_spec_OIII}, \ref{fig:LBQS_spec_OIII} and \ref{fig:2QZJ_spec_OIII} in the Appendix. Extreme velocities are observed across the whole emission-line regions in all three targets.
   }
   \label{fig:Outflows}
\end{figure*}

 Following the same approach as for H$\alpha$, for the H$\beta$ emission line, we fitted a narrow and a broad line component (where the latter describing the emission from the BLR). We fitted the narrow H$\beta$ component as a single Gaussian component, while the broad line component was fitted as Gaussian profile multiplied by broken power-law, similarly to the broad line component of the H$\alpha$ emission line, described above. In agreement with \citet{Canodiaz12} (also see \citealt{Williams17}) and \citet{Carniani15} we do not observe strong FeII emission in our targets and such a component is not detected in our spatially-resolved fitting (see below) and consequently does not impact upon our conclusions of the distribution of the high velocity [O~{\sc iii}] gas.\footnote{Although \citet{Carniani15} do include a weak FeII component for fitting the $H$-band data of LBQS01, as discussed in \S~\ref{sec:OIII_results}, we obtain a qualitatively consistent kinematic [O~{\sc iii}] structure to that presented by these authors in \citet{Carniani16} (i.e., their Figure~3 compared to our Figure~\ref{fig:Outflows}). Therefore, our approach is sufficient for our primary purpose of mapping the high velocity gas and for comparing to this previous work.} Indeed, this is not surprising as the targets were pre-selected to {\em not} have bright FeII emission to make the [O~{\sc iii}] kinematic analyses easier \citep[][]{Carniani15}.

The emission-line profiles were fit with the models described above using the Python \texttt{lmfit} package, excluding spectral channels which were strongly affected by the skylines. To construct a skyline residual mask we extracted a sky spectrum from the object-free (sky only) spatial pixels in the cube and identified the strongest skyline residuals by picking any spectral pixels outside of 1$\sigma$. We note that our results do not change whether we use 1, 3, and 5 $\sigma$ threshold for masking the sky-line residuals. These are shown as a grey spectrum in Figure~\ref{fig:QSO_Halpha_spec}. We estimated the errors using a Monte Carlo approach. With this method, we added random noise (with the same RMS as the noise in the spectra) to the best fit solution from the initial fit and then we redid the fit. We repeated this 500 times to build a distribution of all free parameters. The median of these distributions are consistent with our original best fit parameters from \texttt{lmfit}. The final errors we quote are from the 1$\sigma$ scatter of these distributions. The best fit parameters, and their uncertainties, for the nuclear spectra are presented in Table~\ref{Table:spec}. The best fit models are shown on the emission-line profiles in Figure~\ref{fig:QSO_Halpha_spec}.

\begin{table}
\caption{A table of the basic properties of the H$\alpha$ and [O~{\sc iii}] emission line profiles from the nuclear spectra. (1) Object ID in this paper;
(2) velocity FWHM of the narrow H$\alpha$ component; 
(3) velocity FWHM of the broad line H$\alpha$ component;
(4) Total [O~{\sc iii}] luminosity; 
(5) The W80 (velocity width containing 80\% of the flux) of the [O~{\sc iii}] emission line;
(6) Velocity offset of the broad component of [O~{\sc iii}] to the systematic redshift. 
}
\centering
\resizebox{0.99\columnwidth}{!}{\begin{tabular}{@{}lccccc@{}} 
\hline 
\hline 
(1) & (2)       & (3)      & (4)       & (5)  &  (6)       \\
ID  &Narrow H$\alpha$& Broad line H$\alpha$    & L$_{\rm [O~{\sc III}]}$ & [O~{\sc iii}] W80 & v$_{\rm broad}$\\
    &FWHM (km s$^{-1}$) & FWHM (km s$^{-1}$)&  ($\times 10^{44}$ ergs s$^{-1}$)           &  (km s$^{-1}$)     & (km s$^{-1}$)    \\
\hline    
HB8903     &  $800\pm 78$  & $7490\pm 250$  & $3.2$   & $1260 \pm 110$ & $-110 \pm 80$ \\
LBQS01     &  $450\pm 65$  & $7690\pm 150$  & $2.0$ & $1900 \pm 140$ & $-230 \pm 120$ \\
2QZJ00     &  $620\pm 80$  & $6590\pm 200$  & $2.5$& $1980 \pm 150$ & $-580 \pm 130$  \\
\hline 
\end{tabular}}

\label{Table:spec}
\end{table}

\subsubsection{Regional spectra across the emission-line regions}\label{sec:reg_spec}

For a simple visual representation of the spatial variations of the emission-line profiles, and to verify the results of our more complex analyses of creating emission-line maps (derived below), we extracted spectra from nine square regions across the quasar host galaxies. These regions were defined as a 3$\times$3 grid of $0.5 \times 0.5$ arcsecond squares (i.e., roughly equivalent to the size of the PSF of our observations). The grids were centred on the optical centre of the quasars. The entire grids sample a region of 1.5$\times$1.5 arcseconds and cover most of the emission from the quasar host galaxies.

Spectral line fitting was performed on the spectra extracted from each region using the same models as described in \S\,\ref{sec:Gal1D}. However, we fixed the central wavelength and line-width of the H$\alpha$ and H$\beta$ broad line components to be the same as obtained from the nuclear spectrum (see Figure~\ref{fig:QSO_Halpha_spec}), leaving only the flux of these components as a free parameter. This is a reasonable approach for such point source emission because only the {\em flux} in these broad line components will vary with angular offset, following the PSF. This approach also helps to avoid degeneracies in the fit, especially in the outer regions where the SNR of the spectra is lower. 

\begin{figure*}
    \includegraphics[width=0.8\paperwidth]{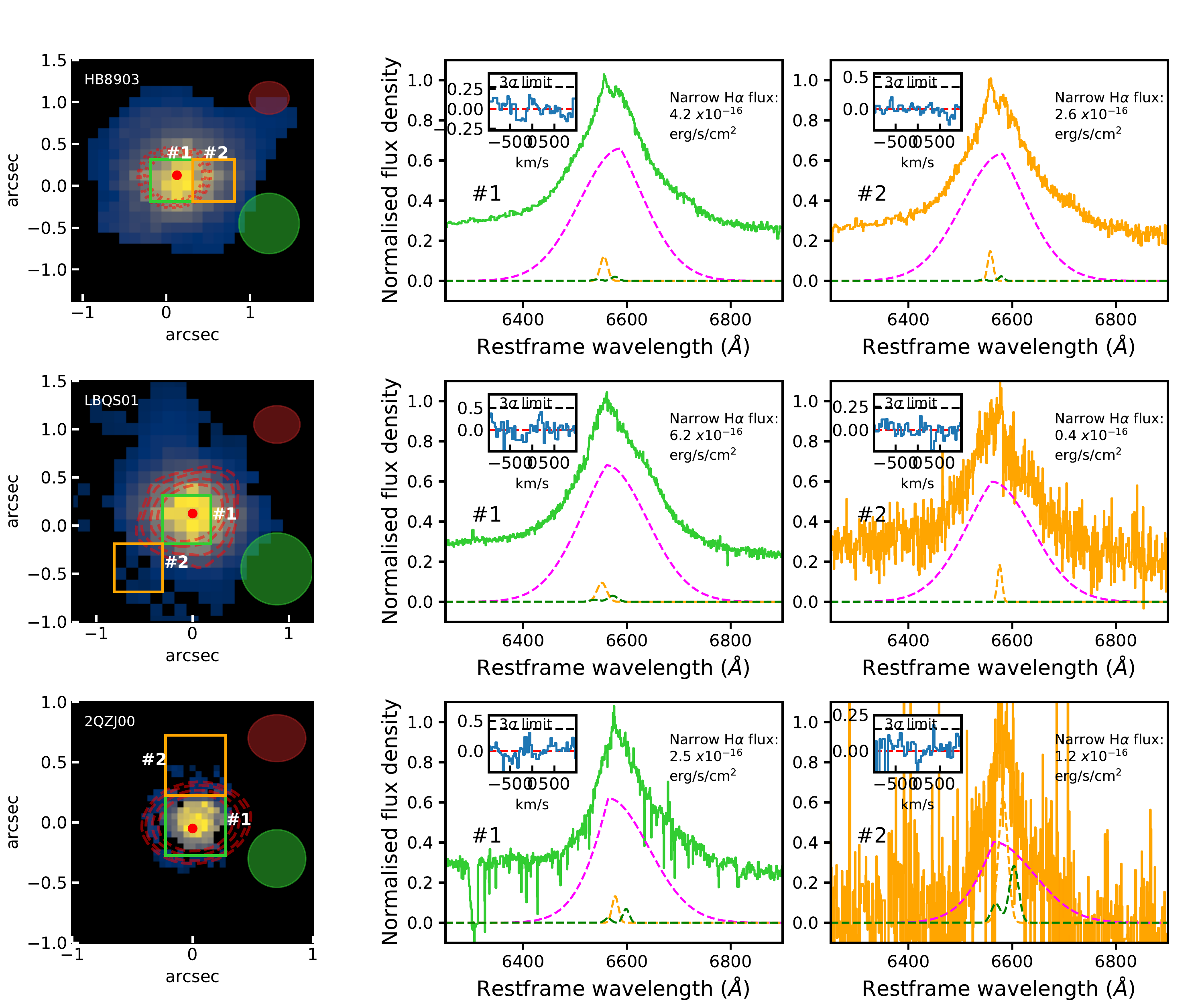}
   \caption{ Left column: Narrow H$\alpha$ (dominated by NLR emission) surface brightness (SB) map created by simultaneously modelling all of the spaxel's spectrum components (see \S\,\ref{sec:mutli-fit}). The ALMA band 7 continuum data is displayed as red contours (2.5, 3, 4, 5 $\sigma$ levels). The green shaded circles show the $K$-band PSF measured from the H$\alpha$ broad-line region. The light green and orange rectangles show the regions from which we extracted spectra in the middle and right columns. Middle column: Spectra extracted from the central region (light green square). Right column: Spectra extracted from the region indicated by the orange square as indicated on the maps. We label the narrow H$\alpha$ flux in each subplot. Overlaid on the H$\alpha$ profiles the yellow and magenta, green dashed curves show the narrow H$\alpha$ (dominated by NLR emission), broad line H$\alpha$, [N~{\sc ii}], respectively. The inset spectrum in each spectral plot are the residual spectra -900 -- 900 km/s around the narrow H$\alpha$ component. We see no evidence for additional residual emission. We find that there is a strong narrow H$\alpha$ emission at the centres of all 3 quasars.}
   \label{fig:Halpha_sum}
\end{figure*}

We present two examples of our regional spectra for each of the objects in Figure~\ref{fig:Outflows} and Figure~\ref{fig:Halpha_sum} and for the [O~{\sc iii}] and H$\alpha$ emisson-line profiles, respectively. We discuss these key spatial regions in \S~\ref{sec:OIII_results} and \S~\ref{sec:SF_morph}. The full set of [O~{\sc iii}] regional spectra are presented in the Appendix in Figures~\ref{fig:HB89_spec_OIII}, \ref{fig:LBQS_spec_OIII} \& \ref{fig:2QZJ_spec_OIII} and the full set of the H$\alpha$ regional spectra are presented in Figures \ref{fig:HB89_spec}, \ref{fig:LBQS_spec} \& \ref{fig:2QZJ_spec}.

\subsubsection{Mapping narrow H$\alpha$ emission}\label{sec:mutli-fit}

To achieve our goal of comparing the location of the rest-frame FIR emission with the emission-line gas we produced two dimensional maps of the narrow H$\alpha$ emission.  That is, all of the remaining H$\alpha$ emission after subtracting the broad-line region emission (see Section~\ref{sec:Gal1D}).

To map the narrow H$\alpha$ emission, the most relevant previous studies \citep[i.e.,][and \citealt{Scholtz20}]{Canodiaz12, Cresci15, Carniani16} used two different approaches: (i) multi-component fitting performed spaxel-by-spaxel and then visualising the distribution of only the narrow H$\alpha$ emission-line component, here-after we call this method ``multi-fit'')
; (ii) subtracting the broad line emission and continuum emission components from the cubes first and then creating a pseudo narrow band image of the residual emission (i.e., an image of the narrow H$\alpha$ emission, as used in; here-after we call this method ``BLR-subtraction''). In principle, these two methods should yield emission-line maps with the same flux and morphology and hence they are complementary methods to verify the integrity of the procedures. In this work we will use the results from the multi-fit method, however, we present the results of the BLR-subtraction method in Appendix \ref{sec:app:QSO-sub} and the maps produced using both methods can be seen in Figure~\ref{fig:Halpha_COG}. We obtain consistent conclusions with both methods.

To increase the SNR of the individual spectra during the production of the maps, we binned the spectra by performing a running median of spaxels with a radius of 0.2 arcseconds. This enhances the SNR while keeping the seeing limited spatial resolution of $\approx$0.4--0.6 arcseconds. This method is sufficient for our purposes of investigating the origin and morphology, on $\gtrsim$4\,kpc scales, of the narrow H$\alpha$ emission \citep[e.g., following][\S~\ref{sec:SF_morph}]{Canodiaz12,Carniani16} and for comparing the spatial distribution of the H$\alpha$ emission with the [O~{\sc iii}] outflows and FIR emission (\S~\ref{sec:FIR_Halpha}). The extracted spaxel-by-spaxel emission-line profiles were fit following \S~\ref{sec:Gal1D}. The final surface brightness maps of the narrow H$\alpha$ component from the spatial fitting are shown in the left column of Figure~\ref{fig:Halpha_sum}. In these maps, we only show values when the SNRs are $>$3. 

We performed multiple checks to verify the spaxel-by-spaxel fitting. Firstly, we compare the total narrow H$\alpha$ flux in the maps to the flux from the total aperture spectra, and we find that the fluxes are consistent within uncertainties on the respective measurements. Furthermore, the distribution and flux measurements we obtain from these maps are consistent with what we observe from visually inspecting the emission-line profiles that were extracted from individual $\approx$0.5$\times$0.5\,arcsecond spatial regions (\S~\ref{sec:reg_spec}; Figure~\ref{fig:Halpha_sum}; also see Figures \ref{fig:HB89_spec}, \ref{fig:LBQS_spec} \& \ref{fig:2QZJ_spec}). We also checked that the surface brightness of the broad line H$\alpha$ emission is well fitted with a 2D Gaussian (as you would expect for the point source BLR emission following the PSF).

We performed two final tests to assess if the H$\alpha$ emission is centrally concentrated and if the emission is spatially resolved: (1) a curve-of-growth analyses and (2) subtracting a model PSF from the narrow H$\alpha$ surface brightness maps. These are described and presented in detail in Appendix\,\ref{sec:app:Hal_sizes} and Appendix\,\ref{sec:app:PSF-sub}. Both methods confirm that the narrow H$\alpha$ emission is centrally concentrated in all three targets. Furthermore, for LBQS01 and 2QZJ00 this emission is consistent with the PSF, i.e., it is spatially unresolved. However, we note that in LBQS01, we detect a faint region of H$\alpha$ emission one arcsecond north of the quasar location. However, given that the bulk of the emission is unresolved, we still designate this target as unresolved for this work. For HB8903 these tests demonstrate that narrow H$\alpha$ emission is spatially resolved, but is still consistent with being centrally concentrated.

\subsubsection{Interpreting the narrow H$\alpha$ components}\label{sec:labelling}

As presented in Section~\ref{sec:Gal1D}, we have so-far only considered a single component to the H$\alpha$ emission that is not part of the broad-line region (i.e., the component we call ``narrow''). However, for this work it is important to establish the relative contribution from AGN NLRs, outflowing gas, and star formation to the total narrow H$\alpha$ flux.

\cite{Carniani16} produced maps of residual narrow H$\alpha$ emission after subtracting a contribution for an AGN NLR (including any possible outflows). They considered any residual emission beyond these components to be attributed to star formation. To search for such residual emission, we created a residual cube (i.e., by subtracting the modelled spectra from the data) and inspected it for any residual emission left by the fitting. We collapse these residual cubes along the spectral regions reported in \citet{Canodiaz12} and \citet{Carniani16} and show them in Figures \ref{fig:HB89_spec}, \ref{fig:LBQS_spec} \& \ref{fig:2QZJ_spec}. We did not detect any significant residual narrow H$\alpha$ emission hence forward refereed to as a narrow star formation component anywhere in the cube. 

\begin{figure*}
    \centering
    \includegraphics[width=0.8\paperwidth]{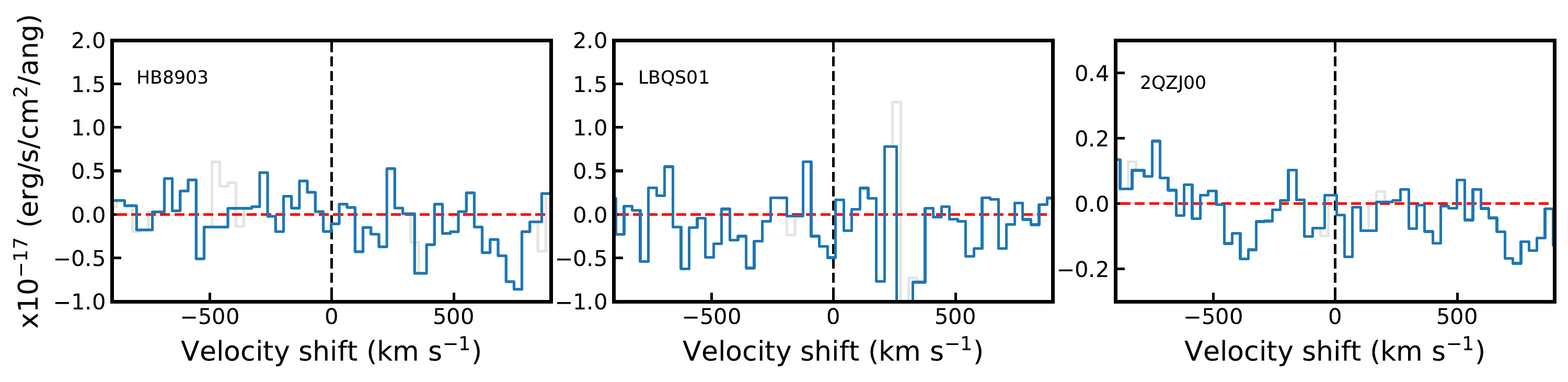}
   \caption{Residual spectra from the K-band H$\alpha$ fitting using the "multi-fit" method. The spectrum is extracted from the same ring shaped region ($0.5<r<0.8$ arcsecond) as done in \citet{Carniani16}. The black dashed line indicates the expected location of the residual "star formation" H$\alpha$ component from \citet{Canodiaz12} and \citet{Carniani16}. The red dashed line shows the zero point of the spectra. We do not detect any additional narrow H$\alpha$ residual emission that could be associated with star formation. 
   }
   \label{fig:Residual}
\end{figure*}

Similarly to \citet{Carniani16}, we also extracted a residual ring aperture spectra of $0.5<r<0.8$ arcsecond around the nucleus. We present these residual spectra in Figure \ref{fig:Residual}, showing $\pm$1000 km s$^{-1}$ around the redshifted H$\alpha$ emission line. We do not detect any additional emission in these residual spectra. We further show the residuals for the individual spatial regions in Figures \ref{fig:HB89_spec}, \ref{fig:LBQS_spec} \& \ref{fig:2QZJ_spec}.

We note that any residual narrow H$\alpha$ emission, beyond the modelled line, will be sensitive to the choice of how to model the emission line. Therefore we investigated multiple approaches into the decomposition of the narrow H$\alpha$ emission line into multiple components associated with an outflow, an AGN NLR and star formation components (see Appendix \ref{sec:app:models}). We found that fitting additional components to our spectra resulted in fits that were not statistically improved and/or they resulted in nonphysical results. 

Based upon all of these tests, we feel confident that our narrow H$\alpha$ maps presented in Figure~\ref{fig:Halpha_sum} are a fair representation of the true morphology of the total narrow H$\alpha$ emission. At the resolution of our observations, all sources are consistent with centrally concentrated narrow H$\alpha$ emission, with only HB8903 being spatially resolved. However, we can only reliably identify a single component to describe the total narrow H$\alpha$ emission dominated by the NLR and we do not identify any residual narrow H$\alpha$ emission that could be associated with star formation. We discuss the implications of this in Section~\ref{sec:SF_morph}. We stressed that despite not detecting the residual narrow "star formation" component as previous studies, we do not claim that there is no unobscured star formation in these objects as it is most likely sub-dominant compared with the NLR H$\alpha$ emission.

\subsubsection{Mapping the ionised outflows traced by [O~{\sc iii}] emission line}\label{sec:mutli-fit-oiii}

Following \citet{Canodiaz12}, \citet{Carniani15} and \citet{Carniani16} we performed spaxel-by-spaxel fitting of the $H$-band IFU data to spatially map the kinematics of the [O~{\sc iii}] emission line. The aim is to establish where the highest velocity [O~{\sc iii}] emitting gas is located. Similarly to the method used to map the narrow H$\alpha$ emission, described above, we binned the spectra by the calculating the running nearby of the nearby spaxels within a radius of 0.2 arcseconds, to enhance the SNR, while keeping the seeing limited spatial resolution of 0.4--0.6 arcseconds. 

We fitted the [O~{\sc iii}]+H$\beta$ spaxel spectra using the same models as for the nuclear spectra (see \S\,\ref{sec:Gal1D}). However, following the same procedure that we employed for the broad line H$\alpha$ (see \S\,\ref{sec:mutli-fit}), we fixed the line-width and the central wavelength of the broad line H$\beta$ component with only the flux as a free parameter. 

The results from the spaxel-by-spaxel fitting were used to spatially map the distribution and kinematics of the [O~{\sc iii}] emission. We focus on maps of: (1) surface brightness; (2) W80 and (3) the velocity offset of the broad component from the systemic. The systematic is defined as the redshift of the narrow [O~{\sc iii}] emission-line component in the nuclear spectra (\S~\ref{sec:Gal1D}). For these maps, we only show values when the SNRs were $>$3. All of these maps are presented in Figures \ref{fig:HB89_spec_OIII}, \ref{fig:LBQS_spec_OIII} and \ref{fig:2QZJ_spec_OIII}, alongside the grids of [O~{\sc iii}] emission-line profiles and their fits (see \S~\ref{sec:reg_spec}). The velocity maps are also shown in Figure~\ref{fig:Outflows}. Reassuringly, a comparison between the emission-line profiles extracted from the spatial grids and these maps demonstrates that the maps are a good description of the data. Similarly to the H$\alpha$ maps, we also inspected the map of the H$\beta$ broad line component to verify that it symmetrically fell off in flux (i.e., it was well described by a 2D Gaussian profile) to provide confidence that the fitting procedures were effective. Again, we also created the residual cube and similarly to the H$\alpha$ fitting, we did not detect any significant residual emission in these cubes. 

For the rest of this work we focus on the maps of W80 (i.e., the overall velocity width of the [O~{\sc iii}] emission line) and the velocity offset (from systemic) of the broad [O~{\sc iii}] emission-line component. This is to broadly follow the previous works of  \citet{Canodiaz12} and \citet{Carniani16}, respectively. Although \citet{Canodiaz12} used a second moment map, instead of W80, to define the emission-line width we verified that the spatial distribution of the high velocity width emission is the same using either definition. We present these maps, alongside example regional spectra in Figure~\ref{fig:Outflows}. We find extreme velocities (i.e., W80$\gtrsim$1000\,km\,s$^{-1}$) across the entirety of the emission-line regions in all three targets, although with spatially-varying kinematic structure. We discuss the interpretation of these maps, and a comparison to the results of the previous works in Section \ref{sec:OIII_results}..

\section{Results and Discussion}

In this study, we are re-assessing the impact of powerful ionised outflows on the host galaxy star formation in three $z\sim2.5$ quasars that have received a lot of attention in the literature because IFU data was used to provide evidence for depleted star formation at the location of [O~{\sc iii}] outflows \citep{Canodiaz12,Carniani16}. In the previous section, we re-analysed the IFU data that was presented in this earlier work and we added new information from our ALMA band 7 observations to trace the rest-frame FIR emission. In \S\,\ref{sec:OIII_results}, we discuss the spatial distribution of ionised outflows. In \S\,\ref{sec:SF_morph} we discuss the morphology and interpretation of the narrow H$\alpha$ emission. In \S\,\ref{sec:FIR_Halpha} we compare the location of the narrow H$\alpha$ emission, rest-frame FIR emission and AGN-driven outflows.  In \S\,\ref{sec:SFR} we investigate the evidence for suppressed galaxy-wide star formation. Finally, in \S\,\ref{sec:qso_implications} we discussed the implications of our results.

\subsection{Mapping of the ionised outflows} \label{sec:OIII_results}

We present our [O~{\sc iii}] kinematic maps in Figure~\ref{fig:Outflows} that were constructed using  the [O~{\sc iii}] emission-line profile models described in \S\ref{sec:Gal1D}. There are many different definitions of ``outflow velocity'' used throughout the literature \citep[see review in][]{Harrison18}. Therefore great care should be taken when interpreting outflow maps across different studies. Here we investigate different approaches to mapping the location of outflows. We were initially motivated to reproduce the broad approaches taken in \citet{Canodiaz12} and \citet{Carniani16} who chose to plot contours of the highest [O~{\sc iii}] emission-line widths and highest velocity offsets of a broad [O~{\sc iii}] component, respectively (see \S~\ref{sec:mutli-fit-oiii}).  

The maps of [O~{\sc iii}] W80, i.e., to characterise the total emission-line widths, are shown in the first column of Figure~\ref{fig:Outflows}. Based on these maps it can be seen that all three objects show extremely broad [O~{\sc iii}] emission-line profiles {\em throughout} the whole emission-line regions with W80 velocities of 900--1500 km\,s$^{-1}$, 1300--2000 km\,s$^{-1}$ and 1600--2000 km\,s$^{-1}$ for HB8903, LBQS01 and 2QZJ00, respectively. These extreme velocities of W80$\gtrsim$1000\,km\,s$^{-1}$, indicate ionised outflows are present across the full extent of the observed emission-line regions \citep[e.g.,][]{Liu13b,Harrison14,Kakkad20}.

In Figure~\ref{fig:Outflows} we have used white dashed contours to highlight the regions with the {\em highest} [O~{\sc iii}] emission-line widths, defined as the top $\approx$10\% of the W80 values. For HB8903 these highest velocities are found to be moderately offset by $\approx$0.3\,arcsecond (i.e., $\approx$2.4\,kpc) to the south-east of the quasar position, for LBQS01 these are found to be $\approx$0.6\,arcsecond (i.e., $\approx$4.8\,kpc) to the west of the quasar position and for 2QZJ00 to be $\approx$0.3\,arcsecond  (i.e., $\approx$2.4\,kpc) to the south-east of the quasar position. We find good agreement with our results and those presented in \citet{Canodiaz12} and \citet{Carniani15} for the location of these highest velocity widths. Following the approach of \citet{Carniani16} (who presented results for LBQS01 and HB8903) we also present maps of the velocity offset (from the systemic) of the broad component of the [O~{\sc iii}] emission line in the second column of Figure~\ref{fig:Outflows}. Similarly as we did for W80, we show the top $\approx$10\% of (blue-shifted) velocities as contours to represent the spatial location of the highest velocities.  We find good agreement with \citet{Carniani16} on the location of the highest velocity offsets.

In Figure~\ref{fig:Outflows} it can already be seen that different conclusions would be drawn on the location of the highest velocities outflows if using W80 compared to broad component velocity offset. This is particularly noticeable for HB8903 where the broadest emission is towards the centre but the highest velocity offsets are to the very east of the emission-line region. Therefore this already highlights that caution should be taken when interpreting such contours as tracing the outflow location; for example, to compare to maps of star formation tracers (see Section~\ref{sec:FIR_Halpha}).

Another approach to map the ionised outflows is to look at the surface brightness distribution of the broad  [O~{\sc iii}] emission-line components, assuming that these trace outflowing gas \citep[e.g.,][]{Rose18,Tadhunter18,Tadhunter19}. This approach traces where the outflowing component has the greatest luminosity and, consequently, is likely to correspond to the majority of the ionised gas mass in the outflow \citep[although using emission lines to trace outflow mass is subject to several challenges;][]{Harrison18}. In Figure~\ref{fig:QSO_hal_alm} we show solid white contours of the surface brightness distribution of the broad [O~{\sc iii}] emission-line components, derived from our spaxel-by-spaxel fitting (see Section~\ref{sec:mutli-fit-oiii}). It can immediately be seen that these contours are centred around the nuclear regions. These results are in agreement with Figure~3 of \citet{Carniani17} who show that the high velocity blue wing of the [O~{\sc iii}] emission of LBQS01 is brightest in the nuclear regions. The same is also observed by \citet{Canodiaz12} for 2QJZ00 where they find that their broad [O~{\sc iii}] emission-line component (i.e., the one tracing the outflow) is brightest in the central regions (see their Figure A.1), even though the highest velocities are located to the south. In summary, an agreement is found across our studies that the broad [O~{\sc iii}] emission-line components peak in the center.

In Figure~\ref{fig:Outflows} and Figure~\ref{fig:QSO_hal_alm} we have shown that different choices to define the spatial distribution of ionised outflow can give very different conclusions on the peak spatial location of these outflows. This has important ramifications for when comparing an outflow location to the spatial distribution of star formation. Figure~\ref{fig:QSO_hal_alm} shows that for the three quasars in this study the high velocity [O~{\sc iii}] emission-line components are most prominent in the centre (in terms of surface brightness), even though the highest velocity widths and highest velocity offsets can be off-centre. We chose to plot contours of the most extreme $\approx$10\% of velocities in Figure~\ref{fig:QSO_hal_alm}, following the broad approach taken by the previous work on these targets \citep[][]{Canodiaz12,Carniani16}. However, we re-iterate that high velocity gas is found across the entirety of the emission line regions (see maps in Figure~\ref{fig:Outflows}) and the high velocity gas is not only confined to the locations inferred by these contours.

\subsection{Interpretation of the narrow H$\alpha$ emission} \label{sec:SF_morph}

Previous works on the three quasars explored in our study presented cavities in residual H$\alpha$ emission (i.e., after subtracting components related to the AGN: BLR+NLR), located at the same location of the highest velocities of the AGN-driven outflows. This was interpreted as evidence that outflows rapidly suppress star formation in the vicinity of the outflows and, potentially, enhancing star formation along the edges of the outflow \citep{Canodiaz12,Carniani16}. Using the analysis from Section~\ref{sec:EL_analyses}, we re-examined the H$\alpha$ emission in these targets.

We created narrow H$\alpha$ maps and we present these in the left column of Figure~\ref{fig:Halpha_sum}. However, as discussed in \S \ref{sec:labelling}, the major source of the emission in our narrow H$\alpha$ maps is the NLR; hence our labelling of this emission as dominated by NLR emission). In all three cases, we see a smooth distribution of the narrow H$\alpha$ emission around the quasars. As discussed at length in  \S\,\ref{sec:mutli-fit} we performed numerous tests to establish that these maps were a fair representation of the data. We found that the emission is consistent with being centrally compact, and is only spatially resolved in the case of HB8903.

Our tests described in \S\,\ref{sec:mutli-fit} showed that only HB8903 has spatially-resolved narrow H$\alpha$ emission (\S~\ref{sec:app:Hal_sizes} \& \ref{sec:app:PSF-sub}) and that we were unable to identify residual emission that could be associated with star formation either in the central regions or outer regions of any of the sources (see Figure~\ref{fig:Halpha_sum} and Figure~\ref{fig:Residual}).  Indeed, the signal-to-noise ratios of the narrow H$\alpha$ components are low (see Figure~\ref{fig:Halpha_sum}) and therefore it is extremely challenging to reliably de-compose these into three components (i.e., an outflow, an AGN narrow line region and star formation emission). This is especially challenging for these Type~1 quasars because the broad-line H$\alpha$ components are so dominant and the narrow components so weak. Although possibilities explored to help with this included tying the properties of the H$\alpha$ emission-line profile to the [O~{\sc iii}] emission-line profile, this still resulted in degenerate and solutions with extreme/unrealistic emission line ratios. It is possible that such degeneracies may be the cause of the non-physical negative surface brightness values seen in the cavities of the H$\alpha$ emission maps used to represent the distribution of star formation in Figure 8 of \citet{Carniani16}.

We further note the challenges in confidently establishing the dominant ionisation mechanism of the H$\alpha$ emission. Local IFU studies demonstrate it is possible to produce AGN un-contaminated maps of star formation by using emission-line ratio diagnostics to spatially decompose the contributions from star formation, shocks and AGN to individual kinematic components when there is sufficient signal-to-noise ratios in multiple lines and with sufficient spatial resolution. However such approaches are not feasible with these data.\footnote{We note that the adaptive optics assisted IFU observations of 2QZJ00 in \citealt{Williams17} find that the overall emission-line ratios are consistent with
AGN ionization across most of the sampled regions in their data; however, they note that most extended, diffuse, emission associated with star formation is likely to be missed due to the use of adaptive optics.}

In summary, within the limits of these data we find that it is not possible to identify H$\alpha$ emission that can be confidently associated with star formation and free from AGN contamination. Therefore, the surface brightness maps of the overall narrow H$\alpha$ emission are likely to be dominated by AGN-related processes.

\subsection{Comparison of the H$\alpha$, FIR continuum and AGN-driven outflows}\label{sec:FIR_Halpha}

\begin{figure}
    \centering
    \includegraphics[width=0.88\columnwidth]{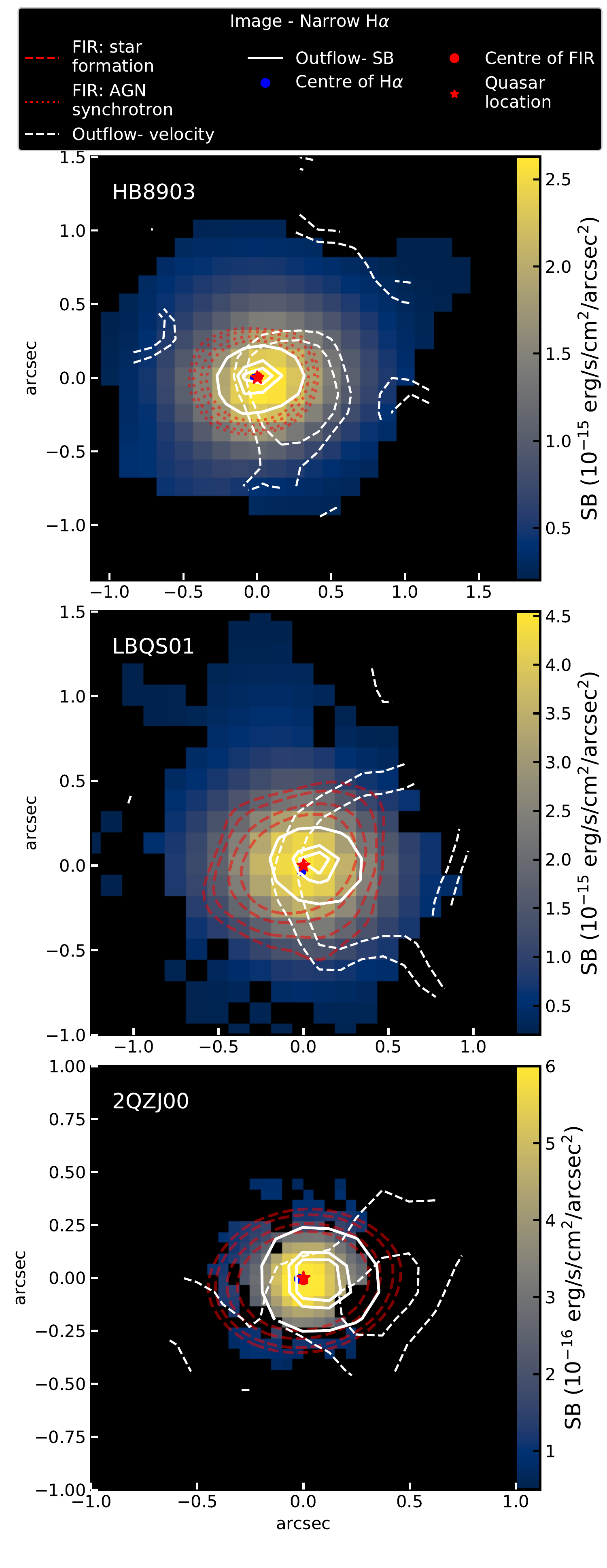}
   \caption{A comparison of the spatial distribution of FIR emission (red contours; levels of 2.5, 3, 4, 5$\sigma$), surface brightness of narrow H$\alpha$ (dominated by NLR emission) emission (background maps; see \S\,\ref{sec:mutli-fit}) and AGN-driven outflows (white contours; \S\,\ref{sec:mutli-fit-oiii}) for: HB8903 (top); LBQS01 (middle) and; 2QZJ00 (bottom). The FIR is attributed to star formation (dashed red contours) for LBQS01 and 2QZJ00, and to AGN radio synchrotron for HB8903 (dotted red contours; \S\,\ref{sec:SED}). The white dashed contours show [O~{\sc iii}]  W80 velocities with values of: 1200 and 1300\,km\,s$^{-1}$ for HB8903; 1800 and 1900\,km\,s$^{-1}$ for LBQS01 and; 1900 and 2000\,km\,s$^{-1}$ for 2QZJ00. The white solid contours represent the distribution of the broad [O~{\sc iii}] ``outflow'' component (levels of 0.68, 0.9 and 0.95 of the peak). The outflows peak in the centre, co-spatial with the FIR emission, 
   but the highest velocities are found off centre.}
   \label{fig:QSO_hal_alm}
\end{figure}

In Figure~\ref{fig:QSO_hal_alm} we compare the spatial distribution of the narrow H$\alpha$ (dominated by NLR emission) emission (background map) and rest-frame FIR emission (red contours). As demonstrated in \S~\ref{sec:SED} the rest-frame FIR emission is tracing dust heated by star formation in LBQS01 and 2QZJ00 (indicated as red dashed contours in the figure), and AGN radio synchrotron emission in HB8903 (indicated as red dotted contours in the figure). The H$\alpha$ emission is most likely dominated by AGN-related processes in all three targets (see \S~\ref{sec:SF_morph}).  The red and blue points show the location of the peak of H$\alpha$ and rest-frame FIR emission, while the red star indicates the location of the quasar. The rest-frame FIR emission is centrally concentrated and we find that the peaks of the H$\alpha$ and rest-frame FIR emission are co-spatial in all three quasars. 

At least for LBQS01 and 2QZJ00, we have shown that there are significant levels of dust in the central $\approx$1.5-5.5\,kpc (\S~\ref{sec:ALMA_analyses}). This is consistent with previous ALMA studies of sub--mm and star-forming galaxies, with and without an AGN, at high redshift which show dust emission extended over a $\approx$1--5\,kpc \citep[e.g.,][Lamperti et~al. in prep;  see Figure \ref{fig:LIR_Size}]{Ikarashi15,Simpson15,Harrison16Alm,Hodge16, Spilker16, Tadaki17,Fujimoto18,Brusa18, Lang19,Schulze19,chen20,Chang20,Scholtz20}. Furthermore, \citet{Carniani17} find centrally concentrated CO(3-2) emission for LBQS01, i.e., co-spatial with the rest-frame FIR emission that we trace here. This would be consistent with obscured star formation at the same location as the molecular gas reservoir. Due to the strong radio emission in HB8903, we are not able to reliably map the distribution of dust with our observations.

We find that the rest-frame FIR emission overlaps with the H$\alpha$ emission cavities that were presented in \citet{Canodiaz12} and \citet{Carniani16}. Therefore, without reliable obscuration corrections, observed maps of H$\alpha$ emission are likely to miss a significant fraction of the total intrinsic emission. Along with the several other complications discussed in \S~\ref{sec:SF_morph} we suggest that the narrow H$\alpha$ maps presented here and in \citet{Canodiaz12} and \citet{Carniani16} are not reliable for mapping the on-going star formation in the host galaxies. 

We discussed different definitions for mapping the the [O~{\sc iii}] outflows in \S\,\ref{sec:OIII_results} and we present these as white contours on Figure~\ref{fig:QSO_hal_alm}, compared to the obscured star formation distribution. With either outflow definition (i.e., using velocity contours or surface brightness contours) we see no evidence that the outflows have had any significant impact on the distribution of star formation. In particular, by mapping the AGN-driven outflows in terms of the surface brightness of the broad component of the [O~{\sc iii}] emission line, shows that the outflow is spatially located at the same location as the obscured star formation. We do see an offset between this obscured star formation and the {\em highest velocities} of the [O~{\sc iii}] emission but this could be just an indication of how the outflows are escaping from the host galaxies \cite[e.g.,][also see \S~\ref{sec:qso_implications}]{Gabor14}. 

In summary, we find no evidence that the ionised outflows have had an instantaneous impact on the distribution of the in-situ star formation. We discuss the limitations and implications of these conclusions in \S~\ref{sec:qso_implications}

\subsection{Global star formation rates of our quasars}\label{sec:SFR}

In the previous section, we saw no evidence that star formation is disturbed at the location of the [O~{\sc iii}] outflows in the three quasars of this study. Nonetheless, high velocity outflows are found across the observed emission-line regions (Figure~\ref{fig:Outflows}). Therefore, it is appropriate to assess if there is any evidence that the {\em global} (i.e., galaxy wide) star formation rates of our quasar host galaxies have been suppressed, potentially by the outflows observed. 

\begin{figure}
    \centering
    \includegraphics[width=0.99\columnwidth]{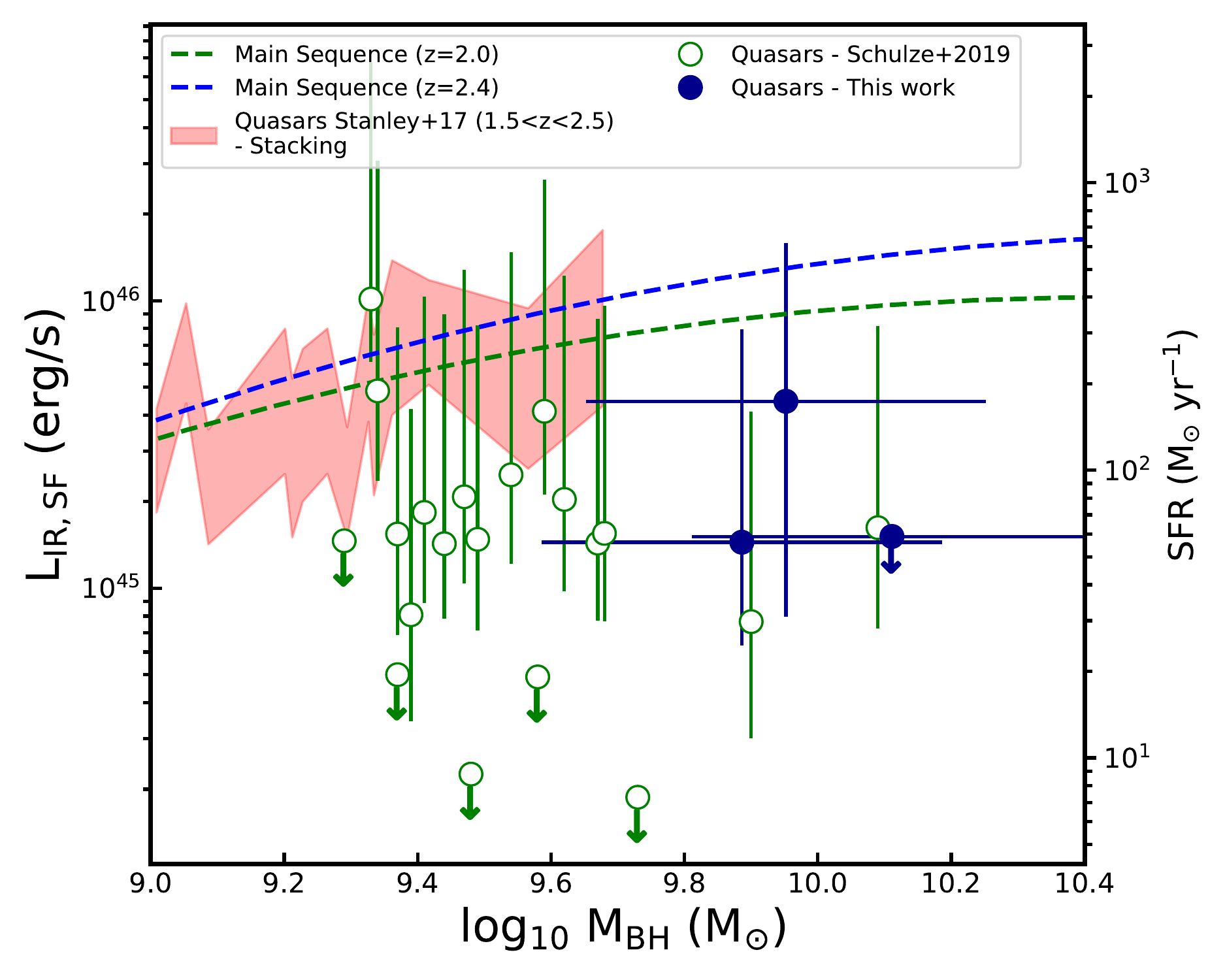}
   \caption{Plot of the total infra-red luminosity due to star formation (L$_{\rm IR,SF}$) as a function of black hole mass of the quasars. The dark blue filled and green hollow points show quasars investigated in this work and quasars from \citet{Schulze19} (z$\sim$ 2), respectively. The red shaded region shows the measured averages and 68 \% confidence interval of quasars for redshift range of 1.5--2.5 from \citet{Stanley17}. The green and blue dashed lines show the expected L$_{\rm IR,SF}$ of main-sequence galaxies \citep{Schreiber15} (recalculated for black-hole mass rather than stellar mass) for z=2 and z=2.4, respectively. There is some evidence that the most powerful quasars scatter to lower SFRs}
   \label{fig:LIR_MBH}
\end{figure}

In Figure~\ref{fig:LIR_MBH} we present the star formation rates (see \S~\ref{sec:SED}) as a function of black hole masses (see \S~\ref{sec:Sample}) for the three quasars. In this figure, we compare our sample to the results of \citet{Stanley17}, who measured the mean star formation rate (also using far-infrared luminosity as their star formation tracer) of $z=$1.5--2.5 quasars as a function of black hole mass; we show the 68 \% confidence interval around the mean values from this study as a red shaded region. To allow for a comparison between our quasar host galaxies and the overall star-forming galaxy population, we also show tracks of the average star-formation rate from \citet{Schreiber15} for star-forming galaxies. To produce these tracks we converted black hole mass to stellar mass following \citealt{Bennert11} and then we calculated the star-formation rate for a given mass at both $z=2.0$ and $z=2.4$ following \citet{Schreiber15}. We note that our comparison to these tracks are dominated by the systematic uncertainties in star formation rates and, because the tracks are almost flat over the mass range of interest, the uncertainties in black hole masses or the conversions to stellar masses have no impact upon our conclusions.

In Figure~\ref{fig:LIR_MBH} it can be seen that the average star formation rates of quasars, taken from \citet{Stanley17}, follow the main sequence of star forming galaxies (in agreement with their conclusions). However, their black hole masses are limited to $<10^{9.7}$\,$\Msol$, compared to the $\approx10^{10}$\,$\Msol$\ of three quasars in this study.  Extrapolating the main sequence into the predicted mass range of our targets would suggest that the quasars in this study have star formation rates a factor of $\approx$10 below the main sequence.

In Figure~\ref{fig:LIR_MBH} we also compare our targets to the quasars from \citet{Schulze19}. Their study observed a sample of 20 very luminous type 1 quasars at $z\sim2$ (log$_{10}$ L$_{\rm Bol}> 46.9$ ergs s$^{-1}$) with ALMA band 7, making them an ideal comparison sample to our study. To make a robust comparison between our studies, we refitted the optical to sub-mm SEDs of the \citet{Schulze19} quasars using the FortesFit code described in \S\,\ref{sec:SED}. We plot the results of this re-analyses as green hollow points in Figure~\ref{fig:LIR_MBH} where it can be seen that the quasars from \citet{Schulze19} lie on or below the main sequence. Previous work has shown that even if the mean star formation rates of AGN host galaxies follow the main sequence, the distributions can scatter to lower values \citep[][]{Mullaney15,Bernhard19}. However, recent work by \citet{Grimmett20} implies that the distribution of the more powerful AGN (such as the quasars in this study) tend towards following the same distribution as main sequence star-forming galaxies. Broadly speaking, the powerful quasars presented here would not follow this trend as they appear to fall below the main sequence galaxies. 

Although caution should be taken when using global star formation rate measurements as evidence for AGN feedback, especially in the absence of specific model predictions \cite[e.g.,][also see \S~\ref{sec:qso_implications}]{Harrison17,Scholtz18}, based on the results presented in Figure~\ref{fig:LIR_MBH} we can not rule out that the host galaxies of the bolometrically luminous (i.e., $L_{\rm bol}\approx10^{47.5}$\,ergs s$^{-1}$) high black hole mass (i.e., $M_{\rm BH}\approx10^{10}$\,M$_{\odot}$) quasars in this study are ``on the path'' to becoming quenched with lower star formation rates than you would expect for the black hole masses. This can also be in agreement with \citet{Carniani17} who find that all three quasars appear to have low molecular gas content compared to main-sequence galaxies at the same redshift using observations of the CO(3-2) emission line.

\subsection{Implications of our results}\label{sec:qso_implications}

We have shown that there is no strong evidence that the powerful ionised outflows in the three quasars in this study have a localised ``in-situ'' rapid impact on the star formation inside their host galaxies (\S~\ref{sec:FIR_Halpha}). Nonetheless, we could not rule out global star formation rate suppression in the host galaxies \citep[see][]{Carniani16}.

Our conclusions are consistent with most IFU observations (with or without adaptive optics) and inteferometric observations of high redshift AGN that do not find strong evidence that outflows are instantaneously shutting down in-situ star formation in the host galaxies \citep[e.g.,][Lamperti et~al., in prep]{Scholtz20,Davies20KMOS}. Our study extends this previous work to more powerful quasars with extreme outflows where the impact might be expected to be most dramatic (Figure~\ref{fig:OIII_Lbol}). However, we caution that our results are only for three powerful quasars and there are samples in the literature which include sources that are even more powerful, and exhibit even more extreme outflows, than those investigated here \citep[][]{Zakamska16,Bischetti17,Perrotta19}. It would now be interesting to assess if the outflows in these systems have any impact on the in-situ star formation. 

The star formation tracer that we have been able to use in this study is far-infrared emission that has a mean stellar population age contribution to the emission from $\approx$5\,Myr, up to $\approx$100\,Myr \citet{Kennicutt12}. Unfortunately, we lack a reliable tracer of star formation that is {\em dominated} by recent star formation episodes (i.e., $\lesssim$10\,Myr), which may be helpful for searching for more immediate changes in star formation. However, with current facilities, it is not clear how to reliably overcome this problem for the most powerful high-redshift quasars where the challenges of removal of AGN contamination to tracers such as H$\alpha$ and ultraviolet are severe, to impossible (see \S~\ref{sec:SF_morph}). Future studies of Type~2 quasars (as opposed to Type~1 quasars) would at least reduce the challenges of subtracting the broad line region contribution to the H$\alpha$ emission and the ultra-violet continuum emission associated with the accretion disk but would still suffer from many of the other challenges we outlined in \S~\ref{sec:SF_morph}.

When searching for observational evidence of feedback by AGN, it is also important to ask if the results are actually in conflict with specific theoretical predictions \citep[e.g.,][]{Harrison17,Scholtz18}. Simulations of individual galaxies show the reality of how outflows would impact upon their host galaxies is complex. For example, simulations show that the level of any negative or positive impact on the host galaxy star formation rates will depend on many factors such as: the relative orientation of an outflow (or jet) with a disk; the gas fraction and distribution of the ISM; and the luminosity of the AGN itself \citep[e.g.,][]{Wagner13,Pontzen17,Zubovas17,Mukherjee18,Costa20}. Furthermore, whilst simulations suggest that {\em some} level of rapid suppression in the star formation rate could be caused by AGN-driven outflows this may typically be by modest factors and only in the central regions even in the case of the powerful quasars \citep[e.g.,][]{Gabor14,Costa18,Costa20}. These simulations show complete destruction of the gas disc is extremely unlikely since outflows tend to become collimated and escape through paths of least resistance. 

In agreement with the aforementioned simulations, using the greater spatial resolution possible for integral field spectroscopy and mm-inteferometry observations of local galaxies, some evidence for {\em modest} positive and negative feedback on {\em small scales} has been observed \citep[e.g.,][]{Croft06,Alatalo15,Cresci15b,Querejeta16,Rosario19,Shin19,Husemann19,Perna20}. These effects are out of reach of what can be observed with the lower (i.e., $\sim$2--5\,kpc) resolution and sensitivity of IFU observations of high-redshift galaxies. However, even from these impressive observations of local galaxies, it is not clear if the global, long-term impact of these effects will be significant. 

 As discussed in \citet{Costa20}, quasars may be able to cause feedback effects without the removal of dense star-forming material. For example, through the ejection of gas from the gaseous halo, which would accelerate the decline of gas inflow from the halo. Such a process would operate on longer $\sim$100\,Myr timescales. Observations of the Sunyaev-Zeldovich effect have provided tantalising evidence that quasars are effective at depositing energy into the circumgalactic environment \citep[e.g.,][]{Crichton16,Lacy19}. Our observations would be consistent with such a mode of AGN feedback that has an impact on star formation over a longer timescale, perhaps through the cumulative effect of quasar episodes \citep[also see][]{McCarthy11,Zubovas12,Gabor14,Scholtz18,Costa20}.

\section{Conclusions}

We have presented new high-resolution ALMA band 7 continuum observations (rest-frame $\lambda \sim 250 \mu$m) of three quasars at $z\sim2.5$ that have previously been presented as showing evidence for suppressed star formation based on cavities in the narrow H$\alpha$ emission at the location of the quasar-driven outflows (traced with [O~{\sc iii}] emission; \citealt{Canodiaz12,Carniani16}). All three quasars are significantly detected by ALMA (SNR$>$25) and we exploited these observations to trace the dust-obscured star formation. Furthermore, we re-analysed the $H$-band and $K$-band archival IFU data. Based on our analyses we find:

\begin{enumerate}
    \item  {\bf FIR emission contains a significant dust-obscured star formation component in two of the three quasars.} This result is based on: (a) the observation that the FIR emission traced by ALMA band 7 observations is spatially extended on scales typical for high-redshift star-forming galaxies (i.e., $\approx$1.5--5.5\,kpc; see Figure~\ref{fig:QSO_ALMA_uv} \& \ref{fig:LIR_Size}) and; (b) our SED decomposition shows that the FIR emission contains a significant star formation heated dust component ($\sim 86\%$) (see \S\,\ref{sec:SED} and Figure~\ref{fig:SED_FIR_radio}); 
    therefore, we conclude that we can use the spatially-resolved maps to trace the obscured star formation in LBQS01 and 2QZJ00 and this star formation is located in the regions of the previously reported cavities in the narrow H$\alpha$ emission (Figure~\ref{fig:QSO_hal_alm}). In the third quasar (HB8903), the FIR emission is dominated by AGN synchrotron emission.
    
    \item {\bf High velocity ionised gas is found across the whole emission-line regions as traced by the [O~{\sc iii}] emission line (W80$\gtrsim$1000 km s$^{-1}$).} This is based on our [O~{\sc iii}] velocity maps (Figure~\ref{fig:Outflows}). Following the approach of the previous work, when the outflow locations are defined based on the top $\approx$10\% of emission-line velocities, they are offset by $\approx$2--5\,kpc from the central quasar. However, based on the surface brightness distribution of the broad component of the [O~{\sc iii}] emission line, the outflows peak in the centre 
    (see Figure \ref{fig:QSO_hal_alm}). Consequently, the relative spatial distribution of an outflow and star formation is sensitive to the choice of outflow definition. 
    
    \item {\bf We do not identify any H$\alpha$ emission that is a reliable tracer of spatially-resolved star formation} despite testing a variety approaches to isolate such emission. This is because we are unable to reliably identify any residual H$\alpha$ emission beyond that associated with the AGN narrow line region and/or outflow in the galaxy-integrated or spatially-resolved spectra (see Figure~\ref{fig:Halpha_sum} and Figure~\ref{fig:Residual}). Furthermore, the H$\alpha$ emission is only spatially resolved in one of the three targets.
    
    \item {\bf We do not observe any localised instantaneous suppression of star formation by the ionised outflows.} Specifically, we do not find any evidence for suppressed star formation co-spatial with the location of the bulk of the ionised outflows, at least for the two targets for which we have been able to trace the star formation with the ALMA data (see Figure~\ref{fig:QSO_hal_alm}).
    
    \item {\bf We cannot rule out global star formation suppression caused by the quasars}. Based on the infrared luminosities we find that the inferred galaxy-wide SFRs of the three quasars are 50--170 $\Msolyr$. Assuming a black hole to stellar mass conversion from \citet{Bennert11}, these are an order of magnitude lower than mass-matched main sequence galaxies at the same redshift. This could also be in agreement with the findings of \citet{Carniani17}, who reported low molecular gas content in these same objects.
\end{enumerate}

 Using the FIR emission as a tracer of star formation, we see no direct evidence that ionised outflows instantaneously suppress in-situ star formation in the nuclear regions of these extreme quasars on $\sim 4$\,kpc scales. Any instantaneous impact by these outflows must be subtle, occurring beyond that which can be detected within the resolution and sensitivity limits of these observations. This is consistent with \citet{Scholtz18} which showed a good agreement between ALMA observations of X-ray AGN and the predictions from EAGLE hydrodynamical simulation. To search for more subtle effects will require deep observations with future high-resolution IFU facilities such as with VLT/ERIS, {\em JWST}/NIRSpec and ELT/HARMONI. We suggest that any impact on star formation from a single outflow event must be limited. However, our results could still be consistent with a more long term impact of the central quasars resulting in a more gradual global suppression; for example, the cumulative effect of outflow episodes could accelerate the decline of gas inflow onto the host galaxy \citep[also see e.g.,][]{Canodiaz12,Gabor14,Carniani16,Costa20}.

\section*{Data Availability}

The raw IFU and ALMA data is available at VLT/SINFONI and ALMA archive services. The reduced data and main analyses code underlying this article are available on Github \href{https://github.com/honzascholtz/Three_QSOs}{here}.

\section*{Acknowledgements}

We thank the referee for useful comments that improved the quality of this work. We thank Sebastien Muller from Nordic ARC Node for helping with UVMultiFit. JS acknowledges support from the Nordic ALMA Regional Centre (ARC) node based
at Onsala Space Observatory. The Nordic ARC node is funded through Swedish Research Council grant
No 2017-00648. We gratefully acknowledge support from the Science and Technology Facilities Council (JS through ST/N50404X/1; DJR and DMA through grant ST/L00075X/1). JS and KK acknowledge support from the Knut and Alice Wallenberg Foundation. C.C.C. acknowledges support from the Ministry of Science and Technology of Taiwan (MOST 109-2112-M-001-016-MY3). This paper makes use of ALMA data: 2017.00112.S, 2013.0.00965.S and 2015.1.00407.S. ALMA is a partnership of ESO (representing its member states), NSF (USA) and NINS (Japan), together with NRC (Canada) and NSC and ASIAA (Taiwan), in cooperation with the Republic of Chile. The Joint ALMA Observatory is operated by ESO, AUI/NRAO and NAOJ. We thank the authors of \citet{Carniani16} for making their K-band cubes public. This work uses data from VLT/SINFONI that were part of programmes ID 077.B-0218(A), 086.B-0579(A) and 091.A-0261(A). 
This work has made use of data from the European Space Agency (ESA) mission {\it Gaia} (\url{https://www.cosmos.esa.int/gaia}), processed by the {\it Gaia} Data Processing and Analysis Consortium (DPAC,
\url{https://www.cosmos.esa.int/web/gaia/dpac/consortium}). Funding for the DPAC has been provided by national institutions, in particular the institutions
participating in the {\it Gaia} Multilateral Agreement.

\bibliographystyle{mnras}
\bibliography{mybib.bib} 

\begin{thebibliography}{}
\makeatletter
\relax
\def\mn@urlcharsother{\let\do\@makeother \do\$\do\&\do\#\do\^\do\_\do\%\do\~}
\def\mn@doi{\begingroup\mn@urlcharsother \@ifnextchar [ {\mn@doi@}
  {\mn@doi@[]}}
\def\mn@doi@[#1]#2{\def\@tempa{#1}\ifx\@tempa\@empty \href
  {http://dx.doi.org/#2} {doi:#2}\else \href {http://dx.doi.org/#2} {#1}\fi
  \endgroup}
\def\mn@eprint#1#2{\mn@eprint@#1:#2::\@nil}
\def\mn@eprint@arXiv#1{\href {http://arxiv.org/abs/#1} {{\tt arXiv:#1}}}
\def\mn@eprint@dblp#1{\href {http://dblp.uni-trier.de/rec/bibtex/#1.xml}
  {dblp:#1}}
\def\mn@eprint@#1:#2:#3:#4\@nil{\def\@tempa {#1}\def\@tempb {#2}\def\@tempc
  {#3}\ifx \@tempc \@empty \let \@tempc \@tempb \let \@tempb \@tempa \fi \ifx
  \@tempb \@empty \def\@tempb {arXiv}\fi \@ifundefined
  {mn@eprint@\@tempb}{\@tempb:\@tempc}{\expandafter \expandafter \csname
  mn@eprint@\@tempb\endcsname \expandafter{\@tempc}}}

\bibitem[\protect\citeauthoryear{{Alaghband-Zadeh}, {Banerji}, {Hewett}  \&
  {McMahon}}{{Alaghband-Zadeh} et~al.}{2016}]{AlaghbandZadeh16}
{Alaghband-Zadeh} S.,  {Banerji} M.,  {Hewett} P.~C.,   {McMahon} R.~G.,  2016,
  \mn@doi [\mnras] {10.1093/mnras/stw682}, \href
  {https://ui.adsabs.harvard.edu/abs/2016MNRAS.459..999A} {459, 999}

\bibitem[\protect\citeauthoryear{{Alatalo} et~al.,}{{Alatalo}
  et~al.}{2015}]{Alatalo15}
{Alatalo} K.,  et~al., 2015, \mn@doi [\apj] {10.1088/0004-637X/798/1/31}, \href
  {https://ui.adsabs.harvard.edu/abs/2015ApJ...798...31A} {798, 31}

\bibitem[\protect\citeauthoryear{{Alexander} \& {Hickox}}{{Alexander} \&
  {Hickox}}{2012}]{Alexander12}
{Alexander} D.~M.,  {Hickox} R.~C.,  2012, \mn@doi [\nar]
  {10.1016/j.newar.2011.11.003}, \href
  {http://adsabs.harvard.edu/abs/2012NewAR..56...93A} {56, 93}

\bibitem[\protect\citeauthoryear{{Balmaverde} \& {Capetti}}{{Balmaverde} \&
  {Capetti}}{2015}]{Balmaverde15}
{Balmaverde} B.,  {Capetti} A.,  2015, \mn@doi [\aap]
  {10.1051/0004-6361/201526496}, \href
  {http://adsabs.harvard.edu/abs/2015A%26A...581A..76B} {581, A76}

\bibitem[\protect\citeauthoryear{{Balmaverde} et~al.,}{{Balmaverde}
  et~al.}{2016}]{Balmaverde16}
{Balmaverde} B.,  et~al., 2016, \mn@doi [\aap] {10.1051/0004-6361/201526694},
  \href {https://ui.adsabs.harvard.edu/abs/2016A&A...585A.148B} {585, A148}

\bibitem[\protect\citeauthoryear{{Baron} et~al.,}{{Baron}
  et~al.}{2018}]{Baron18}
{Baron} D.,  et~al., 2018, \mn@doi [\mnras] {10.1093/mnras/sty2113}, \href
  {https://ui.adsabs.harvard.edu/abs/2018MNRAS.480.3993B} {480, 3993}

\bibitem[\protect\citeauthoryear{{Beckmann} et~al.,}{{Beckmann}
  et~al.}{2017}]{Beckmann17}
{Beckmann} R.~S.,  et~al., 2017, preprint, \href
  {http://adsabs.harvard.edu/abs/2017arXiv170107838B} {} (\mn@eprint {arXiv}
  {1701.07838})

\bibitem[\protect\citeauthoryear{{Bennert}, {Auger}, {Treu}, {Woo}  \&
  {Malkan}}{{Bennert} et~al.}{2011}]{Bennert11}
{Bennert} V.~N.,  {Auger} M.~W.,  {Treu} T.,  {Woo} J.-H.,   {Malkan} M.~A.,
  2011, \mn@doi [\apj] {10.1088/0004-637X/742/2/107}, \href
  {https://ui.adsabs.harvard.edu/abs/2011ApJ...742..107B} {742, 107}

\bibitem[\protect\citeauthoryear{{Bernhard}, {Grimmett}, {Mullaney}, {Daddi},
  {Tadhunter}  \& {Jin}}{{Bernhard} et~al.}{2019}]{Bernhard19}
{Bernhard} E.,  {Grimmett} L.~P.,  {Mullaney} J.~R.,  {Daddi} E.,  {Tadhunter}
  C.,   {Jin} S.,  2019, \mn@doi [\mnras] {10.1093/mnrasl/sly217}, \href
  {https://ui.adsabs.harvard.edu/abs/2019MNRAS.483L..52B} {483, L52}

\bibitem[\protect\citeauthoryear{{Bischetti} et~al.,}{{Bischetti}
  et~al.}{2017}]{Bischetti17}
{Bischetti} M.,  et~al., 2017, \mn@doi [\aap] {10.1051/0004-6361/201629301},
  \href {https://ui.adsabs.harvard.edu/abs/2017A&A...598A.122B} {598, A122}

\bibitem[\protect\citeauthoryear{{Brusa} et~al.,}{{Brusa}
  et~al.}{2015}]{Brusa15}
{Brusa} M.,  et~al., 2015, \mn@doi [\mnras] {10.1093/mnras/stu2117}, \href
  {https://ui.adsabs.harvard.edu/abs/2015MNRAS.446.2394B} {446, 2394}

\bibitem[\protect\citeauthoryear{{Brusa} et~al.,}{{Brusa}
  et~al.}{2018}]{Brusa18}
{Brusa} M.,  et~al., 2018, \mn@doi [\aap] {10.1051/0004-6361/201731641}, \href
  {http://adsabs.harvard.edu/abs/2018A%26A...612A..29B} {612, A29}

\bibitem[\protect\citeauthoryear{{Burgarella} et~al.,}{{Burgarella}
  et~al.}{2013}]{Burgarella13}
{Burgarella} D.,  et~al., 2013, \mn@doi [\aap] {10.1051/0004-6361/201321651},
  \href {https://ui.adsabs.harvard.edu/abs/2013A&A...554A..70B} {554, A70}

\bibitem[\protect\citeauthoryear{{Calzetti}}{{Calzetti}}{2013}]{Calzetti13}
{Calzetti} D.,  2013, {Star Formation Rate Indicators}.
p.~419

\bibitem[\protect\citeauthoryear{{Cano-D{\'{\i}}az}, {Maiolino}, {Marconi},
  {Netzer}, {Shemmer}  \& {Cresci}}{{Cano-D{\'{\i}}az}
  et~al.}{2012}]{Canodiaz12}
{Cano-D{\'{\i}}az} M.,  {Maiolino} R.,  {Marconi} A.,  {Netzer} H.,  {Shemmer}
  O.,   {Cresci} G.,  2012, \mn@doi [\aap] {10.1051/0004-6361/201118358}, \href
  {http://adsabs.harvard.edu/abs/2012A%26A...537L...8C} {537, L8}

\bibitem[\protect\citeauthoryear{{Carniani} et~al.,}{{Carniani}
  et~al.}{2015}]{Carniani15}
{Carniani} S.,  et~al., 2015, \mn@doi [\aap] {10.1051/0004-6361/201526557},
  \href {http://adsabs.harvard.edu/abs/2015A%26A...580A.102C} {580, A102}

\bibitem[\protect\citeauthoryear{{Carniani} et~al.,}{{Carniani}
  et~al.}{2016}]{Carniani16}
{Carniani} S.,  et~al., 2016, \mn@doi [\aap] {10.1051/0004-6361/201528037},
  \href {http://adsabs.harvard.edu/abs/2016A%26A...591A..28C} {591, A28}

\bibitem[\protect\citeauthoryear{{Carniani} et~al.,}{{Carniani}
  et~al.}{2017}]{Carniani17}
{Carniani} S.,  et~al., 2017, \mn@doi [\aap] {10.1051/0004-6361/201730672},
  \href {https://ui.adsabs.harvard.edu/abs/2017A&A...605A.105C} {605, A105}

\bibitem[\protect\citeauthoryear{{Casey}, {Narayanan}  \& {Cooray}}{{Casey}
  et~al.}{2014}]{Casey14}
{Casey} C.~M.,  {Narayanan} D.,   {Cooray} A.,  2014, \mn@doi [\physrep]
  {10.1016/j.physrep.2014.02.009}, \href
  {http://adsabs.harvard.edu/abs/2014PhR...541...45C} {541, 45}

\bibitem[\protect\citeauthoryear{{Chabrier}}{{Chabrier}}{2003}]{Chabrier03}
{Chabrier} G.,  2003, \mn@doi [\pasp] {10.1086/376392}, \href
  {http://adsabs.harvard.edu/abs/2003PASP..115..763C} {115, 763}

\bibitem[\protect\citeauthoryear{{Chambers} et~al.,}{{Chambers}
  et~al.}{2016}]{Panstarrs}
{Chambers} K.~C.,  et~al., 2016, arXiv e-prints, \href
  {https://ui.adsabs.harvard.edu/abs/2016arXiv161205560C} {p. arXiv:1612.05560}

\bibitem[\protect\citeauthoryear{{Chang} et~al.,}{{Chang}
  et~al.}{2020}]{Chang20}
{Chang} Y.-Y.,  et~al., 2020, \mn@doi [\apj] {10.3847/1538-4357/ab595b}, \href
  {https://ui.adsabs.harvard.edu/abs/2020ApJ...888...44C} {888, 44}

\bibitem[\protect\citeauthoryear{{Chen} et~al.,}{{Chen} et~al.}{2017}]{Chen17}
{Chen} C.-C.,  et~al., 2017, \mn@doi [\apj] {10.3847/1538-4357/aa863a}, \href
  {http://adsabs.harvard.edu/abs/2017ApJ...846..108C} {846, 108}

\bibitem[\protect\citeauthoryear{{Chen} et~al.,}{{Chen} et~al.}{2020}]{chen20}
{Chen} C.-C.,  et~al., 2020, \mn@doi [\aap] {10.1051/0004-6361/201936286},
  \href {https://ui.adsabs.harvard.edu/abs/2020A&A...635A.119C} {635, A119}

\bibitem[\protect\citeauthoryear{{Choi}, {Somerville}, {Ostriker}, {Naab}  \&
  {Hirschmann}}{{Choi} et~al.}{2018}]{Choi18}
{Choi} E.,  {Somerville} R.~S.,  {Ostriker} J.~P.,  {Naab} T.,   {Hirschmann}
  M.,  2018, \mn@doi [\apj] {10.3847/1538-4357/aae076}, \href
  {https://ui.adsabs.harvard.edu/abs/2018ApJ...866...91C} {866, 91}

\bibitem[\protect\citeauthoryear{{Cicone}, {Feruglio}, {Maiolino}, {Fiore},
  {Piconcelli}, {Menci}, {Aussel}  \& {Sturm}}{{Cicone}
  et~al.}{2012}]{Cicone12}
{Cicone} C.,  {Feruglio} C.,  {Maiolino} R.,  {Fiore} F.,  {Piconcelli} E.,
  {Menci} N.,  {Aussel} H.,   {Sturm} E.,  2012, \mn@doi [\aap]
  {10.1051/0004-6361/201218793}, \href
  {http://adsabs.harvard.edu/abs/2012A%26A...543A..99C} {543, A99}

\bibitem[\protect\citeauthoryear{{Cicone} et~al.,}{{Cicone}
  et~al.}{2014}]{Cicone14}
{Cicone} C.,  et~al., 2014, \mn@doi [\aap] {10.1051/0004-6361/201322464}, \href
  {http://adsabs.harvard.edu/abs/2014A%26A...562A..21C} {562, A21}

\bibitem[\protect\citeauthoryear{{Cicone}, {Brusa}, {Ramos Almeida}, {Cresci},
  {Husemann}  \& {Mainieri}}{{Cicone} et~al.}{2018}]{Cicone18}
{Cicone} C.,  {Brusa} M.,  {Ramos Almeida} C.,  {Cresci} G.,  {Husemann} B.,
  {Mainieri} V.,  2018, \mn@doi [Nature Astronomy] {10.1038/s41550-018-0406-3},
  \href {https://ui.adsabs.harvard.edu/abs/2018NatAs...2..176C} {2, 176}

\bibitem[\protect\citeauthoryear{{Circosta} et~al.,}{{Circosta}
  et~al.}{2018}]{Circosta18}
{Circosta} C.,  et~al., 2018, \mn@doi [\aap] {10.1051/0004-6361/201833520},
  \href {https://ui.adsabs.harvard.edu/abs/2018A&A...620A..82C} {620, A82}

\bibitem[\protect\citeauthoryear{{Condon}, {Cotton}, {Greisen}, {Yin},
  {Perley}, {Taylor}  \& {Broderick}}{{Condon} et~al.}{1998}]{NVSS}
{Condon} J.~J.,  {Cotton} W.~D.,  {Greisen} E.~W.,  {Yin} Q.~F.,  {Perley}
  R.~A.,  {Taylor} G.~B.,   {Broderick} J.~J.,  1998, \mn@doi [\aj]
  {10.1086/300337}, \href
  {https://ui.adsabs.harvard.edu/abs/1998AJ....115.1693C} {115, 1693}

\bibitem[\protect\citeauthoryear{{Costa}, {Rosdahl}, {Sijacki}  \&
  {Haehnelt}}{{Costa} et~al.}{2018}]{Costa18}
{Costa} T.,  {Rosdahl} J.,  {Sijacki} D.,   {Haehnelt} M.~G.,  2018, \mn@doi
  [\mnras] {10.1093/mnras/sty1514}, \href
  {https://ui.adsabs.harvard.edu/abs/2018MNRAS.479.2079C} {479, 2079}

\bibitem[\protect\citeauthoryear{{Costa}, {Pakmor}  \& {Springel}}{{Costa}
  et~al.}{2020}]{Costa20}
{Costa} T.,  {Pakmor} R.,   {Springel} V.,  2020, \mn@doi [\mnras]
  {10.1093/mnras/staa2321}, \href
  {https://ui.adsabs.harvard.edu/abs/2020MNRAS.497.5229C} {497, 5229}

\bibitem[\protect\citeauthoryear{{Crain} et~al.,}{{Crain}
  et~al.}{2015}]{Crain15}
{Crain} R.~A.,  et~al., 2015, \mn@doi [\mnras] {10.1093/mnras/stv725}, \href
  {http://adsabs.harvard.edu/abs/2015MNRAS.450.1937C} {450, 1937}

\bibitem[\protect\citeauthoryear{{Cresci} \& {Maiolino}}{{Cresci} \&
  {Maiolino}}{2018}]{Cresci18}
{Cresci} G.,  {Maiolino} R.,  2018, \mn@doi [Nature Astronomy]
  {10.1038/s41550-018-0404-5}, \href
  {https://ui.adsabs.harvard.edu/abs/2018NatAs...2..179C} {2, 179}

\bibitem[\protect\citeauthoryear{{Cresci} et~al.,}{{Cresci}
  et~al.}{2015a}]{Cresci15b}
{Cresci} G.,  et~al., 2015a, \mn@doi [\aap] {10.1051/0004-6361/201526581},
  \href {https://ui.adsabs.harvard.edu/abs/2015A&A...582A..63C} {582, A63}

\bibitem[\protect\citeauthoryear{{Cresci} et~al.,}{{Cresci}
  et~al.}{2015b}]{Cresci15}
{Cresci} G.,  et~al., 2015b, \mn@doi [\apj] {10.1088/0004-637X/799/1/82}, \href
  {http://adsabs.harvard.edu/abs/2015ApJ...799...82C} {799, 82}

\bibitem[\protect\citeauthoryear{{Crichton} et~al.,}{{Crichton}
  et~al.}{2016}]{Crichton16}
{Crichton} D.,  et~al., 2016, \mn@doi [\mnras] {10.1093/mnras/stw344}, \href
  {https://ui.adsabs.harvard.edu/abs/2016MNRAS.458.1478C} {458, 1478}

\bibitem[\protect\citeauthoryear{{Croft} et~al.,}{{Croft}
  et~al.}{2006}]{Croft06}
{Croft} S.,  et~al., 2006, \mn@doi [\apj] {10.1086/505526}, \href
  {https://ui.adsabs.harvard.edu/abs/2006ApJ...647.1040C} {647, 1040}

\bibitem[\protect\citeauthoryear{{Dale}, {Helou}, {Magdis}, {Rigopoulou},
  {5MUSES}  \& {HerMES}}{{Dale} et~al.}{2014}]{dale14}
{Dale} D.~A.,  {Helou} G.,  {Magdis} G.,  {Rigopoulou} D.,  {5MUSES}  {HerMES}
  2014, in American Astronomical Society Meeting Abstracts \#223. p. 453.01

\bibitem[\protect\citeauthoryear{{Davies} et~al.,}{{Davies}
  et~al.}{2020a}]{Davies20}
{Davies} R.,  et~al., 2020a, \mn@doi [\mnras] {10.1093/mnras/staa2413}, \href
  {https://ui.adsabs.harvard.edu/abs/2020MNRAS.tmp.2061D} {}

\bibitem[\protect\citeauthoryear{{Davies} et~al.,}{{Davies}
  et~al.}{2020b}]{Davies20KMOS}
{Davies} R.~L.,  et~al., 2020b, \mn@doi [\apj] {10.3847/1538-4357/ab86ad},
  \href {https://ui.adsabs.harvard.edu/abs/2020ApJ...894...28D} {894, 28}

\bibitem[\protect\citeauthoryear{{Di Matteo}, {Springel}  \& {Hernquist}}{{Di
  Matteo} et~al.}{2005}]{DiMatteo05}
{Di Matteo} T.,  {Springel} V.,   {Hernquist} L.,  2005, \mn@doi [\nat]
  {10.1038/nature03335}, \href
  {http://adsabs.harvard.edu/abs/2005Natur.433..604D} {433, 604}

\bibitem[\protect\citeauthoryear{{Dicken}, {Tadhunter}, {Morganti}, {Buchanan},
  {Oosterloo}  \& {Axon}}{{Dicken} et~al.}{2008}]{Dicken08}
{Dicken} D.,  {Tadhunter} C.,  {Morganti} R.,  {Buchanan} C.,  {Oosterloo} T.,
   {Axon} D.,  2008, \mn@doi [\apj] {10.1086/529544}, \href
  {https://ui.adsabs.harvard.edu/abs/2008ApJ...678..712D} {678, 712}

\bibitem[\protect\citeauthoryear{{Dimitrijevi{\'c}}, {Popovi{\'c}}, {Kova{\v
  c}evi{\'c}}, {Da{\v c}i{\'c}}  \& {Ili{\'c}}}{{Dimitrijevi{\'c}}
  et~al.}{2007}]{Dimitrijevic07}
{Dimitrijevi{\'c}} M.~S.,  {Popovi{\'c}} L.~{\v C}.,  {Kova{\v c}evi{\'c}} J.,
  {Da{\v c}i{\'c}} M.,   {Ili{\'c}} D.,  2007, \mn@doi [\mnras]
  {10.1111/j.1365-2966.2006.11238.x}, \href
  {http://adsabs.harvard.edu/abs/2007MNRAS.374.1181D} {374, 1181}

\bibitem[\protect\citeauthoryear{{Dubois}, {Pichon}, {Devriendt}, {Silk},
  {Haehnelt}, {Kimm}  \& {Slyz}}{{Dubois} et~al.}{2013a}]{Dubois13}
{Dubois} Y.,  {Pichon} C.,  {Devriendt} J.,  {Silk} J.,  {Haehnelt} M.,  {Kimm}
  T.,   {Slyz} A.,  2013a, \mn@doi [\mnras] {10.1093/mnras/sts224}, \href
  {https://ui.adsabs.harvard.edu/abs/2013MNRAS.428.2885D} {428, 2885}

\bibitem[\protect\citeauthoryear{{Dubois}, {Gavazzi}, {Peirani}  \&
  {Silk}}{{Dubois} et~al.}{2013b}]{Dubois13b}
{Dubois} Y.,  {Gavazzi} R.,  {Peirani} S.,   {Silk} J.,  2013b, \mn@doi
  [\mnras] {10.1093/mnras/stt997}, \href
  {https://ui.adsabs.harvard.edu/abs/2013MNRAS.433.3297D} {433, 3297}

\bibitem[\protect\citeauthoryear{{Falkendal} et~al.,}{{Falkendal}
  et~al.}{2019}]{Falkendal19}
{Falkendal} T.,  et~al., 2019, \mn@doi [\aap] {10.1051/0004-6361/201732485},
  \href {https://ui.adsabs.harvard.edu/abs/2019A&A...621A..27F} {621, A27}

\bibitem[\protect\citeauthoryear{{Feruglio} et~al.,}{{Feruglio}
  et~al.}{2015}]{Feruglio15}
{Feruglio} C.,  et~al., 2015, \mn@doi [\aap] {10.1051/0004-6361/201526020},
  \href {https://ui.adsabs.harvard.edu/abs/2015A&A...583A..99F} {583, A99}

\bibitem[\protect\citeauthoryear{{Fiore} et~al.,}{{Fiore}
  et~al.}{2017}]{Fiore17}
{Fiore} F.,  et~al., 2017, \mn@doi [\aap] {10.1051/0004-6361/201629478}, \href
  {https://ui.adsabs.harvard.edu/abs/2017A&A...601A.143F} {601, A143}

\bibitem[\protect\citeauthoryear{{Fluetsch} et~al.,}{{Fluetsch}
  et~al.}{2019}]{Fluetsch19}
{Fluetsch} A.,  et~al., 2019, \mn@doi [\mnras] {10.1093/mnras/sty3449}, \href
  {https://ui.adsabs.harvard.edu/abs/2019MNRAS.483.4586F} {483, 4586}

\bibitem[\protect\citeauthoryear{{Fogasy}, {Knudsen}, {Drouart}, {Lagos}  \&
  {Fan}}{{Fogasy} et~al.}{2020}]{Fogasy20}
{Fogasy} J.,  {Knudsen} K.~K.,  {Drouart} G.,  {Lagos} C.~D.~P.,   {Fan} L.,
  2020, \mn@doi [\mnras] {10.1093/mnras/staa472}, \href
  {https://ui.adsabs.harvard.edu/abs/2020MNRAS.493.3744F} {493, 3744}

\bibitem[\protect\citeauthoryear{{F{\"o}rster Schreiber} et~al.,}{{F{\"o}rster
  Schreiber} et~al.}{2009}]{FSchreiber09}
{F{\"o}rster Schreiber} N.~M.,  et~al., 2009, \mn@doi [\apj]
  {10.1088/0004-637X/706/2/1364}, \href
  {http://adsabs.harvard.edu/abs/2009ApJ...706.1364F} {706, 1364}

\bibitem[\protect\citeauthoryear{{F{\"o}rster Schreiber} et~al.,}{{F{\"o}rster
  Schreiber} et~al.}{2018a}]{ForsterSch18b}
{F{\"o}rster Schreiber} N.~M.,  et~al., 2018a, arXiv e-prints, \href
  {http://adsabs.harvard.edu/abs/2018arXiv180704738F} {}

\bibitem[\protect\citeauthoryear{{F{\"o}rster Schreiber} et~al.,}{{F{\"o}rster
  Schreiber} et~al.}{2018b}]{ForsterSch18a}
{F{\"o}rster Schreiber} N.~M.,  et~al., 2018b, \mn@doi [\apjs]
  {10.3847/1538-4365/aadd49}, \href
  {http://adsabs.harvard.edu/abs/2018ApJS..238...21F} {238, 21}

\bibitem[\protect\citeauthoryear{{Freudling}, {Romaniello}, {Bramich},
  {Ballester}, {Forchi}, {Garc{\'\i}a-Dabl{\'o}}, {Moehler}  \&
  {Neeser}}{{Freudling} et~al.}{2013}]{Freudling13}
{Freudling} W.,  {Romaniello} M.,  {Bramich} D.~M.,  {Ballester} P.,  {Forchi}
  V.,  {Garc{\'\i}a-Dabl{\'o}} C.~E.,  {Moehler} S.,   {Neeser} M.~J.,  2013,
  \mn@doi [\aap] {10.1051/0004-6361/201322494}, \href
  {https://ui.adsabs.harvard.edu/abs/2013A&A...559A..96F} {559, A96}

\bibitem[\protect\citeauthoryear{{Fujimoto} et~al.,}{{Fujimoto}
  et~al.}{2018}]{Fujimoto18}
{Fujimoto} S.,  et~al., 2018, \mn@doi [\apj] {10.3847/1538-4357/aac6c4}, \href
  {http://adsabs.harvard.edu/abs/2018ApJ...861....7F} {861, 7}

\bibitem[\protect\citeauthoryear{{Gabor} \& {Bournaud}}{{Gabor} \&
  {Bournaud}}{2014}]{Gabor14}
{Gabor} J.~M.,  {Bournaud} F.,  2014, \mn@doi [\mnras] {10.1093/mnras/stu677},
  \href {https://ui.adsabs.harvard.edu/abs/2014MNRAS.441.1615G} {441, 1615}

\bibitem[\protect\citeauthoryear{{Gaia Collaboration} et~al.,}{{Gaia
  Collaboration} et~al.}{2018}]{Gaiacol18b}
{Gaia Collaboration} et~al., 2018, \mn@doi [\aap]
  {10.1051/0004-6361/201833051}, \href
  {https://ui.adsabs.harvard.edu/abs/2018A&A...616A...1G} {616, A1}

\bibitem[\protect\citeauthoryear{{Gallagher}, {Maiolino}, {Belfiore}, {Drory},
  {Riffel}  \& {Riffel}}{{Gallagher} et~al.}{2019}]{Gallagher19}
{Gallagher} R.,  {Maiolino} R.,  {Belfiore} F.,  {Drory} N.,  {Riffel} R.,
  {Riffel} R.~A.,  2019, \mn@doi [\mnras] {10.1093/mnras/stz564}, \href
  {https://ui.adsabs.harvard.edu/abs/2019MNRAS.485.3409G} {485, 3409}

\bibitem[\protect\citeauthoryear{{Ganguly} \& {Brotherton}}{{Ganguly} \&
  {Brotherton}}{2008}]{Ganguly08}
{Ganguly} R.,  {Brotherton} M.~S.,  2008, \mn@doi [\apj] {10.1086/524106},
  \href {http://adsabs.harvard.edu/abs/2008ApJ...672..102G} {672, 102}

\bibitem[\protect\citeauthoryear{{Genzel} et~al.,}{{Genzel}
  et~al.}{2014}]{Genzel14}
{Genzel} R.,  et~al., 2014, \mn@doi [\apj] {10.1088/0004-637X/796/1/7}, \href
  {http://adsabs.harvard.edu/abs/2014ApJ...796....7G} {796, 7}

\bibitem[\protect\citeauthoryear{{Grimmett}, {Mullaney}, {Bernhard},
  {Harrison}, {Alexander}, {Stanley}, {Masoura}  \& {Walters}}{{Grimmett}
  et~al.}{2020}]{Grimmett20}
{Grimmett} L.~P.,  {Mullaney} J.~R.,  {Bernhard} E.~P.,  {Harrison} C.~M.,
  {Alexander} D.~M.,  {Stanley} F.,  {Masoura} V.~A.,   {Walters} K.,  2020,
  \mn@doi [\mnras] {10.1093/mnras/staa1255}, \href
  {https://ui.adsabs.harvard.edu/abs/2020MNRAS.495.1392G} {495, 1392}

\bibitem[\protect\citeauthoryear{{Hao}, {Kennicutt}, {Johnson}, {Calzetti},
  {Dale}  \& {Moustakas}}{{Hao} et~al.}{2011}]{Hao11}
{Hao} C.-N.,  {Kennicutt} R.~C.,  {Johnson} B.~D.,  {Calzetti} D.,  {Dale}
  D.~A.,   {Moustakas} J.,  2011, \mn@doi [\apj] {10.1088/0004-637X/741/2/124},
  \href {https://ui.adsabs.harvard.edu/abs/2011ApJ...741..124H} {741, 124}

\bibitem[\protect\citeauthoryear{{Harrison}}{{Harrison}}{2017}]{Harrison17}
{Harrison} C.~M.,  2017, \mn@doi [Nature Astronomy] {10.1038/s41550-017-0165},
  \href {http://adsabs.harvard.edu/abs/2017NatAs...1E.165H} {1, 0165}

\bibitem[\protect\citeauthoryear{{Harrison} et~al.,}{{Harrison}
  et~al.}{2012}]{Harrison12}
{Harrison} C.~M.,  et~al., 2012, \mn@doi [\apjl] {10.1088/2041-8205/760/1/L15},
  \href {http://adsabs.harvard.edu/abs/2012ApJ...760L..15H} {760, L15}

\bibitem[\protect\citeauthoryear{{Harrison}, {Alexander}, {Mullaney}  \&
  {Swinbank}}{{Harrison} et~al.}{2014}]{Harrison14}
{Harrison} C.~M.,  {Alexander} D.~M.,  {Mullaney} J.~R.,   {Swinbank} A.~M.,
  2014, \mn@doi [\mnras] {10.1093/mnras/stu515}, \href
  {http://adsabs.harvard.edu/abs/2014MNRAS.441.3306H} {441, 3306}

\bibitem[\protect\citeauthoryear{{Harrison} et~al.,}{{Harrison}
  et~al.}{2016a}]{Harrison16}
{Harrison} C.~M.,  et~al., 2016a, \mn@doi [\mnras] {10.1093/mnras/stv2727},
  \href {http://adsabs.harvard.edu/abs/2016MNRAS.456.1195H} {456, 1195}

\bibitem[\protect\citeauthoryear{{Harrison} et~al.,}{{Harrison}
  et~al.}{2016b}]{Harrison16Alm}
{Harrison} C.~M.,  et~al., 2016b, \mn@doi [\mnras] {10.1093/mnrasl/slw001},
  \href {http://adsabs.harvard.edu/abs/2016MNRAS.457L.122H} {457, L122}

\bibitem[\protect\citeauthoryear{{Harrison}, {Costa}, {Tadhunter},
  {Fl{\"u}tsch}, {Kakkad}, {Perna}  \& {Vietri}}{{Harrison}
  et~al.}{2018}]{Harrison18}
{Harrison} C.~M.,  {Costa} T.,  {Tadhunter} C.~N.,  {Fl{\"u}tsch} A.,  {Kakkad}
  D.,  {Perna} M.,   {Vietri} G.,  2018, \mn@doi [Nature Astronomy]
  {10.1038/s41550-018-0403-6}, \href
  {https://ui.adsabs.harvard.edu/abs/2018NatAs...2..198H} {2, 198}

\bibitem[\protect\citeauthoryear{{Healey}, {Romani}, {Taylor}, {Sadler},
  {Ricci}, {Murphy}, {Ulvestad}  \& {Winn}}{{Healey} et~al.}{2007}]{Healey07}
{Healey} S.~E.,  {Romani} R.~W.,  {Taylor} G.~B.,  {Sadler} E.~M.,  {Ricci} R.,
   {Murphy} T.,  {Ulvestad} J.~S.,   {Winn} J.~N.,  2007, \mn@doi [\apjs]
  {10.1086/513742}, \href
  {https://ui.adsabs.harvard.edu/abs/2007ApJS..171...61H} {171, 61}

\bibitem[\protect\citeauthoryear{{Helfand}, {White}  \& {Becker}}{{Helfand}
  et~al.}{2015}]{First}
{Helfand} D.~J.,  {White} R.~L.,   {Becker} R.~H.,  2015, \mn@doi [\apj]
  {10.1088/0004-637X/801/1/26}, \href
  {https://ui.adsabs.harvard.edu/abs/2015ApJ...801...26H} {801, 26}

\bibitem[\protect\citeauthoryear{{Hirschmann}, {Dolag}, {Saro}, {Bachmann},
  {Borgani}  \& {Burkert}}{{Hirschmann} et~al.}{2014}]{Hirschmann14}
{Hirschmann} M.,  {Dolag} K.,  {Saro} A.,  {Bachmann} L.,  {Borgani} S.,
  {Burkert} A.,  2014, \mn@doi [\mnras] {10.1093/mnras/stu1023}, \href
  {https://ui.adsabs.harvard.edu/abs/2014MNRAS.442.2304H} {442, 2304}

\bibitem[\protect\citeauthoryear{{Hodge} et~al.,}{{Hodge}
  et~al.}{2016}]{Hodge16}
{Hodge} J.~A.,  et~al., 2016, \mn@doi [\apj] {10.3847/1538-4357/833/1/103},
  \href {http://adsabs.harvard.edu/abs/2016ApJ...833..103H} {833, 103}

\bibitem[\protect\citeauthoryear{{Hooper}, {Impey}, {Foltz}  \&
  {Hewett}}{{Hooper} et~al.}{1995}]{Hooper95}
{Hooper} E.~J.,  {Impey} C.~D.,  {Foltz} C.~B.,   {Hewett} P.~C.,  1995,
  \mn@doi [\apj] {10.1086/175673}, \href
  {https://ui.adsabs.harvard.edu/abs/1995ApJ...445...62H} {445, 62}

\bibitem[\protect\citeauthoryear{{Husemann}, {Scharw{\"a}chter}, {Bennert},
  {Mainieri}, {Woo}  \& {Kakkad}}{{Husemann} et~al.}{2016}]{Husemann16}
{Husemann} B.,  {Scharw{\"a}chter} J.,  {Bennert} V.~N.,  {Mainieri} V.,  {Woo}
  J.~H.,   {Kakkad} D.,  2016, \mn@doi [\aap] {10.1051/0004-6361/201527992},
  \href {https://ui.adsabs.harvard.edu/abs/2016A&A...594A..44H} {594, A44}

\bibitem[\protect\citeauthoryear{{Husemann} et~al.,}{{Husemann}
  et~al.}{2019}]{Husemann19}
{Husemann} B.,  et~al., 2019, \mn@doi [\aap] {10.1051/0004-6361/201935283},
  \href {https://ui.adsabs.harvard.edu/abs/2019A&A...627A..53H} {627, A53}

\bibitem[\protect\citeauthoryear{{Ikarashi} et~al.,}{{Ikarashi}
  et~al.}{2015}]{Ikarashi15}
{Ikarashi} S.,  et~al., 2015, \mn@doi [\apj] {10.1088/0004-637X/810/2/133},
  \href {http://adsabs.harvard.edu/abs/2015ApJ...810..133I} {810, 133}

\bibitem[\protect\citeauthoryear{{Jarvis} et~al.,}{{Jarvis}
  et~al.}{2019}]{Jarvis19}
{Jarvis} M.~E.,  et~al., 2019, \mn@doi [\mnras] {10.1093/mnras/stz556}, \href
  {https://ui.adsabs.harvard.edu/abs/2019MNRAS.485.2710J} {485, 2710}

\bibitem[\protect\citeauthoryear{{Kakkad} et~al.,}{{Kakkad}
  et~al.}{2016}]{Kakkad16}
{Kakkad} D.,  et~al., 2016, \mn@doi [\aap] {10.1051/0004-6361/201527968}, \href
  {https://ui.adsabs.harvard.edu/abs/2016A&A...592A.148K} {592, A148}

\bibitem[\protect\citeauthoryear{{Kakkad} et~al.,}{{Kakkad}
  et~al.}{2017}]{Kakkad17}
{Kakkad} D.,  et~al., 2017, \mn@doi [\mnras] {10.1093/mnras/stx726}, \href
  {https://ui.adsabs.harvard.edu/abs/2017MNRAS.468.4205K} {468, 4205}

\bibitem[\protect\citeauthoryear{{Kakkad} et~al.,}{{Kakkad}
  et~al.}{2020}]{Kakkad20}
{Kakkad} D.,  et~al., 2020, arXiv e-prints, \href
  {https://ui.adsabs.harvard.edu/abs/2020arXiv200801728K} {p. arXiv:2008.01728}

\bibitem[\protect\citeauthoryear{{Karouzos}, {Woo}  \& {Bae}}{{Karouzos}
  et~al.}{2016}]{Karouzos16}
{Karouzos} M.,  {Woo} J.-H.,   {Bae} H.-J.,  2016, \mn@doi [\apj]
  {10.3847/0004-637X/819/2/148}, \href
  {https://ui.adsabs.harvard.edu/abs/2016ApJ...819..148K} {819, 148}

\bibitem[\protect\citeauthoryear{{Kennicutt} \& {Evans}}{{Kennicutt} \&
  {Evans}}{2012}]{Kennicutt12}
{Kennicutt} R.~C.,  {Evans} N.~J.,  2012, \mn@doi [\araa]
  {10.1146/annurev-astro-081811-125610}, \href
  {https://ui.adsabs.harvard.edu/abs/2012ARA&A..50..531K} {50, 531}

\bibitem[\protect\citeauthoryear{{Kirkpatrick}, {Sharon}, {Keller}  \&
  {Pope}}{{Kirkpatrick} et~al.}{2019}]{Kirkpatrick19}
{Kirkpatrick} A.,  {Sharon} C.,  {Keller} E.,   {Pope} A.,  2019, \mn@doi
  [\apj] {10.3847/1538-4357/ab223a}, \href
  {https://ui.adsabs.harvard.edu/abs/2019ApJ...879...41K} {879, 41}

\bibitem[\protect\citeauthoryear{{Kormendy} \& {Ho}}{{Kormendy} \&
  {Ho}}{2013}]{Kormendy13}
{Kormendy} J.,  {Ho} L.~C.,  2013, \mn@doi [\araa]
  {10.1146/annurev-astro-082708-101811}, \href
  {http://adsabs.harvard.edu/abs/2013ARA%26A..51..511K} {51, 511}

\bibitem[\protect\citeauthoryear{{Lacy} et~al.,}{{Lacy} et~al.}{2019}]{Lacy19}
{Lacy} M.,  et~al., 2019, \mn@doi [\mnras] {10.1093/mnrasl/sly215}, \href
  {https://ui.adsabs.harvard.edu/abs/2019MNRAS.483L..22L} {483, L22}

\bibitem[\protect\citeauthoryear{{Lang} et~al.,}{{Lang} et~al.}{2019}]{Lang19}
{Lang} P.,  et~al., 2019, \mn@doi [\apj] {10.3847/1538-4357/ab1f77}, \href
  {https://ui.adsabs.harvard.edu/abs/2019ApJ...879...54L} {879, 54}

\bibitem[\protect\citeauthoryear{{Lansbury}, {Jarvis}, {Harrison}, {Alexander},
  {Del Moro}, {Edge}, {Mullaney}  \& {Thomson}}{{Lansbury}
  et~al.}{2018}]{Lansbury18}
{Lansbury} G.~B.,  {Jarvis} M.~E.,  {Harrison} C.~M.,  {Alexander} D.~M.,  {Del
  Moro} A.,  {Edge} A.~C.,  {Mullaney} J.~R.,   {Thomson} A.~P.,  2018, \mn@doi
  [\apjl] {10.3847/2041-8213/aab357}, \href
  {https://ui.adsabs.harvard.edu/abs/2018ApJ...856L...1L} {856, L1}

\bibitem[\protect\citeauthoryear{{Leung} et~al.,}{{Leung}
  et~al.}{2017}]{Leung17}
{Leung} G.~C.~K.,  et~al., 2017, preprint, \href
  {http://adsabs.harvard.edu/abs/2017arXiv170310255L} {} (\mn@eprint {arXiv}
  {1703.10255})

\bibitem[\protect\citeauthoryear{{Liu}, {Zakamska}, {Greene}, {Nesvadba}  \&
  {Liu}}{{Liu} et~al.}{2013}]{Liu13b}
{Liu} G.,  {Zakamska} N.~L.,  {Greene} J.~E.,  {Nesvadba} N. P.~H.,   {Liu} X.,
   2013, \mn@doi [\mnras] {10.1093/mnras/stt1755}, \href
  {https://ui.adsabs.harvard.edu/abs/2013MNRAS.436.2576L} {436, 2576}

\bibitem[\protect\citeauthoryear{{Madau} \& {Dickinson}}{{Madau} \&
  {Dickinson}}{2014}]{Madau14}
{Madau} P.,  {Dickinson} M.,  2014, \mn@doi [\araa]
  {10.1146/annurev-astro-081811-125615}, \href
  {http://adsabs.harvard.edu/abs/2014ARA%26A..52..415M} {52, 415}

\bibitem[\protect\citeauthoryear{{Madau}, {Ferguson}, {Dickinson},
  {Giavalisco}, {Steidel}  \& {Fruchter}}{{Madau} et~al.}{1996}]{Madau96}
{Madau} P.,  {Ferguson} H.~C.,  {Dickinson} M.~E.,  {Giavalisco} M.,  {Steidel}
  C.~C.,   {Fruchter} A.,  1996, \mn@doi [\mnras] {10.1093/mnras/283.4.1388},
  \href {http://adsabs.harvard.edu/abs/1996MNRAS.283.1388M} {283, 1388}

\bibitem[\protect\citeauthoryear{{Mainzer} et~al.,}{{Mainzer}
  et~al.}{2011}]{WISE}
{Mainzer} A.,  et~al., 2011, \mn@doi [\apj] {10.1088/0004-637X/731/1/53}, \href
  {https://ui.adsabs.harvard.edu/abs/2011ApJ...731...53M} {731, 53}

\bibitem[\protect\citeauthoryear{{Maiolino} et~al.,}{{Maiolino}
  et~al.}{2017}]{Maiolino17}
{Maiolino} R.,  et~al., 2017, \mn@doi [\nat] {10.1038/nature21677}, \href
  {https://ui.adsabs.harvard.edu/abs/2017Natur.544..202M} {544, 202}

\bibitem[\protect\citeauthoryear{{Mart{\'\i}-Vidal}, {Vlemmings}, {Muller}  \&
  {Casey}}{{Mart{\'\i}-Vidal} et~al.}{2014}]{UVMultiFit}
{Mart{\'\i}-Vidal} I.,  {Vlemmings} W.~H.~T.,  {Muller} S.,   {Casey} S.,
  2014, \mn@doi [\aap] {10.1051/0004-6361/201322633}, \href
  {https://ui.adsabs.harvard.edu/abs/2014A&A...563A.136M} {563, A136}

\bibitem[\protect\citeauthoryear{{McCarthy}, {Schaye}, {Bower}, {Ponman},
  {Booth}, {Dalla Vecchia}  \& {Springel}}{{McCarthy}
  et~al.}{2011}]{McCarthy11}
{McCarthy} I.~G.,  {Schaye} J.,  {Bower} R.~G.,  {Ponman} T.~J.,  {Booth}
  C.~M.,  {Dalla Vecchia} C.,   {Springel} V.,  2011, \mn@doi [\mnras]
  {10.1111/j.1365-2966.2010.18033.x}, \href
  {http://adsabs.harvard.edu/abs/2011MNRAS.412.1965M} {412, 1965}

\bibitem[\protect\citeauthoryear{{McElroy} et~al.,}{{McElroy}
  et~al.}{2016}]{McElroy16}
{McElroy} R.~E.,  et~al., 2016, \mn@doi [\aap] {10.1051/0004-6361/201629102},
  \href {https://ui.adsabs.harvard.edu/abs/2016A&A...593L...8M} {593, L8}

\bibitem[\protect\citeauthoryear{{Merloni}, {Rudnick}  \& {Di
  Matteo}}{{Merloni} et~al.}{2004}]{Merloni04}
{Merloni} A.,  {Rudnick} G.,   {Di Matteo} T.,  2004, \mn@doi [\mnras]
  {10.1111/j.1365-2966.2004.08382.x}, \href
  {http://adsabs.harvard.edu/abs/2004MNRAS.354L..37M} {354, L37}

\bibitem[\protect\citeauthoryear{{Morganti}, {Tadhunter}  \&
  {Oosterloo}}{{Morganti} et~al.}{2005}]{Morganti05}
{Morganti} R.,  {Tadhunter} C.~N.,   {Oosterloo} T.~A.,  2005, \mn@doi [\aap]
  {10.1051/0004-6361:200500197}, \href
  {https://ui.adsabs.harvard.edu/abs/2005A&A...444L...9M} {444, L9}

\bibitem[\protect\citeauthoryear{{Mukherjee}, {Bicknell}, {Wagner},
  {Sutherland}  \& {Silk}}{{Mukherjee} et~al.}{2018}]{Mukherjee18}
{Mukherjee} D.,  {Bicknell} G.~V.,  {Wagner} A.~Y.,  {Sutherland} R.~S.,
  {Silk} J.,  2018, \mn@doi [\mnras] {10.1093/mnras/sty1776}, \href
  {http://adsabs.harvard.edu/abs/2018MNRAS.479.5544M} {479, 5544}

\bibitem[\protect\citeauthoryear{{Mullaney}, {Alexander}, {Goulding}  \&
  {Hickox}}{{Mullaney} et~al.}{2011}]{Mullaney11}
{Mullaney} J.~R.,  {Alexander} D.~M.,  {Goulding} A.~D.,   {Hickox} R.~C.,
  2011, \mn@doi [\mnras] {10.1111/j.1365-2966.2011.18448.x}, \href
  {http://adsabs.harvard.edu/abs/2011MNRAS.414.1082M} {414, 1082}

\bibitem[\protect\citeauthoryear{{Mullaney}, {Alexander}, {Fine}, {Goulding},
  {Harrison}  \& {Hickox}}{{Mullaney} et~al.}{2013}]{Mullaney13}
{Mullaney} J.~R.,  {Alexander} D.~M.,  {Fine} S.,  {Goulding} A.~D.,
  {Harrison} C.~M.,   {Hickox} R.~C.,  2013, \mn@doi [\mnras]
  {10.1093/mnras/stt751}, \href
  {http://adsabs.harvard.edu/abs/2013MNRAS.433..622M} {433, 622}

\bibitem[\protect\citeauthoryear{{Mullaney} et~al.,}{{Mullaney}
  et~al.}{2015}]{Mullaney15}
{Mullaney} J.~R.,  et~al., 2015, \mn@doi [\mnras] {10.1093/mnrasl/slv110},
  \href {http://adsabs.harvard.edu/abs/2015MNRAS.453L..83M} {453, L83}

\bibitem[\protect\citeauthoryear{{Murphy}, {Chary}, {Dickinson}, {Pope},
  {Frayer}  \& {Lin}}{{Murphy} et~al.}{2011}]{Murphy11}
{Murphy} E.~J.,  {Chary} R.~R.,  {Dickinson} M.,  {Pope} A.,  {Frayer} D.~T.,
  {Lin} L.,  2011, \mn@doi [\apj] {10.1088/0004-637X/732/2/126}, \href
  {https://ui.adsabs.harvard.edu/abs/2011ApJ...732..126M} {732, 126}

\bibitem[\protect\citeauthoryear{{Nagao}, {Marconi}  \& {Maiolino}}{{Nagao}
  et~al.}{2006}]{Nagao06}
{Nagao} T.,  {Marconi} A.,   {Maiolino} R.,  2006, \mn@doi [\aap]
  {10.1051/0004-6361:20054024}, \href
  {https://ui.adsabs.harvard.edu/abs/2006A&A...447..157N} {447, 157}

\bibitem[\protect\citeauthoryear{{Netzer}, {Shemmer}, {Maiolino}, {Oliva},
  {Croom}, {Corbett}  \& {di Fabrizio}}{{Netzer} et~al.}{2004}]{Netzer04}
{Netzer} H.,  {Shemmer} O.,  {Maiolino} R.,  {Oliva} E.,  {Croom} S.,
  {Corbett} E.,   {di Fabrizio} L.,  2004, \mn@doi [\apj] {10.1086/423608},
  \href {https://ui.adsabs.harvard.edu/abs/2004ApJ...614..558N} {614, 558}

\bibitem[\protect\citeauthoryear{{Osterbrock} \& {Ferland}}{{Osterbrock} \&
  {Ferland}}{2006}]{Osterbrock06}
{Osterbrock} D.~E.,  {Ferland} G.~J.,  2006, {Astrophysics of gaseous nebulae
  and active galactic nuclei}

\bibitem[\protect\citeauthoryear{{Perna} et~al.,}{{Perna}
  et~al.}{2018}]{Perna18}
{Perna} M.,  et~al., 2018, \mn@doi [\aap] {10.1051/0004-6361/201833040}, \href
  {https://ui.adsabs.harvard.edu/abs/2018A&A...619A..90P} {619, A90}

\bibitem[\protect\citeauthoryear{{Perna}, {Cresci}, {Brusa}, {Lanzuisi},
  {Concas}, {Mainieri}, {Mannucci}  \& {Marconi}}{{Perna}
  et~al.}{2019}]{Perna19}
{Perna} M.,  {Cresci} G.,  {Brusa} M.,  {Lanzuisi} G.,  {Concas} A.,
  {Mainieri} V.,  {Mannucci} F.,   {Marconi} A.,  2019, \mn@doi [\aap]
  {10.1051/0004-6361/201834193}, \href
  {https://ui.adsabs.harvard.edu/abs/2019A&A...623A.171P} {623, A171}

\bibitem[\protect\citeauthoryear{{Perna} et~al.,}{{Perna}
  et~al.}{2020}]{Perna20}
{Perna} M.,  et~al., 2020, arXiv e-prints, \href
  {https://ui.adsabs.harvard.edu/abs/2020arXiv200903353P} {p. arXiv:2009.03353}

\bibitem[\protect\citeauthoryear{{Perrotta}, {Hamann}, {Zakamska}, {Alexand
  roff}, {Rupke}  \& {Wylezalek}}{{Perrotta} et~al.}{2019}]{Perrotta19}
{Perrotta} S.,  {Hamann} F.,  {Zakamska} N.~L.,  {Alexand roff} R.~M.,  {Rupke}
  D.,   {Wylezalek} D.,  2019, \mn@doi [\mnras] {10.1093/mnras/stz1993}, \href
  {https://ui.adsabs.harvard.edu/abs/2019MNRAS.488.4126P} {488, 4126}

\bibitem[\protect\citeauthoryear{{Planck Collaboration} et~al.,}{{Planck
  Collaboration} et~al.}{2014}]{Planck13}
{Planck Collaboration} et~al., 2014, \mn@doi [\aap]
  {10.1051/0004-6361/201321591}, \href
  {https://ui.adsabs.harvard.edu/abs/2014A%26A...571A..16P} {571, A16}

\bibitem[\protect\citeauthoryear{{Pontzen}, {Tremmel}, {Roth}, {Peiris},
  {Saintonge}, {Volonteri}, {Quinn}  \& {Governato}}{{Pontzen}
  et~al.}{2017}]{Pontzen17}
{Pontzen} A.,  {Tremmel} M.,  {Roth} N.,  {Peiris} H.~V.,  {Saintonge} A.,
  {Volonteri} M.,  {Quinn} T.,   {Governato} F.,  2017, \mn@doi [\mnras]
  {10.1093/mnras/stw2627}, \href
  {https://ui.adsabs.harvard.edu/abs/2017MNRAS.465..547P} {465, 547}

\bibitem[\protect\citeauthoryear{{Querejeta} et~al.,}{{Querejeta}
  et~al.}{2016}]{Querejeta16}
{Querejeta} M.,  et~al., 2016, \mn@doi [\aap] {10.1051/0004-6361/201628674},
  \href {https://ui.adsabs.harvard.edu/abs/2016A&A...593A.118Q} {593, A118}

\bibitem[\protect\citeauthoryear{{Ramos Almeida}, {Acosta-Pulido}, {Tadhunter},
  {Gonz{\'a}lez-Fern{\'a}ndez}, {Cicone}  \& {Fern{\'a}ndez-Torreiro}}{{Ramos
  Almeida} et~al.}{2019}]{RamosAlmeida19}
{Ramos Almeida} C.,  {Acosta-Pulido} J.~A.,  {Tadhunter} C.~N.,
  {Gonz{\'a}lez-Fern{\'a}ndez} C.,  {Cicone} C.,   {Fern{\'a}ndez-Torreiro} M.,
   2019, \mn@doi [\mnras] {10.1093/mnrasl/slz072}, \href
  {https://ui.adsabs.harvard.edu/abs/2019MNRAS.487L..18R} {487, L18}

\bibitem[\protect\citeauthoryear{{Revalski} et~al.,}{{Revalski}
  et~al.}{2018}]{Revalski18}
{Revalski} M.,  et~al., 2018, \mn@doi [\apj] {10.3847/1538-4357/aae3e6}, \href
  {https://ui.adsabs.harvard.edu/abs/2018ApJ...867...88R} {867, 88}

\bibitem[\protect\citeauthoryear{{Rohlfs} \& {Wilson}}{{Rohlfs} \&
  {Wilson}}{1996}]{Rohlfs96}
{Rohlfs} K.,  {Wilson} T.~L.,  1996, {Tools of Radio Astronomy}

\bibitem[\protect\citeauthoryear{{Rosario}}{{Rosario}}{2019}]{Fortes}
{Rosario} D.~J.,  2019, {FortesFit: Flexible spectral energy distribution
  modelling with a Bayesian backbone} (\mn@eprint {ascl} {1904.011})

\bibitem[\protect\citeauthoryear{{Rosario}, {Togi}, {Burtscher}, {Davies},
  {Shimizu}  \& {Lutz}}{{Rosario} et~al.}{2019}]{Rosario19}
{Rosario} D.~J.,  {Togi} A.,  {Burtscher} L.,  {Davies} R.~I.,  {Shimizu}
  T.~T.,   {Lutz} D.,  2019, \mn@doi [\apjl] {10.3847/2041-8213/ab1262}, \href
  {https://ui.adsabs.harvard.edu/abs/2019ApJ...875L...8R} {875, L8}

\bibitem[\protect\citeauthoryear{{Rose}, {Tadhunter}, {Ramos Almeida},
  {Rodr{\'\i}guez Zaur{\'\i}n}, {Santoro}  \& {Spence}}{{Rose}
  et~al.}{2018}]{Rose18}
{Rose} M.,  {Tadhunter} C.,  {Ramos Almeida} C.,  {Rodr{\'\i}guez Zaur{\'\i}n}
  J.,  {Santoro} F.,   {Spence} R.,  2018, \mn@doi [\mnras]
  {10.1093/mnras/stx2590}, \href
  {https://ui.adsabs.harvard.edu/abs/2018MNRAS.474..128R} {474, 128}

\bibitem[\protect\citeauthoryear{{Rupke}, {G{\"u}ltekin}  \&
  {Veilleux}}{{Rupke} et~al.}{2017}]{Rupke17}
{Rupke} D. S.~N.,  {G{\"u}ltekin} K.,   {Veilleux} S.,  2017, \mn@doi [\apj]
  {10.3847/1538-4357/aa94d1}, \href
  {https://ui.adsabs.harvard.edu/abs/2017ApJ...850...40R} {850, 40}

\bibitem[\protect\citeauthoryear{{Scholtz} et~al.,}{{Scholtz}
  et~al.}{2018}]{Scholtz18}
{Scholtz} J.,  et~al., 2018, \mn@doi [\mnras] {10.1093/mnras/stx3177}, \href
  {http://adsabs.harvard.edu/abs/2018MNRAS.475.1288S} {475, 1288}

\bibitem[\protect\citeauthoryear{{Scholtz} et~al.,}{{Scholtz}
  et~al.}{2020}]{Scholtz20}
{Scholtz} J.,  et~al., 2020, \mn@doi [\mnras] {10.1093/mnras/staa030}, \href
  {https://ui.adsabs.harvard.edu/abs/2020MNRAS.492.3194S} {492, 3194}

\bibitem[\protect\citeauthoryear{{Schreiber} et~al.,}{{Schreiber}
  et~al.}{2015}]{Schreiber15}
{Schreiber} C.,  et~al., 2015, \mn@doi [\aap] {10.1051/0004-6361/201425017},
  \href {http://adsabs.harvard.edu/abs/2015A%26A...575A..74S} {575, A74}

\bibitem[\protect\citeauthoryear{{Schulze} et~al.,}{{Schulze}
  et~al.}{2019}]{Schulze19}
{Schulze} A.,  et~al., 2019, \mn@doi [\mnras] {10.1093/mnras/stz1746}, \href
  {https://ui.adsabs.harvard.edu/abs/2019MNRAS.tmp.1696S} {p.~1696}

\bibitem[\protect\citeauthoryear{Schwarz}{Schwarz}{1978}]{Schwarz78}
Schwarz G.,  1978, Ann. Statist., 6, 461

\bibitem[\protect\citeauthoryear{{Segers}, {Schaye}, {Bower}, {Crain},
  {Schaller}  \& {Theuns}}{{Segers} et~al.}{2016}]{Segers16}
{Segers} M.~C.,  {Schaye} J.,  {Bower} R.~G.,  {Crain} R.~A.,  {Schaller} M.,
  {Theuns} T.,  2016, \mn@doi [\mnras] {10.1093/mnrasl/slw111}, \href
  {https://ui.adsabs.harvard.edu/abs/2016MNRAS.461L.102S} {461, L102}

\bibitem[\protect\citeauthoryear{{Shanks} et~al.,}{{Shanks}
  et~al.}{2015}]{Shanks15}
{Shanks} T.,  et~al., 2015, \mn@doi [\mnras] {10.1093/mnras/stv1130}, \href
  {https://ui.adsabs.harvard.edu/abs/2015MNRAS.451.4238S} {451, 4238}

\bibitem[\protect\citeauthoryear{{Shemmer}, {Netzer}, {Maiolino}, {Oliva},
  {Croom}, {Corbett}  \& {di Fabrizio}}{{Shemmer} et~al.}{2004}]{Shemmer04}
{Shemmer} O.,  {Netzer} H.,  {Maiolino} R.,  {Oliva} E.,  {Croom} S.,
  {Corbett} E.,   {di Fabrizio} L.,  2004, \mn@doi [\apj] {10.1086/423607},
  \href {https://ui.adsabs.harvard.edu/abs/2004ApJ...614..547S} {614, 547}

\bibitem[\protect\citeauthoryear{{Shin}, {Woo}, {Chung}, {Baek}, {Cho}, {Kang}
  \& {Bae}}{{Shin} et~al.}{2019}]{Shin19}
{Shin} J.,  {Woo} J.-H.,  {Chung} A.,  {Baek} J.,  {Cho} K.,  {Kang} D.,
  {Bae} H.-J.,  2019, arXiv e-prints, \href
  {https://ui.adsabs.harvard.edu/abs/2019arXiv190700982S} {p. arXiv:1907.00982}

\bibitem[\protect\citeauthoryear{{Silk} \& {Rees}}{{Silk} \&
  {Rees}}{1998}]{Silk98}
{Silk} J.,  {Rees} M.~J.,  1998, \aap, \href
  {https://ui.adsabs.harvard.edu/abs/1998A&A...331L...1S} {331, L1}

\bibitem[\protect\citeauthoryear{{Simpson} et~al.,}{{Simpson}
  et~al.}{2015}]{Simpson15}
{Simpson} J.~M.,  et~al., 2015, \mn@doi [\apj] {10.1088/0004-637X/807/2/128},
  \href {http://adsabs.harvard.edu/abs/2015ApJ...807..128S} {807, 128}

\bibitem[\protect\citeauthoryear{{Skrutskie} et~al.,}{{Skrutskie}
  et~al.}{2006}]{2MASS}
{Skrutskie} M.~F.,  et~al., 2006, \mn@doi [\aj] {10.1086/498708}, \href
  {https://ui.adsabs.harvard.edu/abs/2006AJ....131.1163S} {131, 1163}

\bibitem[\protect\citeauthoryear{{Slone} \& {Netzer}}{{Slone} \&
  {Netzer}}{2012}]{Slone12}
{Slone} O.,  {Netzer} H.,  2012, \mn@doi [\mnras]
  {10.1111/j.1365-2966.2012.21699.x}, \href
  {https://ui.adsabs.harvard.edu/abs/2012MNRAS.426..656S} {426, 656}

\bibitem[\protect\citeauthoryear{{Soltan}}{{Soltan}}{1982}]{Soltan82}
{Soltan} A.,  1982, \mn@doi [\mnras] {10.1093/mnras/200.1.115}, \href
  {https://ui.adsabs.harvard.edu/abs/1982MNRAS.200..115S} {200, 115}

\bibitem[\protect\citeauthoryear{{Spilker} et~al.,}{{Spilker}
  et~al.}{2016}]{Spilker16}
{Spilker} J.~S.,  et~al., 2016, \mn@doi [\apj] {10.3847/0004-637X/826/2/112},
  \href {http://adsabs.harvard.edu/abs/2016ApJ...826..112S} {826, 112}

\bibitem[\protect\citeauthoryear{{Stanley}, {Harrison}, {Alexander},
  {Swinbank}, {Aird}, {Del Moro}, {Hickox}  \& {Mullaney}}{{Stanley}
  et~al.}{2015}]{Stanley15}
{Stanley} F.,  {Harrison} C.~M.,  {Alexander} D.~M.,  {Swinbank} A.~M.,  {Aird}
  J.~A.,  {Del Moro} A.,  {Hickox} R.~C.,   {Mullaney} J.~R.,  2015, \mn@doi
  [\mnras] {10.1093/mnras/stv1678}, \href
  {http://adsabs.harvard.edu/abs/2015MNRAS.453..591S} {453, 591}

\bibitem[\protect\citeauthoryear{{Stanley} et~al.,}{{Stanley}
  et~al.}{2017}]{Stanley17}
{Stanley} F.,  et~al., 2017, \mn@doi [\mnras] {10.1093/mnras/stx2121}, \href
  {https://ui.adsabs.harvard.edu/abs/2017MNRAS.472.2221S} {472, 2221}

\bibitem[\protect\citeauthoryear{{Stanley}, {Harrison}, {Alexander}, {Simpson},
  {Knudsen}, {Mullaney}, {Rosario}  \& {Scholtz}}{{Stanley}
  et~al.}{2018}]{Stanley18}
{Stanley} F.,  {Harrison} C.~M.,  {Alexander} D.~M.,  {Simpson} J.,  {Knudsen}
  K.~K.,  {Mullaney} J.~R.,  {Rosario} D.~J.,   {Scholtz} J.,  2018, \mn@doi
  [\mnras] {10.1093/mnras/sty1044}, \href
  {http://adsabs.harvard.edu/abs/2018MNRAS.478.3721S} {478, 3721}

\bibitem[\protect\citeauthoryear{{Storchi-Bergmann}, {Lopes}, {McGregor},
  {Riffel}, {Beck}  \& {Martini}}{{Storchi-Bergmann}
  et~al.}{2010}]{StorchiBergmann10}
{Storchi-Bergmann} T.,  {Lopes} R.~D.~S.,  {McGregor} P.~J.,  {Riffel} R.~A.,
  {Beck} T.,   {Martini} P.,  2010, \mn@doi [\mnras]
  {10.1111/j.1365-2966.2009.15962.x}, \href
  {https://ui.adsabs.harvard.edu/abs/2010MNRAS.402..819S} {402, 819}

\bibitem[\protect\citeauthoryear{{Sturm} et~al.,}{{Sturm}
  et~al.}{2011}]{Sturm11}
{Sturm} E.,  et~al., 2011, \mn@doi [\apjl] {10.1088/2041-8205/733/1/L16}, \href
  {http://adsabs.harvard.edu/abs/2011ApJ...733L..16S} {733, L16}

\bibitem[\protect\citeauthoryear{{Tadaki} et~al.,}{{Tadaki}
  et~al.}{2017}]{Tadaki17}
{Tadaki} K.-i.,  et~al., 2017, \mn@doi [\apj] {10.3847/1538-4357/834/2/135},
  \href {http://adsabs.harvard.edu/abs/2017ApJ...834..135T} {834, 135}

\bibitem[\protect\citeauthoryear{{Tadhunter} et~al.,}{{Tadhunter}
  et~al.}{2018}]{Tadhunter18}
{Tadhunter} C.,  et~al., 2018, \mn@doi [\mnras] {10.1093/mnras/sty1064}, \href
  {https://ui.adsabs.harvard.edu/abs/2018MNRAS.478.1558T} {478, 1558}

\bibitem[\protect\citeauthoryear{{Tadhunter}, {Holden}, {Ramos Almeida}  \&
  {Batcheldor}}{{Tadhunter} et~al.}{2019}]{Tadhunter19}
{Tadhunter} C.,  {Holden} L.,  {Ramos Almeida} C.,   {Batcheldor} D.,  2019,
  \mn@doi [\mnras] {10.1093/mnras/stz1755}, \href
  {https://ui.adsabs.harvard.edu/abs/2019MNRAS.488.1813T} {488, 1813}

\bibitem[\protect\citeauthoryear{{Vayner}, {Wright}, {Murray}, {Armus},
  {Larkin}  \& {Mieda}}{{Vayner} et~al.}{2017}]{Vayner17}
{Vayner} A.,  {Wright} S.~A.,  {Murray} N.,  {Armus} L.,  {Larkin} J.~E.,
  {Mieda} E.,  2017, \mn@doi [\apj] {10.3847/1538-4357/aa9c42}, \href
  {https://ui.adsabs.harvard.edu/abs/2017ApJ...851..126V} {851, 126}

\bibitem[\protect\citeauthoryear{{Veilleux}, {Cecil}  \&
  {Bland-Hawthorn}}{{Veilleux} et~al.}{2005}]{Veilleux05}
{Veilleux} S.,  {Cecil} G.,   {Bland-Hawthorn} J.,  2005, \mn@doi [\araa]
  {10.1146/annurev.astro.43.072103.150610}, \href
  {http://adsabs.harvard.edu/abs/2005ARA%26A..43..769V} {43, 769}

\bibitem[\protect\citeauthoryear{{Veilleux} et~al.,}{{Veilleux}
  et~al.}{2013}]{Veilleux13}
{Veilleux} S.,  et~al., 2013, \mn@doi [\apj] {10.1088/0004-637X/776/1/27},
  \href {https://ui.adsabs.harvard.edu/abs/2013ApJ...776...27V} {776, 27}

\bibitem[\protect\citeauthoryear{{Veilleux}, {Maiolino}, {Bolatto}  \&
  {Aalto}}{{Veilleux} et~al.}{2020}]{Veilleux20}
{Veilleux} S.,  {Maiolino} R.,  {Bolatto} A.~D.,   {Aalto} S.,  2020, \mn@doi
  [\aapr] {10.1007/s00159-019-0121-9}, \href
  {https://ui.adsabs.harvard.edu/abs/2020A&ARv..28....2V} {28, 2}

\bibitem[\protect\citeauthoryear{{Venturi} et~al.,}{{Venturi}
  et~al.}{2018}]{Venturi18}
{Venturi} G.,  et~al., 2018, \mn@doi [\aap] {10.1051/0004-6361/201833668},
  \href {https://ui.adsabs.harvard.edu/abs/2018A&A...619A..74V} {619, A74}

\bibitem[\protect\citeauthoryear{{Vietri} et~al.,}{{Vietri}
  et~al.}{2020}]{Vietri20}
{Vietri} G.,  et~al., 2020, \mn@doi [\aap] {10.1051/0004-6361/202039136}, \href
  {https://ui.adsabs.harvard.edu/abs/2020A&A...644A.175V} {644, A175}

\bibitem[\protect\citeauthoryear{{Villar-Mart{\'\i}n}, {Arribas}, {Emonts},
  {Humphrey}, {Tadhunter}, {Bessiere}, {Cabrera Lavers}  \& {Ramos
  Almeida}}{{Villar-Mart{\'\i}n} et~al.}{2016}]{VillarMartin16}
{Villar-Mart{\'\i}n} M.,  {Arribas} S.,  {Emonts} B.,  {Humphrey} A.,
  {Tadhunter} C.,  {Bessiere} P.,  {Cabrera Lavers} A.,   {Ramos Almeida} C.,
  2016, \mn@doi [\mnras] {10.1093/mnras/stw901}, \href
  {https://ui.adsabs.harvard.edu/abs/2016MNRAS.460..130V} {460, 130}

\bibitem[\protect\citeauthoryear{{Vogelsberger} et~al.,}{{Vogelsberger}
  et~al.}{2014}]{Vogelsberger14}
{Vogelsberger} M.,  et~al., 2014, \mn@doi [\mnras] {10.1093/mnras/stu1536},
  \href {http://adsabs.harvard.edu/abs/2014MNRAS.444.1518V} {444, 1518}

\bibitem[\protect\citeauthoryear{{Wagner}, {Umemura}  \& {Bicknell}}{{Wagner}
  et~al.}{2013}]{Wagner13}
{Wagner} A.~Y.,  {Umemura} M.,   {Bicknell} G.~V.,  2013, \mn@doi [\apjl]
  {10.1088/2041-8205/763/1/L18}, \href
  {https://ui.adsabs.harvard.edu/abs/2013ApJ...763L..18W} {763, L18}

\bibitem[\protect\citeauthoryear{{Whitaker}, {van Dokkum}, {Brammer}  \&
  {Franx}}{{Whitaker} et~al.}{2012}]{Whitaker12}
{Whitaker} K.~E.,  {van Dokkum} P.~G.,  {Brammer} G.,   {Franx} M.,  2012,
  \mn@doi [\apjl] {10.1088/2041-8205/754/2/L29}, \href
  {http://adsabs.harvard.edu/abs/2012ApJ...754L..29W} {754, L29}

\bibitem[\protect\citeauthoryear{{Whitaker} et~al.,}{{Whitaker}
  et~al.}{2014}]{Whitaker14}
{Whitaker} K.~E.,  et~al., 2014, \mn@doi [\apj] {10.1088/0004-637X/795/2/104},
  \href {http://adsabs.harvard.edu/abs/2014ApJ...795..104W} {795, 104}

\bibitem[\protect\citeauthoryear{{Williams}, {Maiolino}, {Krongold},
  {Carniani}, {Cresci}, {Mannucci}  \& {Marconi}}{{Williams}
  et~al.}{2017}]{Williams17}
{Williams} R.~J.,  {Maiolino} R.,  {Krongold} Y.,  {Carniani} S.,  {Cresci} G.,
   {Mannucci} F.,   {Marconi} A.,  2017, \mn@doi [\mnras]
  {10.1093/mnras/stx311}, \href
  {https://ui.adsabs.harvard.edu/abs/2017MNRAS.467.3399W} {467, 3399}

\bibitem[\protect\citeauthoryear{{Woo}, {Bae}, {Son}  \& {Karouzos}}{{Woo}
  et~al.}{2016}]{Woo16}
{Woo} J.-H.,  {Bae} H.-J.,  {Son} D.,   {Karouzos} M.,  2016, \mn@doi [\apj]
  {10.3847/0004-637X/817/2/108}, \href
  {https://ui.adsabs.harvard.edu/abs/2016ApJ...817..108W} {817, 108}

\bibitem[\protect\citeauthoryear{{Woo}, {Son}  \& {Bae}}{{Woo}
  et~al.}{2017}]{Woo17}
{Woo} J.-H.,  {Son} D.,   {Bae} H.-J.,  2017, \mn@doi [\apj]
  {10.3847/1538-4357/aa6894}, \href
  {https://ui.adsabs.harvard.edu/abs/2017ApJ...839..120W} {839, 120}

\bibitem[\protect\citeauthoryear{{Woo}, {Son}  \& {Rakshit}}{{Woo}
  et~al.}{2020}]{Woo20}
{Woo} J.-H.,  {Son} D.,   {Rakshit} S.,  2020, arXiv e-prints, \href
  {https://ui.adsabs.harvard.edu/abs/2020arXiv200804919W} {p. arXiv:2008.04919}

\bibitem[\protect\citeauthoryear{{Wylezalek} \& {Zakamska}}{{Wylezalek} \&
  {Zakamska}}{2016}]{Wylezalek16}
{Wylezalek} D.,  {Zakamska} N.~L.,  2016, \mn@doi [\mnras]
  {10.1093/mnras/stw1557}, \href
  {https://ui.adsabs.harvard.edu/abs/2016MNRAS.461.3724W} {461, 3724}

\bibitem[\protect\citeauthoryear{{Zakamska} \& {Greene}}{{Zakamska} \&
  {Greene}}{2014}]{Zakamska14}
{Zakamska} N.~L.,  {Greene} J.~E.,  2014, \mn@doi [\mnras]
  {10.1093/mnras/stu842}, \href
  {https://ui.adsabs.harvard.edu/abs/2014MNRAS.442..784Z} {442, 784}

\bibitem[\protect\citeauthoryear{{Zakamska} et~al.,}{{Zakamska}
  et~al.}{2016}]{Zakamska16}
{Zakamska} N.~L.,  et~al., 2016, \mn@doi [\mnras] {10.1093/mnras/stw718}, \href
  {https://ui.adsabs.harvard.edu/abs/2016MNRAS.459.3144Z} {459, 3144}

\bibitem[\protect\citeauthoryear{{Zubovas} \& {Bourne}}{{Zubovas} \&
  {Bourne}}{2017}]{Zubovas17}
{Zubovas} K.,  {Bourne} M.~A.,  2017, \mn@doi [\mnras] {10.1093/mnras/stx787},
  \href {https://ui.adsabs.harvard.edu/abs/2017MNRAS.468.4956Z} {468, 4956}

\bibitem[\protect\citeauthoryear{{Zubovas} \& {King}}{{Zubovas} \&
  {King}}{2012}]{Zubovas12}
{Zubovas} K.,  {King} A.,  2012, \mn@doi [\apjl] {10.1088/2041-8205/745/2/L34},
  \href {http://adsabs.harvard.edu/abs/2012ApJ...745L..34Z} {745, L34}

\makeatother
\end{thebibliography}

\appendix

\section{Verifying SED fitting procedure}\label{sec:app:SED_ver}

\begin{figure}
    \centering
    \includegraphics[width=0.99\columnwidth]{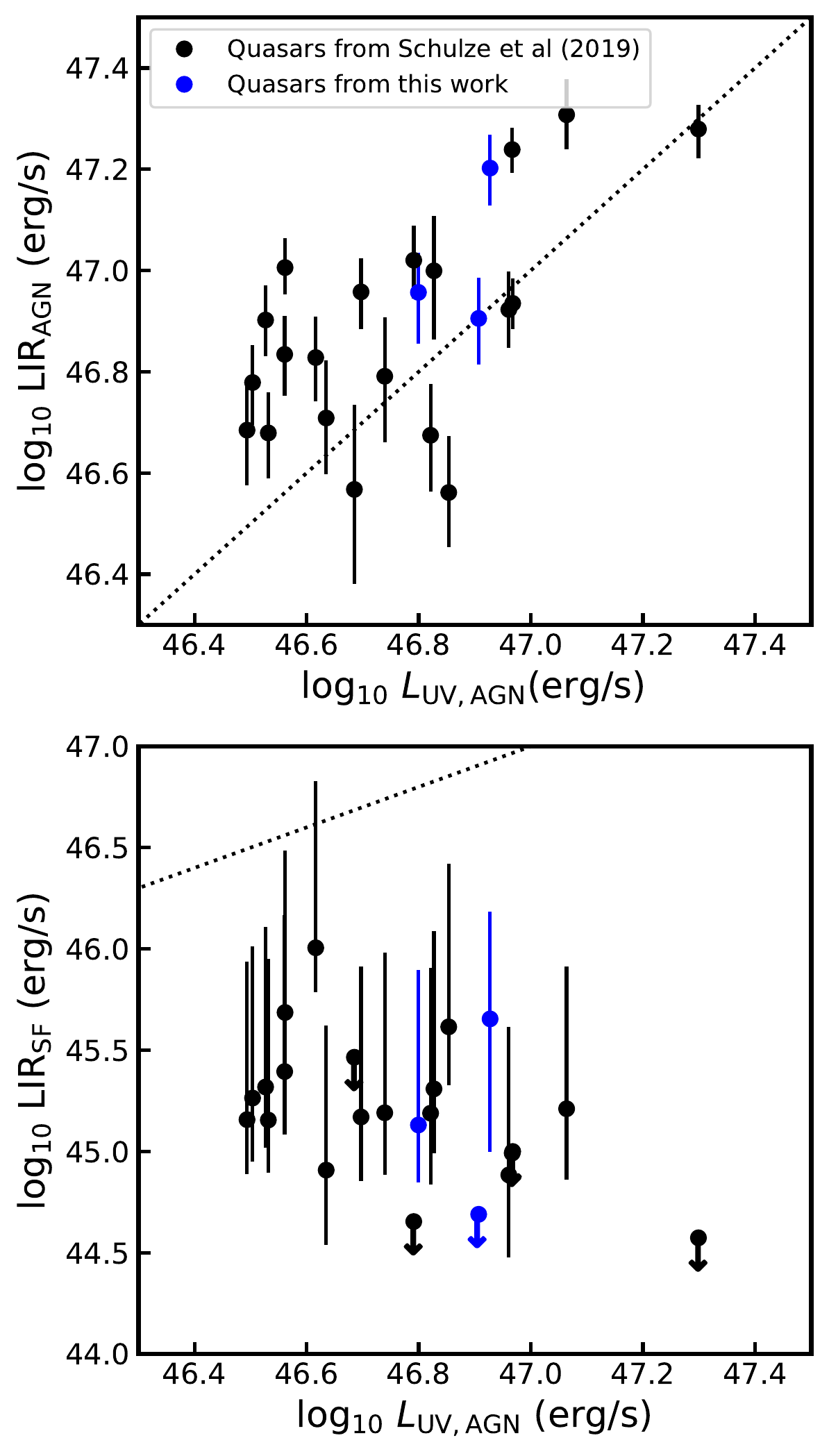}
   \caption{Top panel: Comparison of infra-red luminosity due to AGN ($\mathrel{\rm L_{IR,AGN}}$) versus UV luminosity due to AGN ($L_{\rm UV, AGN}$). Bottom Panel: Comparison of infra-red luminosity due to SF ($\mathrel{\rm L_{IR,SF}}$) versus UV luminosity due to AGN ($L_{\rm UV, AGN}$). The black and blue points represent quasars from \citet{Schulze19} and this work, respectively. The dotted black line in both panels indicate 1:1 ratio. We re-fitted the quasars from \citet{Schulze19} using method described in \S\ref{sec:SED}. While there is a strong correlation between $L_{\rm UV, AGN}$ and $\mathrel{\rm L_{IR,AGN}}$, we see no correlation between $\mathrel{\rm L_{IR,SF}}$ and $L_{\rm UV, AGN}$. This indicates that our SED code correctly removes AGN contribution to the infra-red luminosity. 
   }
   \label{fig:SED_test}
\end{figure}

To further verify the validity of our SED fitting we have explored the derived correlation between FIR luminosity from star formation ($L_{\rm IR,SF}$), FIR in luminosity from AGN ($L_{\rm IR,AGN}$ and UV luminosity from AGN ($L_{\rm UV, AGN}$) in Figure \ref{fig:SED_test}. To provide a larger sample we additionally fitted the photometry of a sample of 20 quasars from \citet{Schulze19} as they are in a similar redshift and bolometric luminosity range as our targets and include ALMA band 7 observations (see \S \ref{sec:SFR} for a discussion on the derived star-formation rates). We find a strong correlation between $L_{\rm IR,AGN}$ and $L_{\rm UV, AGN}$ while a very weak correlation between $L_{\rm IR,SF}$ and $L_{\rm UV, AGN}$. This gives confidence that the star formation component is independent of the AGN luminosity and we are not attributing emission from the AGN to the star formation component at infra-red bands. 

\begin{figure}
    \centering
    \includegraphics[width=0.99\columnwidth]{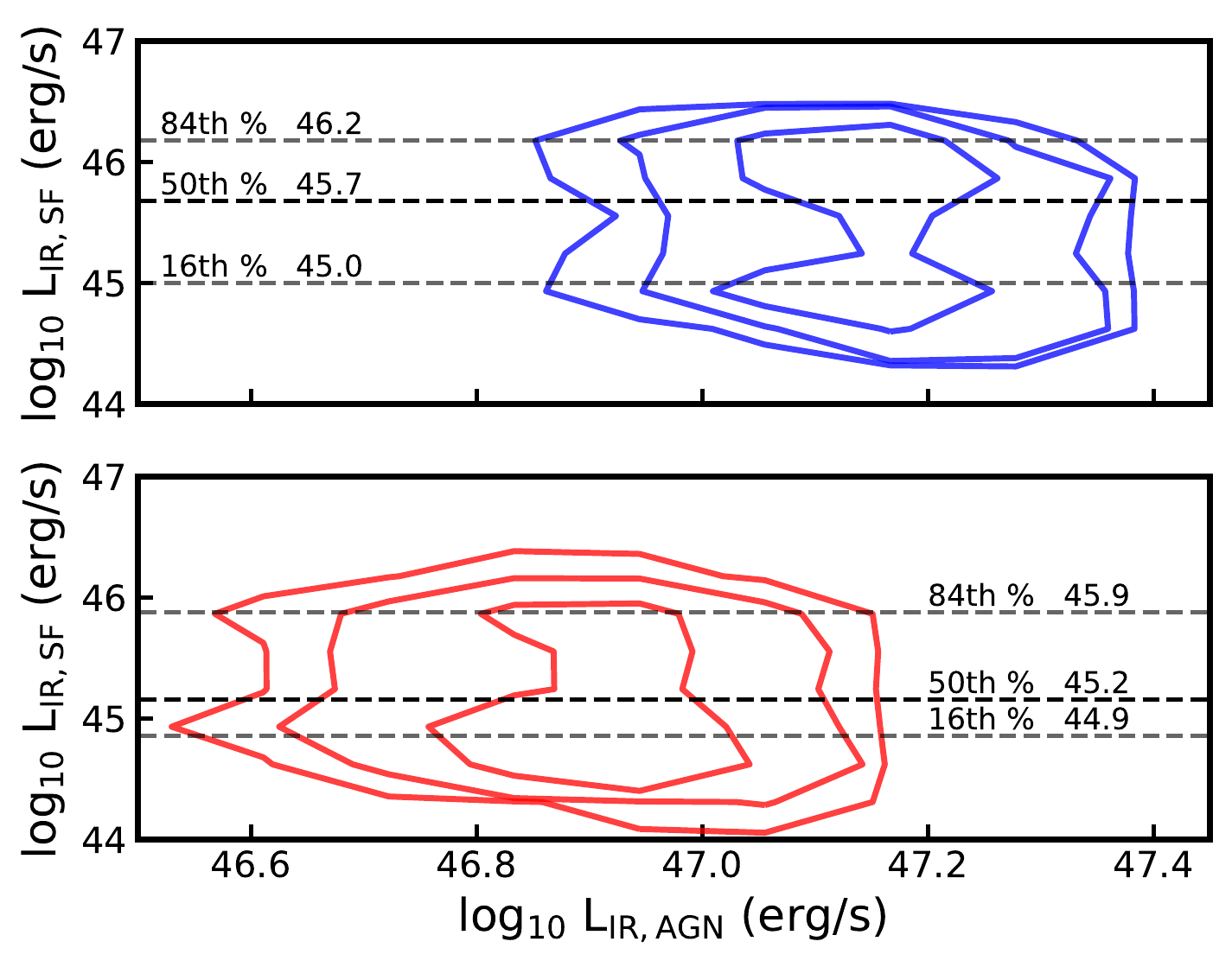}
   \caption{Posterior distribution for $\mathrel{\rm L_{IR,SF}}$ and $\mathrel{\rm L_{IR,AGN}}$ for LBQS01 (top panel) and 2QZJ00 (bottom panel). The dashed lines indicate 16$^{\rm th}$, 50$^{\rm th}$ and 84$^{\rm th}$
   for the $\mathrel{\rm L_{IR,SF}}$ parameter. The contours indicate 1, 2 and 3 $\sigma$ values. With the available photometry we are able to constrain the $\mathrel{\rm L_{IR,SF}}$ within 0.5 dex.}
   \label{fig:PDFs}
\end{figure}

We show the posterior density functions (PDFs) for the $L_{\rm IR,SF}$ and $L_{\rm IR, AGN}$ in Figure \ref{fig:PDFs} for 2QZJ00 and LBQS01. With the mid-IR and ALMA band 7 and 3 photometry we are able to constrain the $L_{\rm IR,SF}$ to 0.5 dex. However, we stress that these PDFs show uncertainties on the $L_{\rm IR,SF}$ and $L_{\rm IR, AGN}$ and not on the hot dust contamination to ALMA band 7 photometry, as described in \S~\ref{sec:SED}.

\section{A detailed assessment of the narrow H$\alpha$ emission}
In this Appendix we perform a series of additional analyses to those already presented in \S~\ref{sec:EL_analyses} to assess the distribution of the narrow H$\alpha$ emission in the IFU data cubes and then compare our results to the previous work that presented these same IFU data. Firstly we consider the total narrow H$\alpha$ emission, i.e.,  all of the emission after subtracting the broad-line region (see Section~\ref{sec:Gal1D}) and assess how reliably we can map this emission (Sections B1--B3). We then assess if it is possible to de-compose the narrow H$\alpha$ emission into multiple components associated with an AGN NLR, any outflows and any star formation emission (Section B4).

\subsection{BLR-subtraction method for mapping narrow H$\alpha$ emission}\label{sec:app:QSO-sub}

Although the multi-component fit of the spatial spectra is our preferred method of creating the surface brightness maps of the narrow H$\alpha$ emission (\S~\ref{sec:EL_analyses}), we repeat with the alternative method of first subtracting the broad-line region and looking for residuals \citep[e.g.,][see \S~\ref{sec:mutli-fit}]{Cresci15}. We refer to this method as the BLR-subtraction method. This method consists of subtracting the BLR and continuum emission from the data cube first, followed by creating a pseudo narrow-band image of the (residual) narrow H$\alpha$ emission.  

Before we fit and subtract the broad line H$\alpha$ emission, we masked the spectral regions containing the narrow H$\alpha$, [N~{\sc ii}] and [S~{\sc ii}] doublets, determined from the spectra show in Figure~\ref{fig:QSO_Halpha_spec} (\S~\ref{sec:Gal1D}). This ensures that the narrow H$\alpha$ and [N~{\sc ii}] line-emission do not artificially boost the broad line H$\alpha$ fit, which would eventually lead to over-subtracting the broad-line  H$\alpha$ emission from the cube. We perform the spaxel-by-spaxel fitting, by only fitting the broad line and continuum models (see \S\,\ref{sec:Gal1D} for more details). We fix the central wavelength and line-width to the same values as obtained in the nuclear spectrum. We then subtract the model broad line H$\alpha$ emission and continuum model, spaxel by spaxel, from the cube to create a residual cube, only containing the narrow emission lines (i.e., the narrow H$\alpha$, [N~{\sc ii}] doublet and [S~{\sc ii}] doublet).

From this residual cube we created a nuclear spectrum from following \S\ref{sec:Gal1D} and fitted this residual total spectrum with models describing only the narrow emission lines. We note that the fluxes of the narrow H$\alpha$ from the nuclear spectra using the original and the residual cubes are consistent within 10\%, showing that our BLR and continuum subtraction performed correctly. We then collapsed the cube along the spectral channels centred on the peak wavelength of the narrow H$\alpha$ emission with a width equivalent to the FWHM of the emission-line profile. We present this pseudo narrow band H$\alpha$ image in the second column of Figure~\ref{fig:Halpha_COG}.  The observed morphology of the narrow H$\alpha$ emission, which is dominated by AGN emission (see Section~\ref{sec:labelling}), is consistent with that seen in our primary `multi-fit' method (\S~\ref{sec:mutli-fit}).

\subsection{Modelling the curves-of-growth of narrow H$\alpha$ emission}\label{sec:app:Hal_sizes}

\begin{figure*}
    \centering
    \includegraphics[width=0.80\paperwidth]{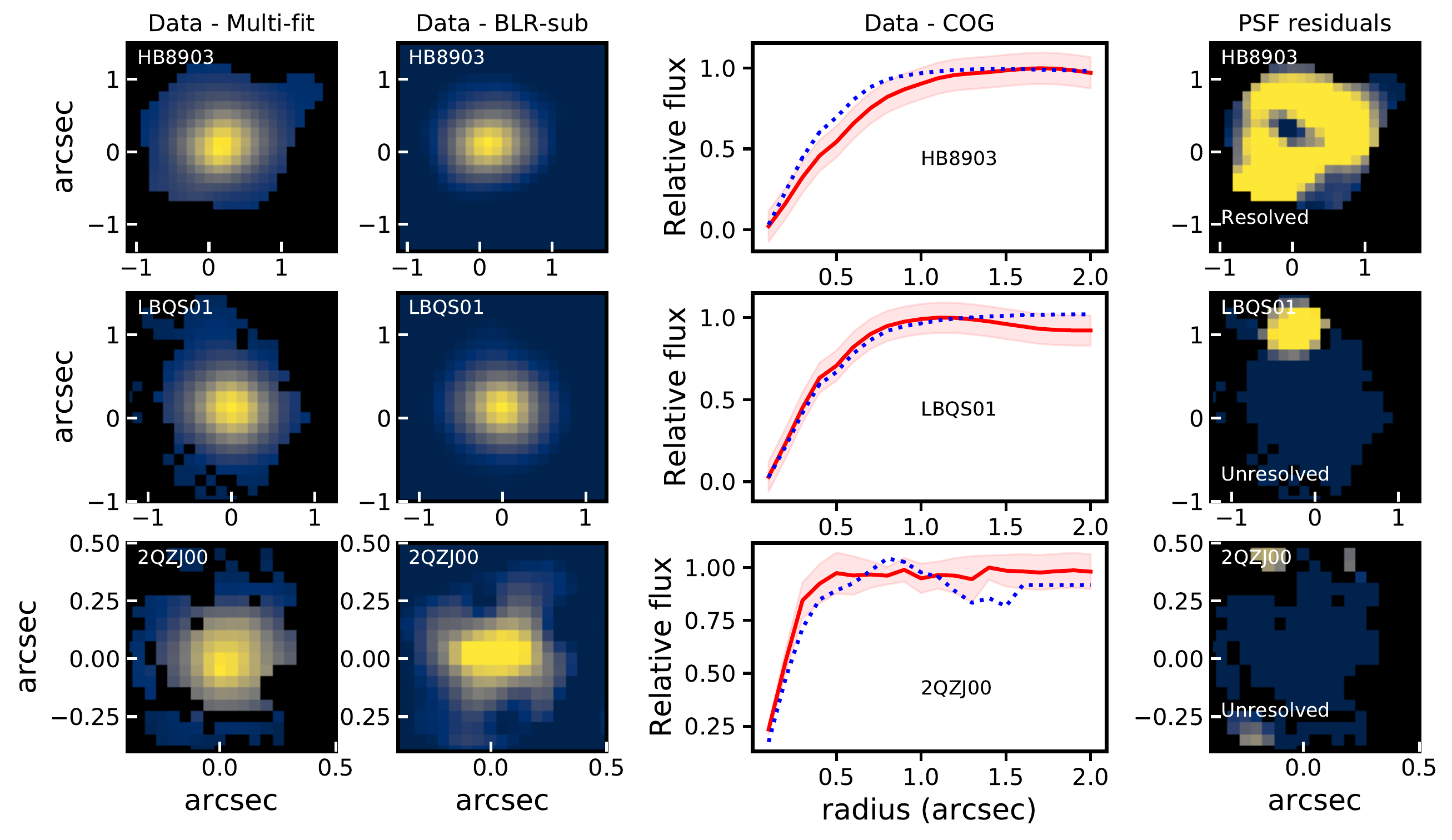}
   \caption{Summary of our analyses on the H$\alpha$ data and modelling the COG for different narrow H$\alpha$ morphologies. Columns from left to right: 
   Column 1: The narrow H$\alpha$ maps from the multi-fit method (see \S\,\ref{sec:mutli-fit}).
   Column 2: The narrow H$\alpha$ maps from the BLR-subtraction method (see \S\,\ref{sec:mutli-fit}). The scaling for the maps is set between 3$\sigma$ and maximum of the map.
   Column 3: The curves of growth for the H$\alpha$ emission. The solid red curve shows the COG for the narrow H$\alpha$ with the shaded region indicating the 1 $\sigma$ uncertainty. The dashed blue curves indicate the COG of the broad line region (BLR). We spatially resolve the narrow H$\alpha$ emission in HB8903.
   Column 4: PSF subtracted narrow H$\alpha$ images. HB8903 shows bright residual emission indicating that this object is resolved in narrow H$\alpha$ emission. The central parts of LBQS01 and 2QZJ00 are unresolved. The scaling of the maps is set between 3-5 $\sigma$. 
   }
   \label{fig:Halpha_COG}
\end{figure*}

In this section, we used curve-of-growth (COG) analyses \citep[see e.g.,][]{Chen17, chen20, Scholtz20} as an additional test of the distribution of the total narrow H$\alpha$ emission. The COG method consists of measuring the total enclosed flux in a series of increasingly large apertures, centred on the central quasar position. We fitted the emission-line profiles with the models described in \S\,\ref{sec:Gal1D}. Similarly to the creation of the emission line maps (see \S\,\ref{sec:mutli-fit}), we lock the central wavelength and line-width of the broad line H$\alpha$ component. We measure both the narrow and broad line H$\alpha$ flux in each aperture. This provides another method for assessing the spatial distribution of the H$\alpha$ emission with higher SNRs than the spaxel-by-spaxel fitting described in \S~\ref{sec:mutli-fit}, albeit resulting in a one-dimensional analysis. The COG of the broad line H$\alpha$ is also used to measure the PSF of the observations. The measured fluxes in the individual curves-of-growth are normalised to the maximum flux measured, for both the narrow and BLR component. As a result if the normalised emission flux grows slower than the normalized flux of the PSF (BLR in this case), the emission is resolved.

The third column of Figure~\ref{fig:Halpha_COG} shows the comparison of the COG for the narrow H$\alpha$ emission (solid red curves) and broad line H$\alpha$ emission (blue dotted curves). To interpolate between the data points, we used linear splines and we then estimated the half-light radii, i.e., the radius containing 50 \% of the total flux. We calculated the objects {\em intrinsic} sizes of the emission line regions ($r_{e}$), by subtracting off the size of the PSF (measured from the broad line emission) in quadrature. Uncertainties on the final H$\alpha$ sizes are calculated by considering the full range of possible radii for the 1$\sigma$ range of fluxes at each radii. We spatially resolved narrow H$\alpha$ in only one of the quasars (HB8903; i.e., the PSF subtracted size was not consistent with zero for this target) with a size of $r_{e}=1.9\pm 0.8$ kpc (FWHM $4.3\pm1.8$ kpc). 

\subsection{Subtraction of a PSF model from the narrow H$\alpha$ maps}\label{sec:app:PSF-sub}

As a final investigation into the distribution of the total narrow H$\alpha$ emission, we test if the observed surface brightness maps (Figure~\ref{fig:Halpha_sum}) are spatially resolved by subtracting a PSF model. This compliments the one-dimensional COG analyses presented in the previous section by performing a similar test in two dimensions. 

We used the 2D Gaussian profile to the broad line region H$\alpha$ emission to characterise the PSF (see \S~\ref{sec:mutli-fit}). We then subtract the PSF normalised to the peak surface brightness of the narrow H$\alpha$ map. We present these PSF subtracted narrow H$\alpha$ maps in fourth column of Figure~\ref{fig:Halpha_COG}. For LBQS01, we see a small patch of residual emission to the north region, however, the centre (majority) part of the emission is not resolved. In 2QZJ00, we do not see any significant residual emission, confirming that the narrow H$\alpha$ emission for this target is not resolved. In contrast, we do see significant residual emission in the HB8903. This confirms our COG analyses (\S~\ref{sec:app:Hal_sizes}) that the narrow H$\alpha$ emission is spatially resolved for this one target. We note that the zero surface brightness emission at the centre of the residual map in Figure~\ref{fig:Halpha_COG}, is due to normalising the PSF to the peak of the narrow H$\alpha$ emission. This map should not be interpreted as an intrinsic distribution of the narrow H$\alpha$ emission, it merely demonstrates there is residual emission beyond that expected for a PSF.

We note that in LBQS01 there is faint, but well detected ($>5 \sigma$) residual emission 1 arcsecond north from the quasars location. However, the bulk of the emission is still unresolved, hence are classification of this objects as unresolved.

\subsection{Exploring more complex models for the narrow H$\alpha$ emission line profiles}\label{sec:app:models}

He we investigate more detailed modelling of the narrow H$\alpha$ emission-line profiles. So far in this Appendix we have treated the narrow H$\alpha$ emission as a single component and have not attempted to decompose for contributions from AGN and star formation. This is relevant when looking to physically interpret any H$\alpha$ emission as associated with star formation (see \S~\ref{sec:mutli-fit}). We attempted variety of different models to both the nuclear and ring spectra (see \S~\ref{sec:Gal1D} and  \S~\ref{sec:mutli-fit}) to try to decompose the AGN narrow line region, outflow and star formation components (see \citealt{Canodiaz12,Carniani16}). In each case we calculated BIC to assess if additional components were required (see \S \ref{sec:ALMA_analyses}). We fitted the following models to the narrow H$\alpha$ emission-line profiles:
\begin{enumerate}
    \item Two Gaussian components, where the second component is forced to be narrow H$\alpha$ (FWHM$<500$ km s$^{-1}$) to describe the potential star formation component proposed in the previous studies (see \citealt{Canodiaz12,Carniani16}). The fitted additional narrow component had negligible fluxes (i.e., $\sim 10^{-23}$ ergs s$^{-1}$ cm$^{-2}$ compared to the rms of $\sim 5^{-19}$ ergs s$^{-1}$ cm$^{-2}$) and are essentially not detected.
    
    \item We fitted an additional Gaussian component to each of H$\alpha$ and [N~{\sc ii}] to represent an outflow (as seen in [O~{\sc iii}]). The resulting AGN-driven outflow component had an unrealistic H$\alpha$/[N~{\sc ii}] ratio of 1000. We still obtained nonphysical flux ratios when we tied the velocity widths and velocity centroids to be the same as the outflow components seen in [O~{\sc iii}].
\end{enumerate}

In summary, we conclude that a single Gaussian component to describe the narrow H$\alpha$ emission is sufficient. Due to the degeneracies of the complex models it is not possible to obtain reliable constraints on any additional components.

\section{Regional [OIII] and H$\alpha$ spectra}
\begin{figure*}
    \includegraphics[width=0.9\paperwidth]{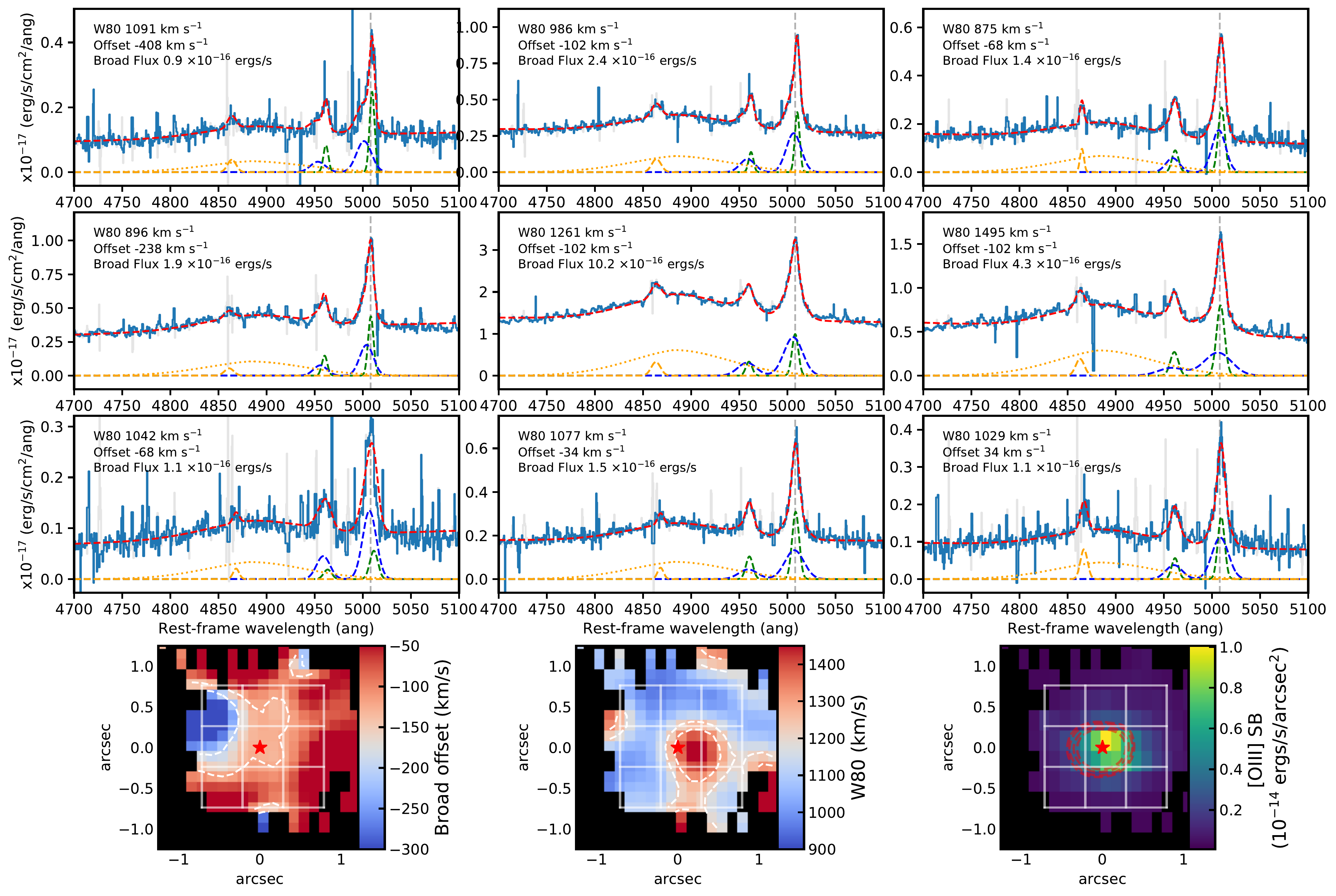}
   \caption{Analyses of the [O~{\sc iii}] emission line profiles for HB8903. 
   Bottom images: (Left) Map of the velocity of the broad component; (Middle:) Map of the W80 (velocity width containing 80 \% of the flux); (Right:) total [O~{\sc iii}] surface brightness map (SB). The red contours show the ALMA band 7 continuum images (2.5, 3, 4, 5 $\sigma$ levels). The red star indicates the centre of the AGN continuum. White grid indicates the regions from which we extracted spectra on top.
   Top spectra: [O~{\sc iii}] + H$\beta$ emission line profiles extracted from the white grids shown on the maps. The data is shown as blue solid line with grey lines indicating the spectral regions significantly affected by the sky-lines. The narrow and broad [O~{\sc iii}], H$\beta$ components and the total fit are shown as green, blue, orange and red lines. The grey vertical dashed line shows the velocity of the quasar. The W80 and velocity offset of [O~{\sc iii}] broad component are indicated on the spectra for easier comparison with the maps. The error on the velocity of the broad component and W80 is $\sim 80$ kms$^{-1}$ in both the maps and regional spectra. 
   }
   \label{fig:HB89_spec_OIII}
\end{figure*}

\begin{figure*}
    \includegraphics[width=0.9\paperwidth]{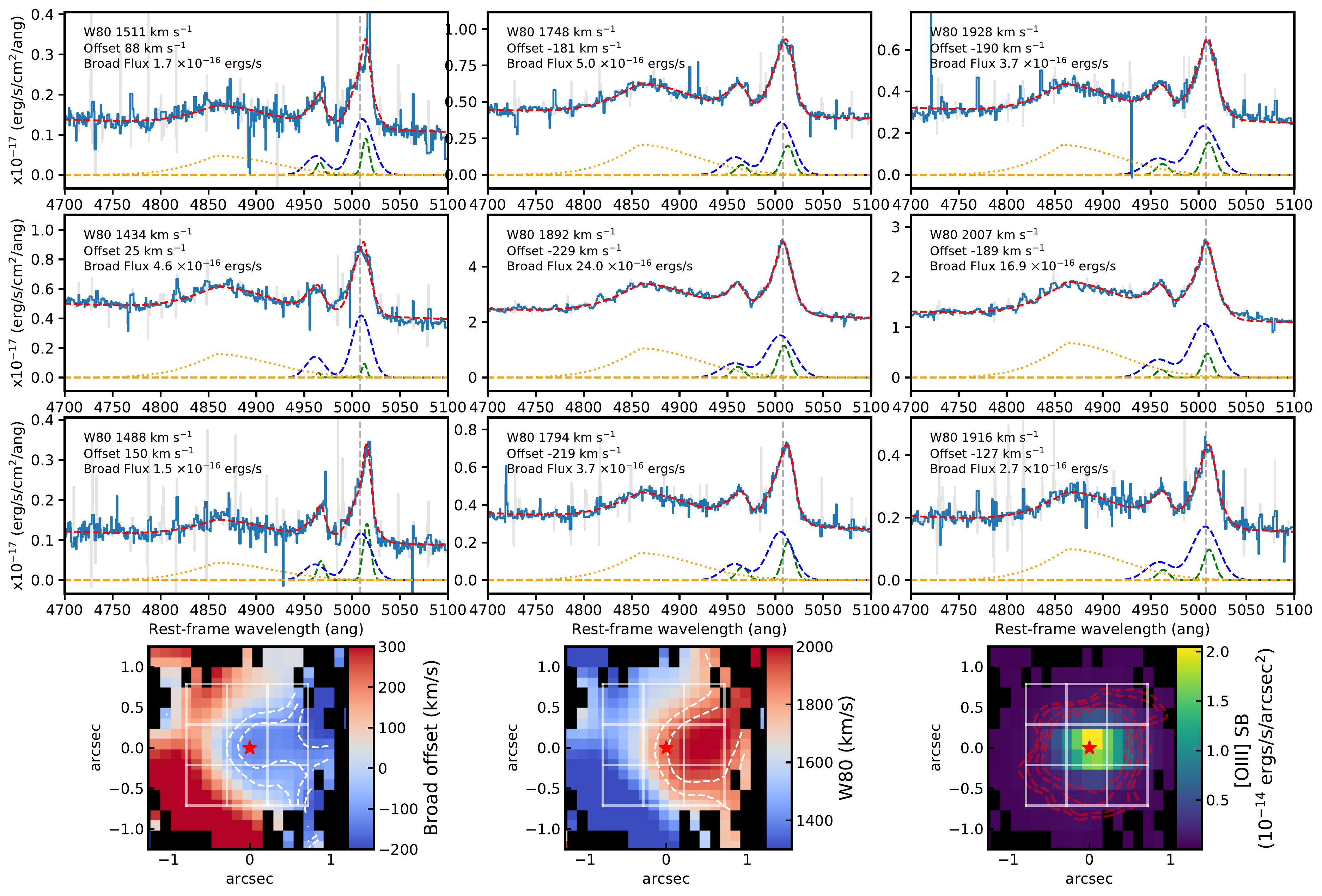}
   \caption{Analyses of the [O~{\sc iii}] emission line profiles for LBQS01. 
   Bottom images: (Left) Map of the velocity of the broad component; (Middle:) Map of the W80 (velocity width containing 80 \% of the flux); (Right:) total [O~{\sc iii}] surface brightness map (SB). The red contours show the ALMA band 7 continuum images (2.5, 3, 4, 5 $\sigma$ levels). The red star indicates the centre of the AGN continuum. White grid indicates the regions from which we extracted spectra on top.
   Top spectra: [O~{\sc iii}] + H$\beta$ emission line profiles extracted from the white grids shown on the maps. The data is shown as blue solid line with grey lines indicating the spectral regions significantly affected by the sky-lines. The narrow and broad [O~{\sc iii}], H$\beta$ components and the total fit are shown as green, blue, orange and red lines. The grey vertical dashed line shows the velocity of the quasar. The W80 and velocity offset of [O~{\sc iii}] broad component are indicated on the spectra for easier comparison with the maps. The error on the velocity of the broad component and W80 is $\sim 80$ kms$^{-1}$ in both the maps and regional spectra. 
   }
   \label{fig:LBQS_spec_OIII}
\end{figure*}

\begin{figure*}
    \includegraphics[width=0.9\paperwidth]{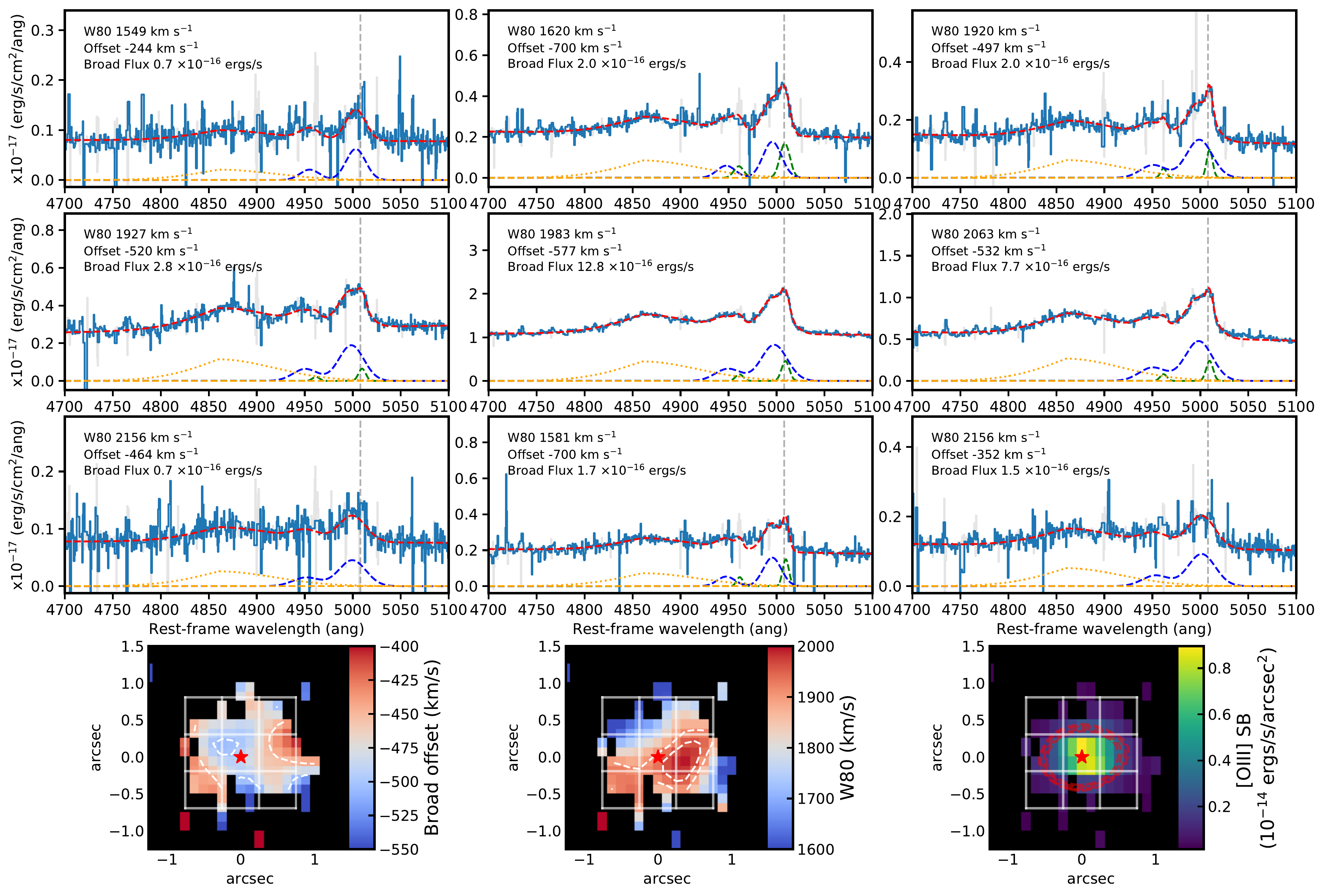}
   \caption{Analyses of the [O~{\sc iii}] emission line profiles for 2QZJ00. 
   Bottom images: (Left) Map of the velocity of the broad component; (Middle:) Map of the W80 (velocity width containing 80 \% of the flux); (Right:) total [O~{\sc iii}] surface brightness map (SB). The red contours show the ALMA band 7 continuum images (2.5, 3, 4, 5 $\sigma$ levels). The red star indicates the centre of the AGN continuum. White grid indicates the regions from which we extracted spectra on top.
   Top spectra: [O~{\sc iii}] + H$\beta$ emission line profiles extracted from the white grids shown on the maps. The data is shown as blue solid line with grey lines indicating the spectral regions significantly affected by the sky-lines. The narrow and broad [O~{\sc iii}], H$\beta$ components and the total fit are shown as green, blue, orange and red lines. The grey vertical dashed line shows the velocity of the quasar. The W80 and velocity offset of [O~{\sc iii}] broad component are indicated on the spectra for easier comparison with the maps. The error on the velocity of the broad component and W80 is $\sim 80$ kms$^{-1}$ in both the maps and regional spectra. 
   }
   \label{fig:2QZJ_spec_OIII}
\end{figure*}

\begin{figure*}
    \centering
    \includegraphics[width=0.9\paperwidth]{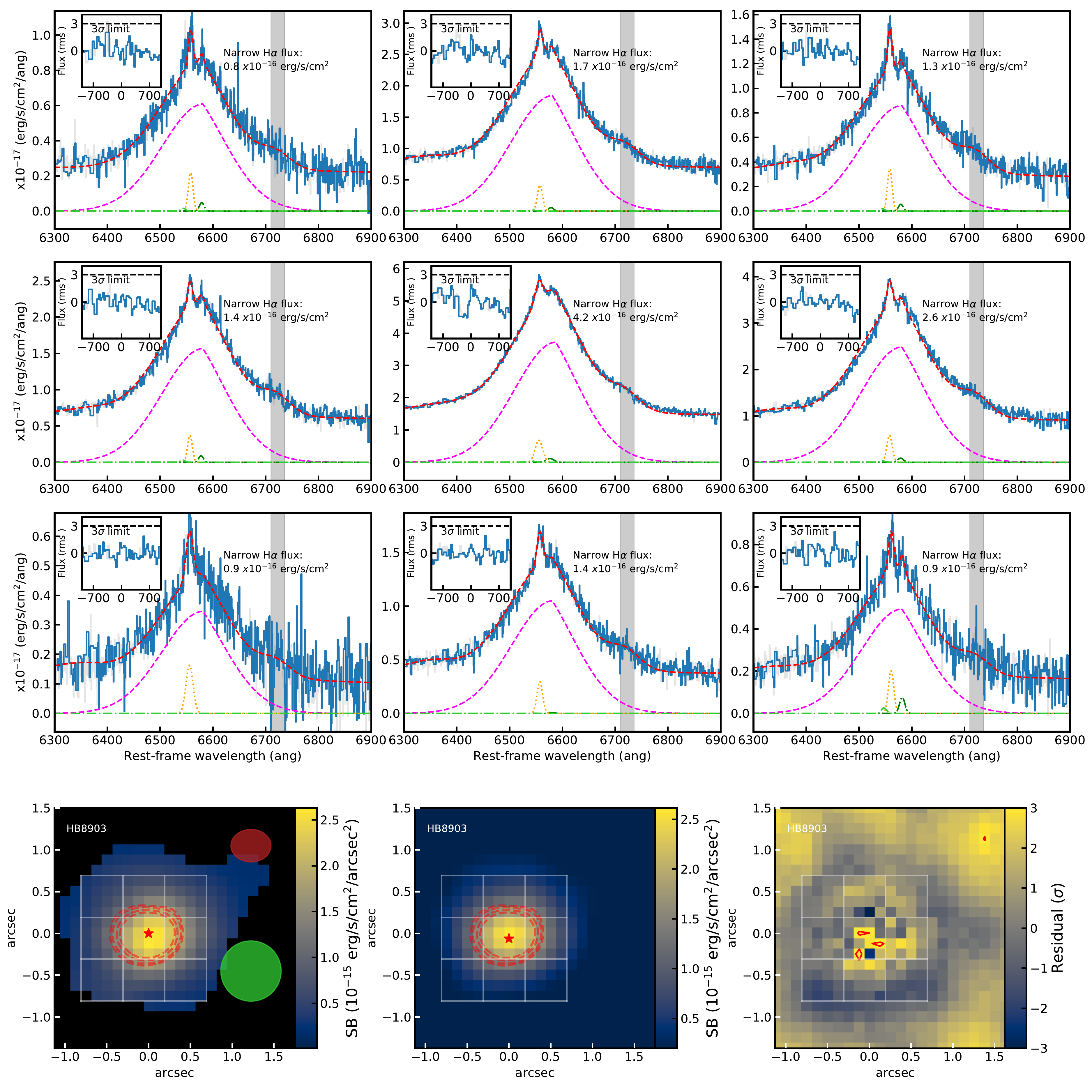}
    \centering
   \caption{Narrow H$\alpha$ (dominated by NLR emission) surface brightness (SB) maps and region spectra for HB8903 from our analyses.
   Bottom left image: Narrow H$\alpha$ image created by simultaneously modelling all the spaxel's spectrum components. Bottom center image: narrowband image of the narrow H$\alpha$ image created after subtracting the broad line H$\alpha$. The ALMA band 7 continuum data is displayed as red contours (2.5, 3, 4, 5 $\sigma$ levels). In the center image, the red circle indicates the ALMA PSF and the green circle shows the H$\alpha$ PSF determined from the observations determined from the broad line, while the white squares show regions from which we extracted spectra on the right. Bottom right image: Image form collapsing the residuals of the spaxel-by-spaxel fitting along spectral channels specified by \citet{Carniani16}. The red solid contours indicate any emission at 3, and 5 $\sigma$.
   Spectra on the top: H$\alpha$ and [N~{\sc ii}] spectra extracted from the regions corresponding to the white squares in the images on the left. We show the narrow H$\alpha$ flux in each subplot to allow comparison between the maps and the spectra. For the emission-line profiles, the curves refer to the same components as in Figure~\ref{fig:QSO_Halpha_spec}. The inset plot in each spectra shows the residual spectrum around the region were previous studies detected residual narrow H$\alpha$ emission.
   The grey shaded area shows the location of the [S~{\sc ii}]. 
   }
   \label{fig:HB89_spec}
\end{figure*}

\begin{figure*}
    \includegraphics[width=0.9\paperwidth]{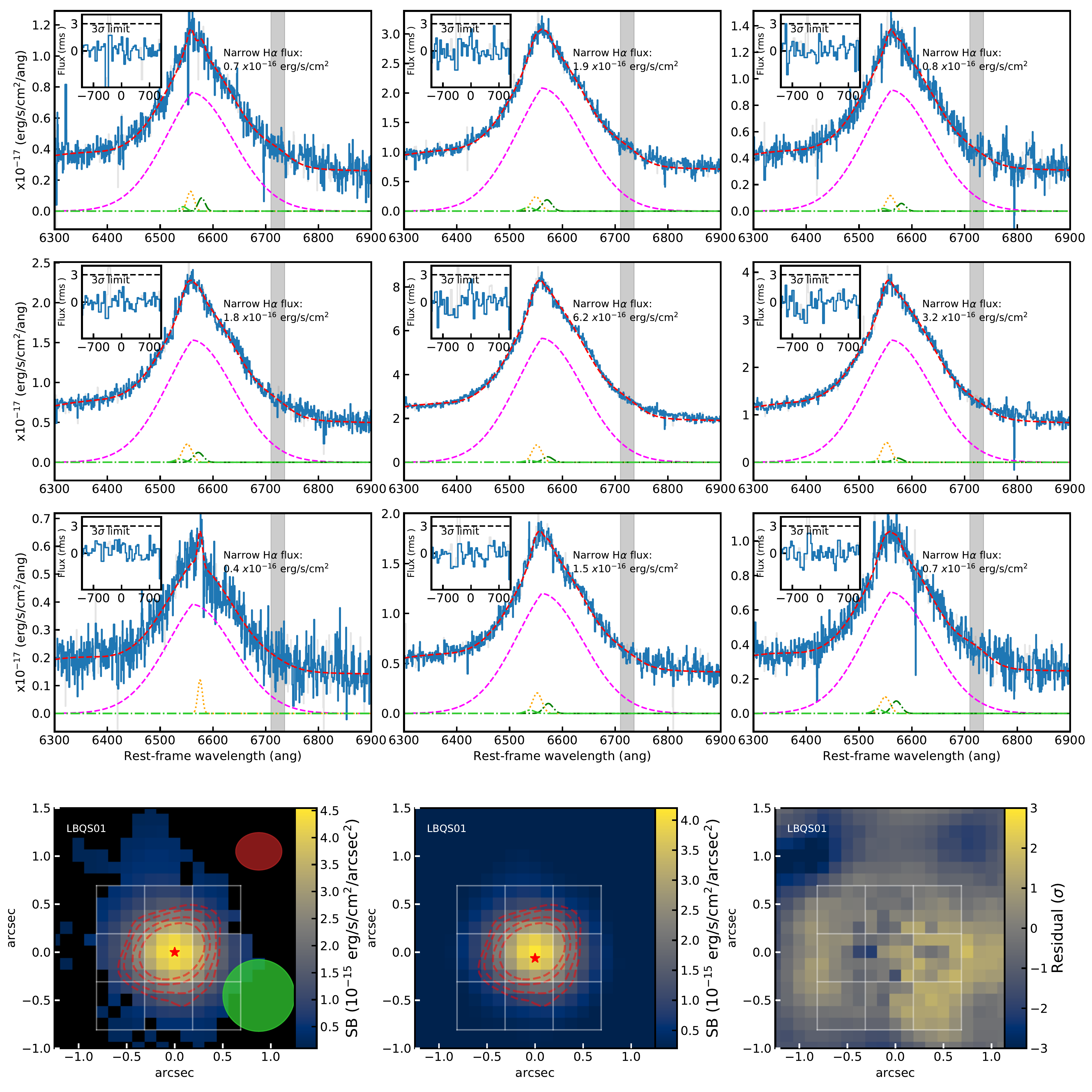}
   \caption{Narrow H$\alpha$ (dominated by NLR emission) surface brightness (SB) maps and region spectra for LBQS01 from our analyses.
   Bottom left image: Narrow H$\alpha$ image created by simultaneously modelling all the spaxel's spectrum components. Bottom center image: narrowband image of the narrow H$\alpha$ image created after subtracting the broad line H$\alpha$. The ALMA band 7 continuum data is displayed as red contours (2.5, 3, 4, 5 $\sigma$ levels). In the center image, the red circle indicates the ALMA PSF and the green circle shows the H$\alpha$ PSF determined from the observations determined from the broad line, while the white squares show regions from which we extracted spectra on the right. Bottom right image: Image form collapsing the residuals of the spaxel-by-spaxel fitting along spectral channels specified by \citet{Carniani16}. The red solid contours indicate any emission at 3, and 5 $\sigma$.
   Spectra on the top: H$\alpha$ and [N~{\sc ii}] spectra extracted from the regions corresponding to the white squares in the images on the left. We show the narrow H$\alpha$ flux in each subplot to allow comparison between the maps and the spectra. For the emission-line profiles, the curves refer to the same components as in Figure~\ref{fig:QSO_Halpha_spec}. The inset plot in each spectra shows the residual spectrum around the region were previous studies detected residual narrow H$\alpha$ emission.
   The grey shaded area shows the location of the [S~{\sc ii}]. 
   }
   \label{fig:LBQS_spec}
\end{figure*}

\begin{figure*}
    \includegraphics[width=0.9\paperwidth]{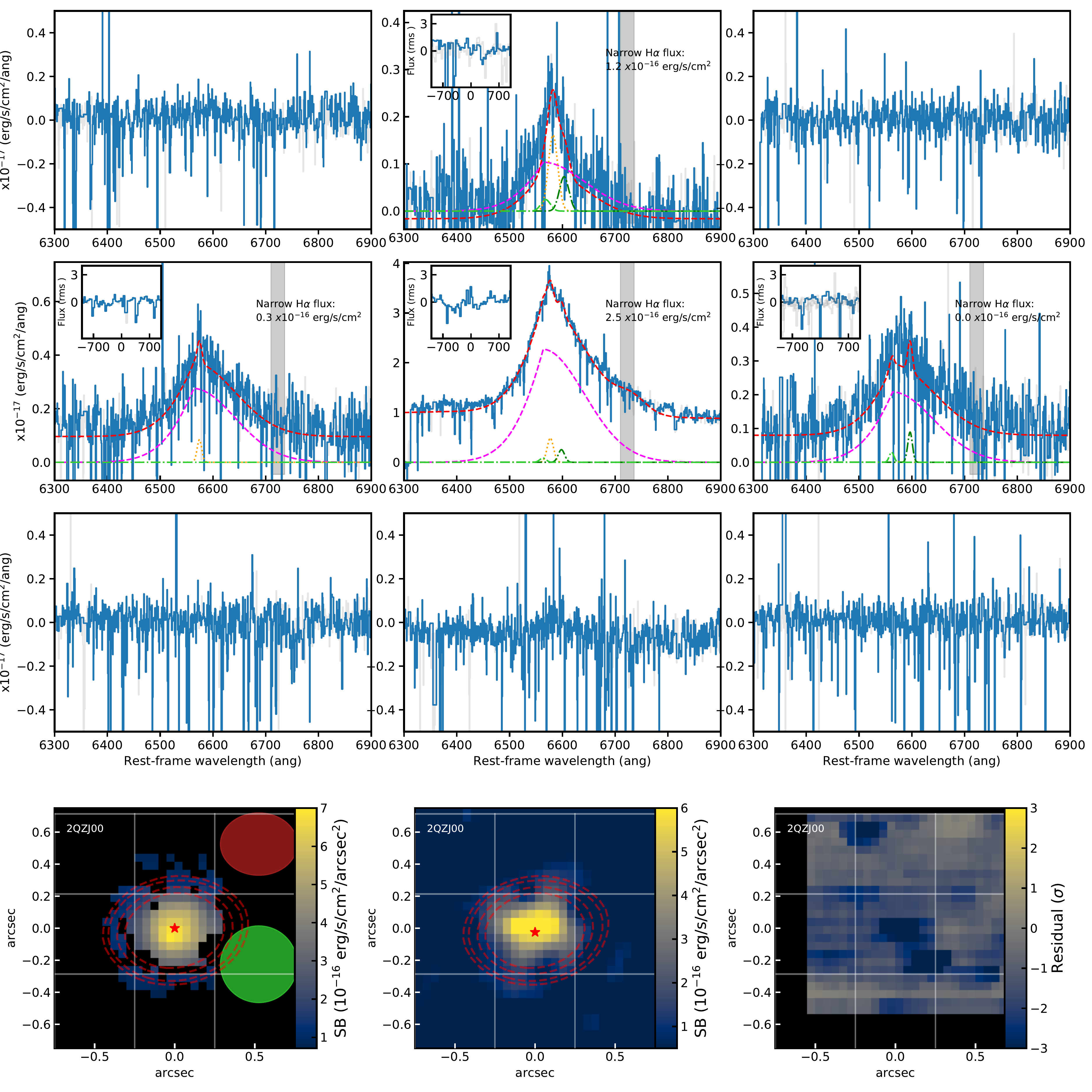}
   \caption{Narrow H$\alpha$ (dominated by NLR emission) surface brightness (SB) maps and region spectra for 2QZJ00 from our work.
   Bottom left image: Narrow H$\alpha$ image created by simultaneously modelling all the spaxel's spectrum components. Bottom center image: narrowband image of the narrow H$\alpha$ image created after subtracting the broad line H$\alpha$. The ALMA band 7 continuum data is displayed as red contours (2.5, 3, 4, 5 $\sigma$ levels). In the center image, the red circle indicates the ALMA PSF and the green circle shows the H$\alpha$ PSF determined from the observations determined from the broad line, while the white squares show regions from which we extracted spectra on the right. Bottom right image: Image form collapsing the residuals of the spaxel-by-spaxel fitting along spectral channels around the fitted narrow H$\alpha$ emission. The red solid contours indicate any emission at 3, and 5 $\sigma$.
   Spectra on the top: H$\alpha$ and [N~{\sc ii}] spectra extracted from the regions corresponding to the white squares in the images on the left. We show the narrow H$\alpha$ flux in each subplot to allow comparison between the maps and the spectra. For the emission-line profiles, the curves refer to the same components as in Figure~\ref{fig:QSO_Halpha_spec}. The inset plot in each spectra shows the residual spectrum around the region were previous studies detected residual narrow H$\alpha$ emission.
   The grey shaded area shows the location of the [S~{\sc ii}]. 
   }
   \label{fig:2QZJ_spec}
\end{figure*}

In this appendix we show the emission-line profiles, and our fits to these profiles, extracted from the spatial grids described in Section~\ref{sec:reg_spec}. In Figures~\ref{fig:HB89_spec_OIII}, \ref{fig:LBQS_spec_OIII} and  \ref{fig:2QZJ_spec_OIII} we show these for the [O~{\sc iii}] and H$\beta$ emission lines for targets HB8903, LBQS01 and 2QZJ00, respectively. In Figures~\ref{fig:HB89_spec}, \ref{fig:LBQS_spec} and  \ref{fig:2QZJ_spec} we show these for the H$\alpha$ and [N~{\sc ii}] emission lines for targets HB89, LBQS01 and 2QZJ00, respectively. 

\section{Photometry used in the SED fitting}\label{sec:app:photometry}
In this appendix, we provide a table of the photometry used to performed our SED analyses presented in \S~\ref{sec:SED}. These optical through to radio photometric magnitudes and flux density measurements are provided in Table~\ref{Table:Phot}.

\begin{landscape}
\begin{table}
\caption{A table of the photometry used in the SED fitting (see Section~\ref{sec:SED}). 
(1) ID of the object in this work; 
(2-6) VST-ATLAS u, g, r, i and z bands
(7-11) PanSTARRS g, r, i,z and bands
(12-14) 2MASS J, H and K bands
(15-18) WISE Band 1, 2, 3 and 4 bands
(19) ALMA Band 7 continuum flux;
(20) ALMA Band 3 continuum flux;
(21)--(23) 1.4\,GHz, 4.8\,GHz and 8.4\,GHz flux densities, respectively.
}
\centering
\resizebox{0.99\paperwidth}{!}{\begin{tabular}{@{}lcccccccccccccccccccccc@{}} 
\hline 
\hline 
(1) & (2)     & (3)     & (4)     & (5)     &  (6)    &    (7)   & (8) & (9) & (10) & (11) & (12) & (13) & (14) & (15) & (16) & (17) & (18) & (19) & (20) & (21)  & (22) & (23)        \\
ID  & ATLAS u & ATLAS g & ATLAS r & ATLAS i & ATLAS z & Pan g & Pan r & Pan i & Pan z & Pan y &     2MASS J & 2MASS H & 2MASS K       &Wise W1& Wise W2 & Wise W3 & Wise W4 & ALMA B7 & ALMA B3  & Radio (1.4 GHz) &  Radio (4.8 GHz) & Radio (8.4 GHz) \\
$\lambda_{\rm obs}$ &

0.365$\mu$m & 0.464$\mu$m & 0.658$\mu$m & 0.806$\mu$m & 0.900$\mu$m & 0.489$\mu$m & 0.620$\mu$m & 0.753$\mu$m& 0.867$\mu$m& 0.962$\mu$m & 1.2$\mu$m &1.6$\mu$m & 2.4$\mu$m  & 3.3$\mu$m &  4.6$\mu$m & 11.5$\mu$m & 22$\mu$m  & 870$\mu$m        &   3 mm   &    21 cm & 0.72 m & 1.44 m  \\
    & (mag) & (mag)   &  (mag)  & (mag) &  (mag) & (mag) & (mag)   &  (mag)  & (mag) &  (mag)  & (mag)   &  (mag)  & (mag) &  (mag)  & (mag)   &  (mag)  & (mag)   & (mJy)   & ($\mu$Jy)    &  (mJy)  & (mJy) & (mJy)  \\
\hline    
HB8903  & $17.88\pm 0.01$ & $17.34\pm 0.003$ & $17.44\pm 0.004$ & $17.400\pm 0.004$& $17.10\pm 0.01$ &- &- &-&-&- & $16.1\pm 0.1$  & $15.6\pm 0.1$  & $14.6\pm 0.1$ & 14.4$\pm$0.03 & $13.3\pm0.2$  & $9.8\pm 0.03$  & $7.8\pm 0.1$ & $1.53\pm 0.03$ & $5697 \pm 18 $& 29.8  & 148  & 107.6\\

LBQS01 &- &- &-&-&-& $17.71\pm 0.004$ & $17.31\pm 0.001$ &$17.26\pm 0.001$ &$17.12\pm 0.004$& $16.93\pm 0.005$ & $15.9\pm 0.07$ & $15.2\pm 0.1$  & $14.9\pm 0.1$ & 14.1$\pm$0.03 & $13.2\pm0.03$  & $9.7\pm0.05$     & $7.6\pm0.2$    & $1.57\pm 0.02$  & $170 \pm 14 $ & $< 1.0$& - & 0.31$\pm$0.06  \\

2QZJ00 &- &- &-&-&-& $17.65\pm 0.003$ & $17.57\pm 0.002$ & $17.65\pm 0.002$ &$17.47\pm 0.004$& $17.29\pm 0.007$ & $16.4\pm 0.10$ & $15.9\pm 0.1$  & $15.3\pm 0.1$ & 14.8$\pm$0.03 & $13.9\pm 0.04$ & $10.4 \pm 0.1$ & $8.2 \pm 0.3$ & $1.06 \pm 0.03$ & $163 \pm 16$ & $< 1.0$  & - &-\\

\hline 
\end{tabular}}

\label{Table:Phot}
\end{table}
\end{landscape}


\bsp    
\label{lastpage}
\end{document}